\documentclass[longauth]{aa}

\usepackage{natbib}
\usepackage[utf8]{inputenc}
\usepackage{amsmath}
\usepackage{siunitx}
\DeclareSIUnit\parsec{pc}
\usepackage{xspace}
\usepackage{graphicx}
\usepackage{lscape}
\usepackage{longtable}
\usepackage{color}

\newcommand{\ab}{\ensuremath{\text{ab}}\xspace}

\newcommand{\bdtruc}{BD~+17~4708\xspace}


\newcommand{\PaperI}{B13}
\newcommand{\ktreize}{K13}
\newcommand{\PaperII}{M14}

\newcommand{\mstar}{\ensuremath{m_{\text{B}}^\star}\xspace}

\newcommand{\sigmaint}{\ensuremath{\sigma_{\text{coh}}}\xspace}
\newcommand{\xun}{\ensuremath{X_1}\xspace}
\newcommand{\xzero}{\ensuremath{X_0}\xspace}
\newcommand{\col}{\ensuremath{\mathcal{C}}\xspace}

\newcommand{\calpar}{\ensuremath{{\vec{\kappa}}}\xspace}
\newcommand{\lcpar}{\ensuremath{{\vec{\eta}}}\xspace}
\newcommand{\muv}{\ensuremath{{\vec{\mu}}}\xspace}
\newcommand{\A}{\ensuremath{{\tens{A}}}\xspace}

\newcommand{\cref}[1][]{\ensuremath{{c_{\text{ref}#1}}}\xspace}

\newcommand{\zp}[1][]{\ensuremath{{Z_{#1}}}\xspace}

\DeclareMathOperator{\logdec}{\log_{10}}

\DeclareMathOperator{\diag}{diag}

\graphicspath{{fig/}}

\newcommand{\igraph}[2][]{\includegraphics[width=#1\linewidth]{#2}}

\newcommand{\LCDM}{\ensuremath{\Lambda}CDM\xspace}

\makeatletter
\let\jnl@style=\rm
\def\ref@jnl#1{{\jnl@style#1}}
\def\aj{\ref@jnl{AJ}}                   % Astronomical Journal
\def\araa{\ref@jnl{ARA\&A}}             % Annual Review of Astron and Astrophys
\def\apj{\ref@jnl{ApJ}}                 % Astrophysical Journal
\def\apjl{\ref@jnl{ApJ}}                % Astrophysical Journal, Letters
\def\apjs{\ref@jnl{ApJS}}               % Astrophysical Journal, Supplement
\def\ao{\ref@jnl{Appl.~Opt.}}           % Applied Optics
\def\apss{\ref@jnl{Ap\&SS}}             % Astrophysics and Space Science
\def\aap{\ref@jnl{A\&A}}                % Astronomy and Astrophysics
\def\aapr{\ref@jnl{A\&A~Rev.}}          % Astronomy and Astrophysics Reviews
\def\aaps{\ref@jnl{A\&AS}}              % Astronomy and Astrophysics, Supplement
\def\azh{\ref@jnl{AZh}}                 % Astronomicheskii Zhurnal
\def\baas{\ref@jnl{BAAS}}               % Bulletin of the AAS
\def\jrasc{\ref@jnl{JRASC}}             % Journal of the RAS of Canada
\def\memras{\ref@jnl{MmRAS}}            % Memoirs of the RAS
\def\mnras{\ref@jnl{MNRAS}}             % Monthly Notices of the RAS
\def\pra{\ref@jnl{Phys.~Rev.~A}}        % Physical Review A: General Physics
\def\prb{\ref@jnl{Phys.~Rev.~B}}        % Physical Review B: Solid State
\def\prc{\ref@jnl{Phys.~Rev.~C}}        % Physical Review C
\def\prd{\ref@jnl{Phys.~Rev.~D}}        % Physical Review D
\def\pre{\ref@jnl{Phys.~Rev.~E}}        % Physical Review E
\def\prl{\ref@jnl{Phys.~Rev.~Lett.}}    % Physical Review Letters
\def\pasp{\ref@jnl{PASP}}               % Publications of the ASP
\def\pasj{\ref@jnl{PASJ}}               % Publications of the ASJ
\def\qjras{\ref@jnl{QJRAS}}             % Quarterly Journal of the RAS
\def\skytel{\ref@jnl{S\&T}}             % Sky and Telescope
\def\solphys{\ref@jnl{Sol.~Phys.}}      % Solar Physics
\def\sovast{\ref@jnl{Soviet~Ast.}}      % Soviet Astronomy
\def\ssr{\ref@jnl{Space~Sci.~Rev.}}     % Space Science Reviews
\def\zap{\ref@jnl{ZAp}}                 % Zeitschrift fuer Astrophysik
\def\nat{\ref@jnl{Nature}}              % Nature
\def\iaucirc{\ref@jnl{IAU~Circ.}}       % IAU Cirulars
\def\aplett{\ref@jnl{Astrophys.~Lett.}} % Astrophysics Letters
\def\apspr{\ref@jnl{Astrophys.~Space~Phys.~Res.}}
\def\bain{\ref@jnl{Bull.~Astron.~Inst.~Netherlands}} 
\def\fcp{\ref@jnl{Fund.~Cosmic~Phys.}}  % Fundamental Cosmic Physics
\def\gca{\ref@jnl{Geochim.~Cosmochim.~Acta}}   % Geochimica Cosmochimica Acta
\def\grl{\ref@jnl{Geophys.~Res.~Lett.}} % Geophysics Research Letters
\def\jcp{\ref@jnl{J.~Chem.~Phys.}}      % Journal of Chemical Physics
\def\jgr{\ref@jnl{J.~Geophys.~Res.}}    % Journal of Geophysics Research
\def\jqsrt{\ref@jnl{J.~Quant.~Spec.~Radiat.~Transf.}}
\def\memsai{\ref@jnl{Mem.~Soc.~Astron.~Italiana}}
\def\nphysa{\ref@jnl{Nucl.~Phys.~A}}   % Nuclear Physics A
\def\physrep{\ref@jnl{Phys.~Rep.}}   % Physics Reports
\def\physscr{\ref@jnl{Phys.~Scr}}   % Physica Scripta
\def\planss{\ref@jnl{Planet.~Space~Sci.}}   % Planetary Space Science
\def\procspie{\ref@jnl{Proc.~SPIE}}   % Proceedings of the SPIE

\makeatother

\newcommand{\ntotc}{\ensuremath{740}\xspace}
\newcommand{\ncfaiiic}{\ensuremath{55}\xspace}
\newcommand{\nriesshstc}{\ensuremath{9}\xspace}
\newcommand{\ncfaic}{\ensuremath{7}\xspace}
\newcommand{\ncfaiic}{\ensuremath{15}\xspace}
\newcommand{\nsnlsc}{\ensuremath{239}\xspace}
\newcommand{\ncalantololoc}{\ensuremath{17}\xspace}
\newcommand{\nsdssc}{\ensuremath{374}\xspace}
\newcommand{\nlowzc}{\ensuremath{11}\xspace}
\newcommand{\ncspc}{\ensuremath{13}\xspace}

\newcommand{\nsnlssdssc}{\ensuremath{613}\xspace}
\newcommand{\bfomegamlcdm}{$0.295 \pm 0.034$\xspace}

\newcommand{\bfw}{$-1.018 \pm 0.057$\xspace}
\newcommand{\bfwall}{$-1.027 \pm 0.055$\xspace}
\newcommand{\bfwksnlssdss}{$-0.996 \pm 0.069$\xspace}
\newcommand{\diffomsigma}{1.8}

\newcommand{\wtrois}{ $-1.093 \pm 0.078$ \xspace}
\newcommand{\omtrois}{ $0.228 \pm 0.038$ \xspace}
\newcommand{\ews}{6.9}
\newcommand{\deltam}{ $-0.061 \pm 0.012$ \xspace}
\newcommand{\hoall}{$68.50 \pm 1.27$\xspace}
\newcommand{\bfwwaw}{$-0.957 \pm 0.124$\xspace}
\newcommand{\bfwwawa}{$-0.336 \pm 0.552$\xspace}
\newcommand{\omshiftvarm}{-0.002\xspace}
\usepackage[colorlinks,citecolor=blue,urlcolor=blue,linkcolor=blue,draft]{hyperref}
\title{Improved cosmological constraints from a joint analysis of the SDSS-II and SNLS supernova samples.}
\titlerunning{Joint cosmological analysis of the SNLS and SDSS SNe Ia.}
\authorrunning{M.~Betoule et al.}

\author{
M.~Betoule\inst{1}
\and
R.~Kessler\inst{2,3}
\and
J.~Guy\inst{1,4}
\and
J.~Mosher\inst{5}
\and
D.~Hardin\inst{1}
\and
R.~Biswas\inst{6}
\and
P.~Astier\inst{1}
\and
P.~El-Hage\inst{1}
\and
M.~Konig\inst{1}
\and
S.~Kuhlmann\inst{6}
\and
J.~Marriner\inst{7}
\and
R.~Pain\inst{1}
\and
N.~Regnault\inst{1}
\and
C.~Balland\inst{1}
\and
B.~A.~Bassett\inst{8,9,10}
\and
P.~J.~Brown\inst{11}
\and
H.~Campbell\inst{12,13}
\and
R.~G.~Carlberg\inst{14}
\and
F.~Cellier-Holzem\inst{1}
\and
D.~Cinabro\inst{15}
\and
A.~Conley\inst{16}
\and
C.~B.~D'Andrea\inst{13}
\and
D.~L.~DePoy\inst{11}
\and
M.~Doi\inst{17,18,19}
\and
R.~S.~Ellis\inst{20}
\and
S.~Fabbro\inst{21}
\and
A.~V.~Filippenko\inst{22}
\and
R.~J.~Foley\inst{23,24}
\and
J.~A.~Frieman\inst{7,3}
\and
D.~Fouchez\inst{25}
\and
L.~Galbany\inst{26,27}
\and
A.~Goobar\inst{28}
\and
R.~R.~Gupta\inst{5,6}
\and
G.~J.~Hill\inst{29}
\and
R.~Hlozek\inst{30}
\and
C.~J.~Hogan\inst{7,3}
\and
I.~M.~Hook\inst{31,32}
\and
D.~A.~Howell\inst{33,34}
\and
S.~W.~Jha\inst{35}
\and
L.~Le~Guillou\inst{1}
\and
G.~Leloudas\inst{28,36}
\and
C.~Lidman\inst{37}
\and
J.~L.~Marshall\inst{11}
\and
A.~M\"oller\inst{38}
\and
A.~M.~Mour\~ao\inst{39}
\and
J.~Neveu\inst{38}
\and
R.~Nichol\inst{13}
\and
M.~D.~Olmstead\inst{40}
\and
N.~Palanque-Delabrouille\inst{38}
\and
S.~Perlmutter\inst{4}
\and
J.~L.~Prieto\inst{11}
\and
C.~J.~Pritchet\inst{21}
\and
M.~Richmond\inst{41}
\and
A.~G.~Riess\inst{42,43}
\and
V.~Ruhlmann-Kleider\inst{38}
\and
M.~Sako\inst{5}
\and
K.~Schahmaneche\inst{1}
\and
D.~P.~Schneider\inst{44}
\and
M.~Smith\inst{45}
\and
J.~Sollerman\inst{46}
\and
M.~Sullivan\inst{47}
\and
N.~A.~Walton\inst{12}
\and
C.~J.~Wheeler\inst{48}
}
\institute{
LPNHE, CNRS/IN2P3, Universit\'e Pierre et Marie Curie Paris 6, Universit\'e Denis Diderot Paris 7, 4 place Jussieu, 75252 Paris Cedex 05, France
\and
Department of Astronomy and Astrophysics, University of Chicago, 5640 South Ellis Avenue, Chicago, IL 60637
\and
Kavli Institute for Cosmological Physics, University of Chicago, 5640 South Ellis Avenue Chicago, IL 60637
\and
LBNL, 1 Cyclotron Rd, Berkeley, CA 94720, USA
\and
Department of Physics and Astronomy, University of Pennsylvania, 209 South 33rd Street, Philadelphia, PA 19104, USA
\and
Argonne National Laboratory, 9700 South Cass Avenue, Lemont, IL 60439, USA
\and
Center for Particle Astrophysics, Fermi National Accelerator Laboratory, P.O. Box 500, Batavia, IL 60510, USA
\and
African Institute for Mathematical Sciences, 6-8 Melrose Road, Muizenberg, Cape Town, South Africa
\and
South African Astronomical Observatory, Observatory, Cape Town, South Africa
\and
Department of Maths and Applied Maths, University of Cape Town, Rondebosch 7701, South Africa
\and
George P. and Cynthia Woods Mitchell Institute for Fundamental Physics and Astronomy, and Department of Physics and Astronomy, Texas A \& M University, College Station, TX 77843-4242
\and
Institute of Astronomy, Madingley Road, Cambridge CB4 0HA, UK
\and
Institute of Cosmology \& Gravitation, University of Portsmouth, Portsmouth PO1 3FX, UK
\and
Department of Astronomy and Astrophysics, University of Toronto, 50 St. George Street, Toronto ON M5S 3H4, Canada
\and
Department of Physics and Astronomy, Wayne State University, Detroit, MI, 48202, USA
\and
Center for Astrophysics and Space Astronomy 389-UCB, University of Colorado, Boulder, CO 80309, USA
\and
Institute of Astronomy, Graduate Shool of Science, The Univ. of Tokyo, 2-21-1 Osawa, Mitaka, 181-0015, Japan
\and
Research center for the early universe, Graduate School of Science, The University of Tokyo, 7-3-1 Hongo, Bunkyo-ku, Tokyo, 113-0033, Japan
\and
Kavli Institute for the Physics and Mathematics of the Universe, 5-1-5 Kashiwanoha, Kashiwa, 277-8583, Japan
\and
Department of Astrophysics, California Institute of Technology, MS 249-17, Pasadena, CA 91125, USA
\and
Department of Physics and Astronomy, University of Victoria, PO Box 3055 STN CSC, Victoria BC V8T 1M8, Canada
\and
University of California, Berkeley, CA 94720 USA
\and
Astronomy Department, University of Illinois at Urbana-Champaign, 1002 W.\ Green Street, Urbana, IL 61801 USA
\and
Department of Physics, University of Illinois Urbana-Champaign, 1110 W.\ Green Street, Urbana, IL 61801 USA
\and
CPPM, Aix-Marseille Université, CNRS/IN2P3, Marseille, France
\and
Institut de F\'isica d’Altes Energies, Universitat Aut\`onoma de Barcelona, E-08193 Bellaterra (Barcelona), Spain
\and
Departamento de Astronom\'ia, Universidad de Chile, Casilla 36-D, Santiago, Chile
\and
The Oskar Klein Centre, Department of Physics, Stockholm University, AlbaNova, 10691 Stockholm, Sweden
\and
McDonald Observatory, University of Texas at Austin, 2515 Speedway, Stop C1402, Austin, TX 78712-1206, USA
\and
Department of Astrophysical Sciences, Princeton University Peyton Hall, 4 Ivy Lane, Princeton NJ 08544
\and
Department of Physics (Astrophysics), University of Oxford, Denys Wilkinson Building, Keble Road, Oxford OX1 3RH, UK
\and
INAF - Osservatorio Astronomico di Roma, via Frascati 33, 00040 Monteporzio (RM), Italy.
\and
Las Cumbres Observatory Global Telescope Network, 6740 Cortona Dr., Suite 102, Goleta, CA 93117
\and
Department of Physics, University of California, Santa Barbara, Broida Hall, Mail Code 9530, Santa Barbara, CA 93106-9530
\and
Department of Physics and Astronomy, Rutgers, The State University of New Jersey, Piscataway, NJ 08854, USA
\and
Dark Cosmology Centre, Niels Bohr Institute, University of Copenhagen, Juliane Maries Vej 30, 2100 Copenhagen, Denmark
\and
Australian Astronomical Observatory, PO Box 915, North Ryde, NSW 1670, Australia.
\and
CEA, Centre de Saclay, Irfu/SPP, F-91191 Gif-sur-Yvette, France
\and
CENTRA - Centro Multidisciplinar de Astrof\'isica and Dep. F\'isica, Instituto Superior T\'ecnico, Universidade de Lisboa, Portugal
\and
Department of Physics and Astronomy, University of Utah, Salt Lake City, UT 84112, USA
\and
School of Physics and Astronomy, Rochester Institute of Technology, Rochester, NY 14623, USA
\and
Department of Physics and Astronomy, Johns Hopkins University, Baltimore, MD 21218, USA
\and
Space Telescope Science Institute, 3700 San Martin Drive, Baltimore, MD 21218, USA
\and
Department of Astronomy and Astrophysics and the Institute for Gravitation and the Cosmos, The Pennsylvania State University, University Park, PA 16802
\and
Department of Physics, University of the Western Cape, Cape Town, 7535, South Africa
\and
The Oskar Klein Centre, Department of Astronomy, AlbaNova, SE-106 91 Stockholm, Sweden
\and
School of Physics and Astronomy, University of Southampton, Southampton, SO17 1BJ, UK
\and
Department of Astronomy, University of Texas at Austin, Austin, TX 78712, USA
}
\usepackage[T1]{fontenc}
\usepackage{lmodern}
\usepackage{txfonts}
\defcitealias{B12}{\PaperI} 
\defcitealias{A13}{A13} 
\defcitealias{mosher}{\PaperII}
\defcitealias{2013ApJ...764...48K}{\ktreize}
\defcitealias{2010A&A...523A...7G}{G10}
\defcitealias{2011ApJS..192....1C}{C11}

\begin{document}

\abstract{} 
{ We present cosmological constraints from a joint analysis of type Ia
  supernova (SN Ia) observations obtained by the SDSS-II and SNLS
  collaborations.  The data set includes several low-redshift samples
  ($z<0.1$), all three seasons from the SDSS-II ($0.05 < z < 0.4$),
  and three years from SNLS ($0.2 <z < 1$), and it totals \ntotc
  spectroscopically confirmed type~Ia supernovae with high quality
  light curves.}
{We followed the methods and assumptions of the SNLS three-year data
  analysis except for the following important improvements: 1) the
  addition of the full SDSS-II spectroscopically-confirmed SN Ia
  sample in both the training of the SALT2 light-curve model and in
  the Hubble diagram analysis (\nsdssc SNe),
  2) intercalibration of the SNLS and SDSS surveys and reduced
  systematic uncertainties in the photometric calibration, performed
  blindly with respect to the cosmology analysis, and
  3) a thorough investigation of systematic errors associated with the
  SALT2 modeling of SN Ia light curves.  }
{We produce recalibrated SN Ia light curves and associated distances
  for the SDSS-II and SNLS samples.\thanks{Table~\ref{table:lcfit} is
    only available in electronic form at the CDS via anonymous ftp to
    cdsarc.u-strasbg.fr (130.79.128.5) or via
    \url{http://cdsweb.u-strasbg.fr/cgi-bin/qcat?J/A+A/}} The large
  SDSS-II sample provides an effective, independent, low-z anchor for
  the Hubble diagram and reduces the systematic error from calibration
  systematics in the low-z SN sample. For a flat \LCDM cosmology, we
  find $\Omega_m =$\bfomegamlcdm (stat+sys), a value consistent with
  the most recent CMB measurement from the \emph{Planck} and
  \emph{WMAP} experiments. Our result is $\diffomsigma\sigma$
  (stat+sys) different than the previously published result of SNLS
  three-year data. The change is due primarily to improvements in the
  SNLS photometric calibration. When combined with CMB constraints, we
  measure a constant dark-energy equation of state parameter $w=$\bfw
  (stat+sys) for a flat universe. Adding BAO distance measurements
  gives similar constraints: $w=$\bfwall. Our supernova measurements
  provide the most stringent constraints to date on the nature of dark
  energy.}
{}


\maketitle

\section{Introduction}
\label{sec:intro}

The accelerating expansion of the universe was discovered fifteen
years ago by measuring accurate distances to distant type Ia
supernovae
\citep{riess_observational_1998,perlmutter_measurements_1999}. The
reason for the acceleration remains unknown, and the term ``dark
energy'' is used to describe the
phenomenon.
Understanding the nature of dark energy is currently one
of the major goals of fundamental physics, and this drives a large
experimental effort in observational cosmology. While a cosmological
constant may be the simplest explanation for the accelerating
expansion, alternatives such as dynamical dark energy or modified
gravity (see, \emph{e.g.},  \citealt{2013LRR....16....6A} for a recent
review) can be tested through their effects in either the late-time
expansion history or the growth of structures in the universe.
By precisely mapping the distance-redshift relation up to redshift
$z\approx1$, type Ia supernovae remain, at this stage, the most
sensitive probe of the late-time expansion history of the universe. 

This goal motivated large-scale systematic searches for SNe Ia in the
past decade. High-redshift (up to $z\sim1$) programs include the
Supernova Legacy Survey (SNLS, \citealt{astier_supernova_2006},
\citealt{2011ApJ...737..102S}), the ESSENCE project
\citep{2007ApJ...666..694W}, and the Pan-STARRS survey
\citep{2012ApJ...750...99T,2013arXiv1310.3824S,2013arXiv1310.3828R}.
Intermediate redshifts ($0.05 < z < 0.4$) were targeted by the SDSS-II
supernovae survey \citep{2008AJ....135..338F,kessler_first-year_2009,
  2009ApJ...703.1374S, 2010MNRAS.401.2331L, 2013ApJ...763...88C}.
Nearby ($z< 0.1$) programs include the Harvard-Smithsonian Center for
Astrophysics survey (CfA, \citealt{2009ApJ...700..331H}), the Carnegie
Supernova Project (CSP,
\citealt{2010AJ....139..519C,2010AJ....139..120F,2011AJ....142..156S}),
the Lick Observatory Supernova Search
\citep[LOSS,][]{2013MNRAS.433.2240G}, and the Nearby Supernova Factory
\citep[SNF,][]{2002SPIE.4836...61A}.  At $z>1$, supernova discovery
and follow-up have been carried out with the Hubble Space Telescope
(\emph{HST}) by several groups
\citep{2007ApJ...659...98R,2012ApJ...746...85S}. 
With a total of about one thousand SNe Ia discovered and
spectroscopically confirmed, these second-generation surveys provide a
measurement of luminosity-distance ratios with a high statistical
precision 
between $z\approx0.01$ and $z\approx 0.7$.

As noted in the recent studies
\citep{2011ApJS..192....1C,2012ApJ...746...85S,2013arXiv1310.3824S},
the accuracy of cosmological constraints is currently limited by
systematic measurement uncertainties, particularly the uncertainty in
the band-to-band and survey-to-survey relative flux calibration. Since
relative flux measurements are the heart of this technique, this
situation is not surprising, especially if one considers the large
number of surveys and instruments involved. Significant efforts have
been undertaken to overcome this limitation
\citep{ivezi_sloan_2007,R09,2012ApJ...750...99T}.
In addition, several other sources of systematic uncertainty have been
identified in SN Ia analyses. The most important concerns are
potential biases related to model assumptions in light-curve fitting
techniques
\citep{2007ApJ...664L..13C,kessler_first-year_2009,2013ApJ...764...48K,2013arXiv1306.4050S},
and variation in the average luminosity of SNe Ia with the properties
of their host galaxies
\citep{Sullivan2010,Kelly2010,Lampeitl2010,Gupta2011,2012arXiv1211.1386J}.
Comprehensive discussions of the various effects associated with the
SN Ia Hubble diagram can be found in \citet{kessler_first-year_2009}
and \citet[hereafter C11]{2011ApJS..192....1C}.

This paper is part of an SNLS-SDSS collaborative effort called the
``joint light-curve analysis'' (hereafter JLA). The JLA was initiated
in 2010 to address the most important limitations identified in
previous analyses. More specifically, the effort was 
primarily directed at 1) improving the accuracy of the photometric
calibration of both surveys, 2) more rigorously determining
uncertainties in the SN Ia light-curve models,
and 3) including the full SDSS-II SNe Ia spectroscopic sample in both
the light-curve training and cosmology analysis.\footnote{For the C11
  analysis, only the first SDSS-II season of data was available and
  used.} 
The SDSS-II spectroscopic sample is part of the final release of the
SDSS-II supernova survey \citep{SDSSRELEASE}. The resulting
improvements in the SDSS and SNLS photometric calibration accuracy are
described in \citet[hereafter \PaperI]{B12}. Improvements in the SALT2
model and its uncertainties are described here and in \citet[hereafter
\PaperII]{mosher}. These improvements were made without regard to
their affect on the derivation of cosmological parameters from our
data.  In particular, the recalibration was completed in October 2012,
before its impact on cosmology was determined.

The main goal of the present paper is to provide stronger cosmological
constraints from a new analysis of the nearby, SDSS-II and SNLS
three-year samples using the full SDSS-II sample and the reductions in
systematic uncertainties that resulted from the JLA analyses.
Otherwise, we follow closely the approach described in the analysis of
the first three years of SNLS
\citep{2010A&A...523A...7G,2011ApJS..192....1C,2011ApJ...737..102S}.
Improvements in calibration understanding apply only to the SDSS-II
and SNLS SNe. Therefore, focusing on the control of systematics, we
restrict ourselves to adding only the last two seasons of the SDSS-II
to the SN data samples that were used in the C11 analysis.

The outline of the paper is as follows. The data samples are briefly
presented in \S\ref{sec:data}. We describe the joint recalibration of
the SNLS and SDSS photometry in Sect.~\ref{sec:supern-surv-sampl}. We
summarize improvements in the systematic uncertainties and validation
of the distance estimates based on the SALT2 model in
Sect.~\ref{sec:joint-training-light}. We detail the construction of a
low-systematic-error joint Hubble diagram in
Sect.~\ref{sec:hubble-diagram-its}. A determination of $\Omega_m$ for
a flat \LCDM universe from supernovae alone is described in
Sect.~\ref{sec:lcdm-constr-from}. We explain, in this section, the
relative impact of each change relative to the C11 analysis. We also
compare our measurement with the independent measurement provided by
the \emph{Planck} CMB experiment. 
Section~\ref{sec:dark-energy-constr} uses additional astrophysical
probes in combination with SNe Ia to break degeneracies and constrain
dark energy in more generic models. In particular, we include precise
measurements of the cosmic microwave background (CMB) and baryon
acoustic oscillations (BAO). 
We summarize the main results and discuss prospects for improvement in
Sect.~\ref{sec:conclusion}.

\section{Data samples}
\label{sec:data}
\newcommand{\SDSS}{SDSS-II}

In this paper, we present a new compilation of SN Ia light curves
including data from the full three years of the SDSS survey. The rest
of our sample is taken from the compilation assembled in
\citet{2011ApJS..192....1C}, hereafter referred to as the ``C11
compilation'', comprising SNe from SNLS, HST and several nearby
experiments. This extended sample of \ntotc SNe Ia is referred to as
the "JLA" sample.

 \subsection{The \SDSS\ SN~Ia sample}
\label{subsec:sdss_data}

The data release of the SDSS-II supernova survey \citep{SDSSRELEASE}
delivers light curves for 10,258 variable and transient sources, as
well as host galaxy identification for thousands of transients,
photometric classifications for the candidates with good multicolor
light curves, dedicated spectroscopic observations for a subset of 889
transients, and host galaxy redshifts obtained using spectra from the
original SDSS spectrograph, the SDSS-III BOSS spectrograph, and the
telescopes used to obtain SDSS SN spectra. These observations resulted
in the largest sample of supernova candidates ever compiled with 4607
likely supernovae, 500 of which have been confirmed as SNe Ia by the
spectroscopic follow-up. Our JLA sample includes a selection of
$\nsdssc$ SNe Ia from this spectroscopic sample. Here we give a brief
summary of the survey, photometry and
calibration.

The \SDSS\ Supernova Survey used the SDSS camera
\citep{1998AJ....116.3040G} on the SDSS 2.5 m telescope
\citep{2000AJ....120.1579Y,2006AJ....131.2332G} at the Apache Point
Observatory (APO) to search for SNe in the northern fall seasons
(September 1 through November 30) of 2005 to 2007.  This survey
scanned a region centered on the celestial equator in the Southern
Galactic hemisphere (designated stripe~82) that is 2.5$^{\circ}$ wide
and spans right ascensions of 20$^{\rm h}$ to 4$^{\rm h}$, covering a
total area of 300~deg$^2$ with a typical cadence of observations of
once every four nights.  Images were taken in five broad passbands,
$ugriz$ \citep{fukugita_sloan_1996,2010arXiv1002.3701D}, with 55
second exposures and processed through the PHOTO photometric pipeline
\citep{2001ASPC..238..269L}.  Within 24~hr of collecting the data, a
dedicated computing cluster at APO was used to search the images for
SN candidates.  Spectra of selected SN candidates were observed in a
program involving about a dozen telescopes: the Hobbey-Eberly
Telescope (HET), the Apache Point Observatory 3.5m Telescope (APO),
the Subaru Telescope, the 2.4-m Hiltner Telescope at the
Michigan-Dartmouth-MIT Observatory (MDM), the European Southern
Observatory (ESO) New Technology Telescope (NTT), the Nordic Optical
Telescope (NOT), the Southern African Large Telescope (SALT), the
William Herschel Telescope (WHT), the Telescopio Nazionale Galileo
(TNG), the Keck I Telescope, and the Magellan Telescope. Details of
the SDSS-II SN Survey are given in
\citet{2008AJ....135..338F} and \citet{2008AJ....135..348S}, and the procedures
for spectroscopic identification and redshift determinations are
described in \citet{2008AJ....135.1766Z}. Some subsamples of the
spectra have been subjected to more detailed analyzes
\citep{2011A&A...526A..28O,2011arXiv1101.1565K,2012AJ....143..113F}. The
determination of host galaxy redshifts for the BOSS data is
described in \citet{2013arXiv1308.6818O}.

The SN photometry for \SDSS\ is based on Scene Model Photometry (SMP)
described in \citet{holtzman_sloan_2008}.  The basic approach of SMP
is to simultaneously model the ensemble of survey images covering an
SN candidate location as a time-varying point source (the SN) and sky
background plus a time-independent galaxy background and nearby
calibration stars, all convolved with a time-varying point-spread
function (PSF).  The fitted parameters are SN position, SN flux for
each epoch and passband, and the host-galaxy intensity distribution in
each passband.  The galaxy model for each passband is a $20 \times 20$
grid of CCD pixels (approximately $8\arcsec \times 8\arcsec$), and
each of the $15\times15 \text{(pixels)}\times 5 \text{(passbands)} = 1125$ galaxy
intensities is an independent fit parameter.  As there is no pixel
resampling or convolution of the observed images, the procedure
yields reliable statistical error estimates.

The calibration is based on the catalog in \citet{ivezi_sloan_2007},
and the resulting SN fluxes returned by SMP are in the native SDSS
system. Here we use true AB magnitudes obtained by applying the small
AB offsets in Table 23 of B13.  As part of this JLA effort, a
declination-dependent calibration adjustment has been applied as
described in B13 and Sect.~\ref{sec:sdss} below.

\subsection{The C11 compilation}
\label{subsec:snls_data}

The C11 compilation includes 242 spectroscopically confirmed SNe~Ia
from the first three seasons of the five-year SNLS survey. The survey
covered four 1~deg$^2$ fields using the MegaCam imager on the 3.6~m
Canada-France-Hawaii Telescope (CFHT). Images were taken in four
passbands similar to those used by the SDSS: $g_M, r_M, i_M, z_M$,
where the subscript $M$ denotes the MegaCam system.  Each field and
passband was repeatedly imaged four or five times per lunation, with exposure
times of $\sim 1$~hr (see \citealt{2006AJ....131..960S} for details)
in order to discover SNe at redshifts up to $z\sim 1$.  The SNLS
images were rapidly processed to discover live transients.
About 1000 supernovae were discovered in the redshift range $0.2 < z <
1$, and 420 of them have been confirmed as a type Ia by massive
spectroscopic follow-up programs
\citep{2005ApJ...634.1190H,2008A&A...477..717B,2008ApJ...674...51E,2009A&A...507...85B,2011MNRAS.410.1262W}.

The rest of the compilation is dominated by low-$z$ ($z <0.08$) SNe
from the third release (Hicken et al. 2009) of photometric data acquired
at the F. L. Whipple Observatory of the Harvard-Smithsonian Center for
Astrophysics (CfA3). The data were acquired between 2001 and 2008
using three different CCD cameras (Keplercam, Minicam and 4Shooter2) and
photometry in the natural systems ($UBVRI$ or $UBVri$) is provided.
We also include high quality photometric data from the first release
(Contreras et al. 2010) of the Carnegie Supernova Project (CSP). Those
data were acquired by the SWOPE instrument at the Las Campanas
Observatory. We make use of the photometry available in the natural
SWOPE system ($ugriBV$).  The low-$z$ part of the compilation is
complemented with older data from various origins (mostly Altavilla et
al. 2004; Hamuy et al. 1996; Jha et al. 2006; Riess et
al. 1999). Those data are calibrated against the Landolt (1992)
photometric standards and color corrected to the Landolt $UBVRI$
system. This last step introduces additional uncertainties in
the photometry that needs to be taken into account (see
Appendix~\ref{sec:syst-induc-color}).
 
Finally, the C11 compilation includes photometry of 14 very high
redshift ($0.7 < z < 1.4$) SNe Ia from space-based observations with
the HST (Riess et al. 2007). The observations were obtained with the
Advanced Camera for Surveys (ACS in wide-field mode) and camera 2 of
the NICMOS instrument.

\section{Joint photometric calibration of the SNLS and the SDSS-II
  surveys}
\label{sec:supern-surv-sampl}

\subsection{The SDSS-II and SNLS supernova surveys}
\label{sec:overv-supern-surv}

The SDSS-II and SNLS experiments provide a large fraction of
the currently available SNe Ia sample (\nsnlssdssc out of \ntotc SNe
Ia in our sample). 
Both experiments were part of large photometric and spectroscopic
surveys, with the photometric component conducted in rolling-search
mode using a single, well-characterized photometric instrument.

The similarity (in design) and complementarity (in redshift) of the
two surveys motivated the attempt to combine efforts for a joint
analysis of the data. While B13 gives a detailed description of the
recalibration of the SDSS-II and SNLS surveys, a brief summary is
given below along with the details of the calibration transfer to the
photometry of supernovae. We then describe consistent estimates of
calibration uncertainties for the full sample.

\subsection{Calibration of photometric measurements}
\label{sec:flux-interpr-phot}

Many of the stars surrounding supernovae in the science fields are
non-variable at the mmag level and can be used as flux references. The
photometry of supernovae is made relative to those stars, referred to
as ``tertiary standards''.

The photometry for the SDSS and SNLS samples was performed completely
independently but with methods that were similar. The description of
the SN differential photometry technique applied to the SNLS data is
given in \citet[Sect. 5, hereafter A13]{A13}.The algorithm has been
validated using semi-artificial sources introduced in real images and
has demonstrated to accurately recover the supernovae flux relative to
surrounding stars with a systematic uncertainty about 1.5~mmag.  The
SDSS photometry is described in \citet{holtzman_sloan_2008} and has
also been tested with artificial sources and null sources using
pre-explosion epochs of real SN.

The photometry methods deliver instrumental fluxes of supernovae and
tertiary standards in consistent but arbitrary units. The
interpretation of those ``instrumental'' fluxes $\phi$ then relies on
the following model:
\begin{equation}
  \label{eq:2}
  \phi 10^{-0.4  \zp_b} = 
  \frac{\int_\lambda \lambda T_b(\lambda) S_\text{SN}(\lambda) d\lambda}
  {\int_\lambda\lambda T_b(\lambda) S_\text{ref} (\lambda) d\lambda}  
\end{equation}
where $S_\text{SN}(\lambda)$ is the supernova spectral energy distribution
(SED) as a function of wavelength, $T_b(\lambda)$ is the effective
instrument transmission in photometric band $b$,
$S_\text{ref}(\lambda)$ is the SED reference which defines
the magnitude system, and $\zp_b$ is the calibration constant (zero-point) 
which anchors the magnitude system to physical units. The
precise determination of $\zp_b$ and $T_b$ is the purpose of survey
calibration.

Currently, our model for $T_b(\lambda)$ is built from laboratory or
\emph{in situ} transmission
measurements of the CCD and filter passbands, combined with on site
measurements of the mean atmospheric absorption. The determination of
$\zp_b$ relies on observations of flux standards. This role is
currently played by spectrophotometric standard stars. SN Ia studies
rely on the most accurate set of standards available, which were
established using the \emph{HST} STIS instrument, and are obtained from 
the CALSPEC database
\citep{bohlin_absolute_2004,2010AJ....139.1515B}.

SDSS-II and SNLS had independent calibration strategies, both relying
on observations obtained with intermediate instruments. SNLS
calibration relied on observations of the \bdtruc primary standard in
the Landolt photometric system \citep{2007AJ....133..768L}, while in
SDSS, the tertiary stars were compared to the HST solar analog
standard stars using a dedicated monitor telescope
\citep{2006AN....327..821T}.  The joint calibration analysis of the
SNLS and SDSS surveys \citepalias{B12} resulted in improvements in our
understanding of survey instruments and calibration accuracy. Detailed
comparisons of the two instrument responses led to two revisions:
\begin{enumerate}
\item The effective transmission curves $T_r(\lambda)$ and
  $T_i(\lambda)$ of MegaCam in $r$ and $i$ bands were revised. This
  revision resulted in a 3~nm shift of the central wavelength toward
  the red, which is larger than the previously estimated uncertainty 
  in the MegaCam passbands (about 1~nm).
\item A 2\% non-uniformity of the SDSS monitor telescope photometric
  response has been corrected. This non-uniformity had virtually no
  impact on the calibration transfer; however, the uniformity of the
  SN survey was affected.
\end{enumerate}
The calibration accuracy was further improved thanks to two additional
sets of observations conducted with MegaCam at the CFHT. The first set
of MegaCam observations was in the SDSS-II and SNLS science fields,
and was dedicated to the direct cross-calibration of the two
surveys. Analysis of this cross-calibration sample with the above
corrections shows that the photometry of the two instruments is
  uniform at the 3~mmag level. It also demonstrates the relative
  agreement of their calibrations at the $5$~mmag level in $riz$ and
  10~mmag in $g$. The second set was dedicated to direct observations
of three primary \emph{HST} standards (see Table~\ref{tab:stisvisit}),
with the goal of reducing the number of steps in the calibration chain
to a minimum. Combining these new observations with the previous
calibration data from the SNLS and SDSS results in a redundant and
consistent picture of the SNLS and SDSS calibrations with reliable
uncertainties. In the most sensitive bands ($g$, $r$, $i$), the
uncertainties introduced in the calibration transfer are now typically
smaller than the uncertainty in the \emph{HST} flux standards
($\sim3$~mmag); in other words, our calibration is now limited by the
precision in the CALSPEC flux calibration. A detailed review of the
current error budget is provided at the end of this section
(Sect.~\ref{sec:covar-matr-calibr}).

\subsection{Recalibration of SDSS-II and SNLS supernova light curves}
\label{sec:recal-sdss-snls}

The joint calibration resulted the large set of calibrated tertiary
standard stars (published in \citetalias{B12}) for the SDSS-II and SNLS science fields. 
We now turn to the
transfer of this calibration to the supernovae photometry.
 
\subsubsection{SNLS}
\label{sec:snls}
We rely on the SNLS3 photometry of supernovae published in
\citet[hereafter G10]{2010A&A...523A...7G}. The SNLS3 photometry used 
the PSF ``resampled photometry'' method (RSP) described in
\citetalias{A13}. Here we improve the calibration transfer from
tertiary stars to supernovae that accounts for the differences between
aperture photometry (used for the standards stars\footnote{Part
    of MegaCam calibration exposures were taken out-of-focus to avoid
    saturation of the brightest standards, making the direct use of
    PSF photometry in the calibration impracticable in practice.})
and PSF photometry (for SNe). 
According to the analysis of \citetalias[Sect. 8]{A13}, we
implement two corrections to address the following issues:
\begin{enumerate}
\item The actual shape of the PSF varies with wavelength. This
  variation is not taken into account by the PSF photometry method. As
  a consequence, the effective throughput of the photometry varies
  with wavelength (see, \emph{e.g.},
  \citetalias[Sect.~3.2]{2010A&A...523A...7G}). In practice, natural
  PSF and aperture magnitudes correspond to slightly different
  photometric systems.
\item Individual measurements with aperture photometry are contaminated
  by local background structure \citepalias[Sect. 4.3.4]{B12}.
\end{enumerate}
The procedure is otherwise similar to that of
\citetalias{2010A&A...523A...7G}.

Taking the two effects into account, the calibration equation for the
PSF zero-points reads:
\begin{equation}
  \label{eq:6}
  \zp = \langle m_\text{ap} + 2.5 \log_{10} (\phi_\text{psf} + N_\text{pix} \hat s) + \gamma_2 (g-i) + \gamma_1 \rangle
\end{equation}
where $m_\text{ap}$ are the aperture magnitudes published in
\citetalias{B12}, $\phi_\text{psf}$ are the instrumental PSF fluxes, $N_\text{pix}
\hat s$ accounts for the effective contamination of apertures of size
$N_\text{pix}$ (about $800$ pixels) by background residual levels $\hat s$
(estimated for each star along with PSF fluxes), $g-i$ are the AB
colors of tertiary stars, and $\gamma_1$ and $\gamma_2$ are the
coefficients of a linear color transformation between the aperture and
PSF systems. The $\gamma_1$ and $\gamma_2$ coefficients are given in
Table~\ref{tab:ct}.
\begin{table}
  \centering
\caption{Coefficient of synthetic color transformation\tablefootmark{a} between aperture and PSF MegaCam magnitudes.}
\label{tab:ct}
\begin{tabular*}{\linewidth}{@{\extracolsep{\fill}}l@{\extracolsep{\fill}}r@{\extracolsep{\fill}}r@{\extracolsep{\fill}}r@{\extracolsep{\fill}}r}
\hline\hline
&$g$&$r$&$i$&$z$\\
\hline
$\gamma_1\quad (mmag)$&$-1.54\pm0.14$&$0.01\pm0.02$&$0.57\pm0.05$&$-0.31\pm0.03$\\
$\gamma_2\times 10^{3}$&$-3.76\pm0.11$&$-0.63\pm0.01$&$-1.31\pm0.04$&$0.69\pm0.03$\\
\hline
\end{tabular*}
\tablefoot{\\\tablefoottext{a}{See Eq.~(\ref{eq:6}). Quoted uncertainties are 1$\sigma$.}}
\end{table}

\begin{table*}
  \centering
  \caption{Selection of SNLS tertiary stars entering the determination of zero-points.}
  \label{tab:tertiarycuts}
  \begin{tabular}{lrrrr}
    \hline
    \hline
    & $g$ & $r$ & $i$ & $z$ \\
    \hline
    Magnitude range & $17 < g < 21$ & $17.5 < r < 21$ & $17.5 < i < 21$ & $16.5 < z < 20$ \\
    Color range & $0.25 < g-i < 2.75$ & $0.25 < g-i < 2.75$ &  $0.25 < g-i < 2.75$ & $0.25 < g-i < 2.75$ \\
    PSF fit quality\tablefootmark{a} & $\chi_\text{psf}^2/dof<20$ & $\chi_\text{psf}^2/dof<20$ & $\chi_\text{psf}^2/dof<20$ & $\chi_\text{psf}^2/dof<20$ \\
    Repeatability of aperture flux\tablefootmark{b} & $\chi_\text{ap}^2/dof<5$& $\chi_\text{ap}^2/dof<5$& $\chi_\text{ap}^2/dof<5$& $\chi_\text{ap}^2/dof<5$\\
    \hline
  \end{tabular}
  \tablefoot{\tablefoottext{a}{Errors in the PSF model are not taken into account in the computation of the $\chi^2$. As a result, for bright stars the $\chi^2$ can be large}. Nevertheless, catastrophic $\chi^2$ are more likely to be related to measurements affected by undetected saturation which should be discarded. This cut removes about 0.8\% of the tertiary stars.
    \tablefoottext{b}{The $\chi^2$ is built from an accurate model of aperture photometry errors (see \citetalias[Eq. 15]{B12}). Large $\chi^2$ are related either to variable stars or to problems in the aperture photometry (\emph{e.g.}, apertures with varying contamination).}}
\end{table*}

\begin{figure}
  \centering
  \igraph{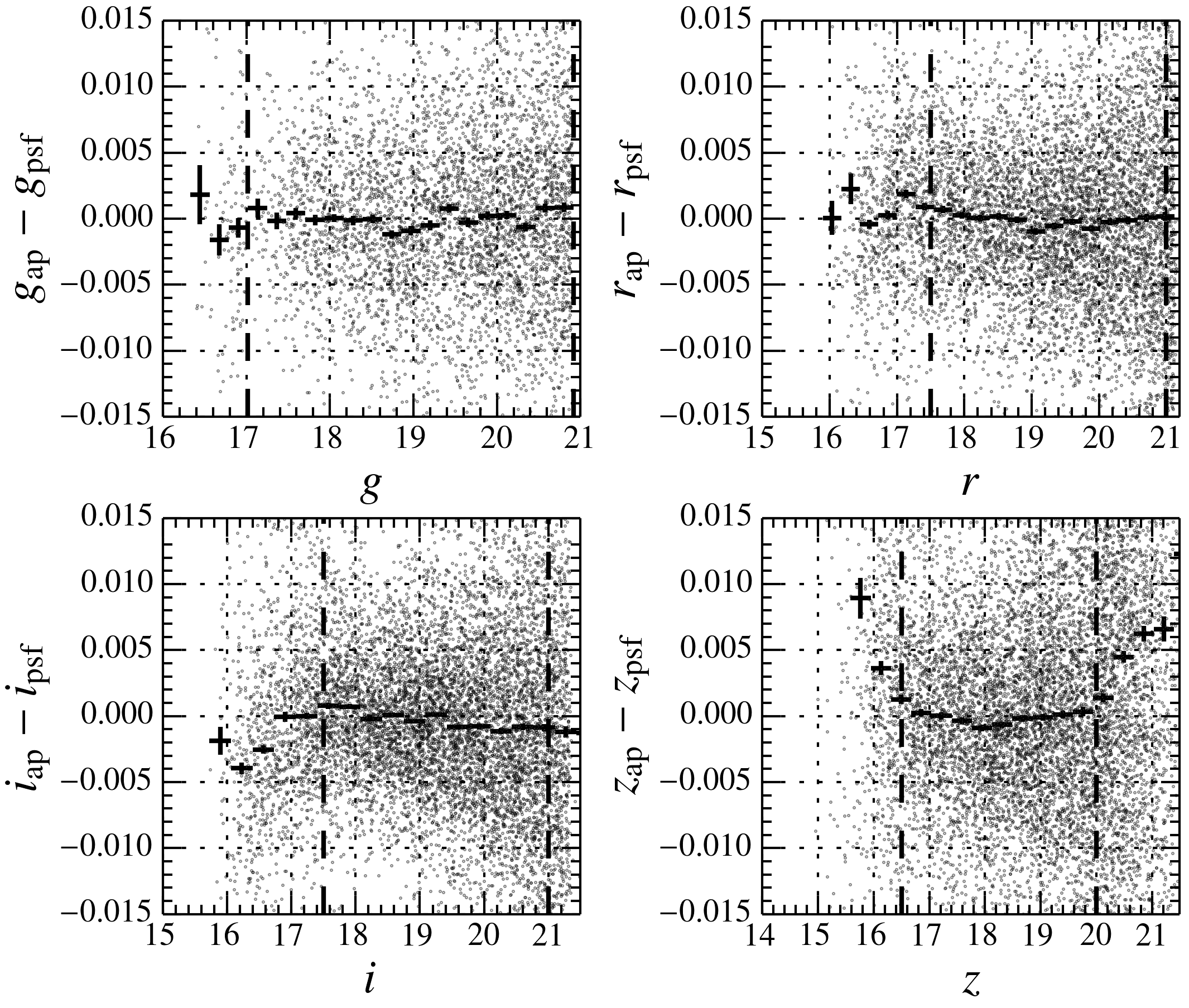}
  \caption{The difference between aperture and PSF photometry as a
    function of the SNLS tertiary standard star magnitude. The
    aperture photometry from \citetalias{B12} has been corrected for
    residual contamination using an estimate of the local background
    level obtained with PSF photometry. The magnitude range used in
    the zero-point fit (vertical dashed lines) is chosen so that the
    aperture catalog is expected to be free from selection bias
    \citepalias[Fig.~12]{B12}.}
  \label{fig:linearity}
\end{figure}

The weighted ensemble average of all the tertiary stars in a given
band, field, and CCD is used to compute the zero-point.  A few quality
cuts are applied to the tertiary stars prior to the fit, and are
summarized in Table~\ref{tab:tertiarycuts}. In particular we select
stars in a magnitude range where the aperture catalog is not affected
by selection bias \citepalias[Fig.~12]{B12} and in a color range where
the aperture to PSF transformations are accurate. We discard
potentially variable stars by applying a cut on the $\chi^2$ of
repeated measurements. Outliers are rejected iteratively at
$2.5\sigma$. Residuals for the zero-point fit are displayed as a
function of magnitude in Figure~\ref{fig:linearity} and show that the
non-linearities noted in \citetalias[Fig. 5]{2010A&A...523A...7G} are
now corrected at the mmag level by the aperture contamination
correction. The cuts leave about 25 stars, on average, per CCD and
field for the determination of each zero-point. The typical
statistical uncertainty on the zero-point determination is about
1~mmag, which is small compared to the systematic uncertainties.  The
systematic uncertainty associated with the color transformation and
aperture corrections amounts to 0.1, 0.8 , 1.0, 0.9~mmag in $griz$,
respectively. In addition, A13 quotes a 1.5~mmag systematic accounting
for potential bias in the RSP photometry method.

With respect to the previous release of SNLS light curves
\citepalias{2010A&A...523A...7G}, the $griz$ zero-points are shifted
on average by $-12.9$, $-0.9$, $1.3$ and $-17.9$~mmag. The main
contribution to this change is the recalibration of the tertiary
standard catalog, the new transfer procedure described above
contributing only noticeably in $g$ and $i$. The changes in the
calibration of the SNLS tertiaries involved the correction of a sign
error (described in \citetalias[Sect.~10.4]{B12}) in addition to the
revision of MegaCam $r$ and $i$ transmission curves and the new
calibration data described above (Sect.~\ref{sec:flux-interpr-phot}).
Note that the new calibration relies on the observations of 5
\emph{HST} standards (three observed directly) rather than just a single
one\footnote{\bdtruc, which is somewhat peculiar, as discussed in
\citealt[Sect. 11]{R09}.} as in C11.  The change in $g$ zero-point is
the most significant ($3 \sigma$), as the various contributions
act in the same direction in this band. The other zero-point
changes are well within the previously quoted uncertainties.

\subsubsection{SDSS}
\label{sec:sdss}

The photometry of SDSS-II supernovae and their calibration is
described in \cite{holtzman_sloan_2008}. The differential supernova
photometry is based on a comparison of the PSF magnitude of the
supernova and the PSF magnitudes of nearby tertiary standard stars and
is insensitive to errors in the PSF. Chromatic effects in the SDSS PSF
are expected to be negligible. The joint calibration study
\citepalias{B12} calibrates directly the PSF magnitudes of SDSS
tertiary stars. Direct measurements of HST standard stars by the
monitor telescope and by MegaCam, transformed to the SDSS survey
telescope, allow a computation of the zero-points for each SDSS
filter. The HST solar analog standard stars are used to minimize any
possible error in color terms in the transformation from the monitor
to the survey telescope.  The zero-points thus obtained are slightly
different than the nominal SDSS zero-points, resulting in small
corrections to the reported magnitudes.  Since the SDSS photometry is
obtained as a difference between the SN and stellar magnitudes, the
correction of the PSF catalogs described in \citetalias{B12} can be
readily applied to SN photometry. The correction takes the following
form:
\begin{equation}
  \label{eq:5}
  m' = m - f(\delta) - \delta_\ab 
\end{equation}
where $f$ is the uniformity correction that is a function of
declination $\delta$ as given in \citetalias[Fig.~23]{B12}, and
$\delta_\ab$ is the average offset of corrected SDSS magnitudes to the
AB system given in \citetalias[Table~23]{B12}. Recalibration of
SDSS-II light curves is carried out by applying Eq.~(\ref{eq:5}) to
the natural SN magnitudes. Once this correction is applied, the
average calibration of SDSS-II SNe is shifted by $-31$, $-4$, $0$, $0$
and $-6$~mmag in $u$, $g$, $r$, $i$ and $z$. The $gri$ changes are
well within expected uncertainties (about 6~mmag). The $u$ and $z$
band photometry were not used in previous analyses of the SDSS SN
spectroscopic sample
\citep{2009PASP..121.1028K,2011ApJS..192....1C}. A full description of
the recalibrated light curves can be found in \citet{SDSSRELEASE}.

\subsection{Uncertainties in the photometric calibration of light curves}
\label{sec:covar-matr-calibr}

The interpretation of supernova measurements is affected by
uncertainties on the $T_b(\lambda)$ and $\zp_b$ terms of
Eq.~(\ref{eq:2}). We parameterize the uncertainty on $T_b$ 
by a single parameter: the shift of the mean wavelength
$\lambda^\text{eff}_b$ (interpreted as a global shift of the transmission
function).
 The vector of calibration parameters is
thus $\calpar = (\zp_b, \lambda^\text{eff}_b)$ with $b$ running over all
the photometric bands of all the instruments. We summarize the uncertainties
affecting those parameters in a single covariance matrix assembled as
follows.

In all cases, our primary flux calibration reference is the \emph{HST} system
as defined in the CALSPEC 2011 release. We thus consider the following
sources of uncertainty:
\begin{enumerate}
\item The uncertainty on the color of the primary flux reference,
  \emph{i.e.}, the definition of the \emph{HST} system itself. \label{item:4}
\item The STIS measurement error affecting the transfer of the primary
  \emph{HST} calibrators to the CALSPEC secondary standards.\label{item:5}
\item The error in the ground-based photometric measurements of the
  CALSPEC standards.\label{item:6}
\item The error in the calibration transfer from the CALSPEC standards
  to the tertiary standards.\label{item:7}
\item Extra sources of systematic uncertainties introduced by the
  SN photometry method.\label{item:8}
\item Uncertainties in the instrument response curves.\label{item:9}
\end{enumerate}
Items~\ref{item:4}--~\ref{item:5} constitute an external source of
uncertainty (affecting the CALSPEC spectra) which is common to all
surveys. Items~\ref{item:6}--\ref{item:8} affect the accuracy of the
calibration transfer from the primary CALSPEC standards to the
supernova measurements. Item~\ref{item:6} is thought to be well
estimated and understood in all cases. Item~\ref{item:7}
and~\ref{item:8} are survey specific. In the case of SNLS and SDSS
they are thought to be well controlled \citepalias{B12}. The control
of Item~\ref{item:9} also varies from one survey to another; however,
its importance in the calibration is typically of second order.

Note that we do not consider uncertainties on the absolute flux scale
of the primary flux reference as such uncertainties would only affect
the overall normalization, which is marginalized away in our Hubble
diagram analysis.

\subsubsection{Global uncertainties on the flux standards}
\label{sec:glob-uncert-flux}

We follow the assumptions made in \citetalias{B12} for modeling the
uncertainty on CALSPEC spectra. Namely, we assume a global 0.5\% slope
uncertainty (1~$\sigma$) over the range
\SI{3000}{\angstrom}--\SI{10000}{\angstrom} for the uncertainty in the
white-dwarf system color.\footnote{The white dwarf system was recently
  redefined by a change in the modeling of the white dwarf
  fluxes (see
  \url{http://www.stsci.edu/hst/observatory/crds/calspec.html}). The
  change induced in color (about 0.003 in $g-z$ color) is consistent with our estimate of the  uncertainty.}

In addition, we consider the STIS measurement error for individual
spectra, which is accurately measured from the repeated observations
of the monitoring star AGK +81 266. We assume that the measurement
uncertainty on composite spectra decreases as the square root of the
number of STIS visits. When the number of visits is not the same for
the two grisms considered, we conservatively use the smaller number of
visits. The number of visits per star for CALSPEC spectra version 003
is given in Table~\ref{tab:stisvisit}.

\begin{table}
  \centering
  \caption{Number of STIS visits for the observed CALSPEC standards.}
  \label{tab:stisvisit}
  \begin{tabular}{p{1cm}*{5}{c}}
    \hline
    \hline
    Star & \bdtruc & P041C & P177D\tablefootmark{b} & P330E\tablefootmark{b} & SNAP2\tablefootmark{b} \\
    \hline
    \# visit\tablefootmark{a} & 3 & 2 & 2 & 3 & 1\\
    \hline
  \end{tabular}
  \tablefoot{\tablefoottext{a}{Mininum for the grism G430L and G750L.}
    \tablefoottext{b}{Those three stars were observed directly with MegaCam.}}
\end{table}

\subsubsection{Uncertainties in the SNLS and SDSS calibration}
\label{sec:uncert-snls-sdss}

Uncertainties affecting the calibration transfer from \emph{HST}
standard to the SNLS and SDSS tertiary stars are taken from
\citetalias[Table~22]{B12}. We account for correlations between bands
introduced by the cross-calibration of the two surveys, for the
correlation between filter wavelength shifts $\lambda^\text{eff}_b$, and
the definition of the AB zero-point $\zp_b$. For SNLS only, we add in
quadrature the systematic uncertainty associated with the calibration
transfer from tertiary standards to supernova light curves discussed
above (Sect.~\ref{sec:recal-sdss-snls}).

\subsubsection{Uncertainties in the calibration of the low-$z$ sample}
\label{sec:uncert-calibr-other}

All low-$z$ experiments are calibrated against secondary photometric
standards: either the Smith \citep{2002AJ....123.2121S} or Landolt
\citep{landolt_ubvri_1992} equatorial standards. In both cases,
photometry in the secondary system for the F-subdwarf \bdtruc CALPSEC
spectrophotometric standard is available
\citep{2002AJ....123.2121S,2007AJ....133..768L}, and can be used to
anchor the photometry to the \emph{HST} flux scale. We must account for
measurement uncertainties of \bdtruc in both systems as a correlated
source of uncertainties for all low-$z$ samples.

A complete review of internal calibration uncertainties was made in
C11. We follow their prescriptions except for uncertainties in the CSP
and CfAIII samples that are revised according to
\citet{2011AJ....142..156S}, \citet{2012AJ....144...17M}, and a
comparison study described in appendix~\ref{sec:consistency-low-z}. We
also revisit the $U$ band calibration uncertainty in the low-$z$
sample as described in appendix~\ref{sec:syst-induc-color}.
Table~\ref{tab:DL} summarizes internal calibration uncertainties
attributed to the low-$z$ sample.

\subsubsection{Uncertainties in the calibration of \emph{HST} supernovae}
\label{sec:uncert-calibr-hst}

We use the interpretation of \cite{2007ApJ...659...98R} \emph{HST} SN
calibration described in \citetalias[\S
2.4]{2011ApJS..192....1C}. Given the small statistical weight of the
\emph{HST} sample included in this analysis, it was not necessary to
propagate recent improvements in the treatment of the NICMOS
non-linearity \citep{2012ApJ...746...85S}.

\subsubsection{Full covariance matrix of calibration uncertainties}
\label{sec:full-covar-matr}

We summarize the calibration uncertainties, accounting for correlated
effects between bands and surveys, into a covariance matrix of the
calibration parameter vector \calpar.  The full covariance matrix is
released with the data (see Appendix~\ref{sec:data-release}). The
square roots of the diagonal elements of the covariance matrix are
given in Table~\ref{tab:sig}.  For the SDSS-II and SNLS surveys that
dominate our sample, the accuracy of the average calibration is at the
5~mmag level, primarily due to the uncertainty in the CALPSEC flux
standards.

\begin{table*}
  \centering
  \caption{Internal calibration uncertainties for the low-$z$ samples split by photometric systems.}\label{tab:DL}
  \begin{tabular*}{\textwidth}{@{\extracolsep{\fill}}l|*{5}{@{\extracolsep{\fill}}c}|*{5}{@{\extracolsep{\fill}}c}|*{5}{@{\extracolsep{\fill}}c}|*{6}{@{\extracolsep{\fill}}c}}
    \hline\hline
    Instrument & \multicolumn{5}{c|}{Standard\tablefootmark{h}} & \multicolumn{5}{c|}{KEPLERCAM\tablefootmark{h}} & \multicolumn{5}{c|}{4Shooter\tablefootmark{h}} & \multicolumn{6}{c}{SWOPE\tablefootmark{h}}\\
    Band & $U$ & $B$ & $V$ & $R$ & $I$ & $U$ & $B$ & $V$ & $r$ & $i$ & $U$ & $B$ & $V$ & $R$ & $I$ & $u$ & $g$ & $r$ & $i$ & $B$ & $V$ \\
    \hline
    $\sigma(\lambda_\text{eff})$\tablefootmark{a} (nm) & 2.5\tablefootmark{c} & 1.2\tablefootmark{c} & 1.2\tablefootmark{c} & 2.5\tablefootmark{c} & 2.5\tablefootmark{c} 
                                                      & 2.5\tablefootmark{c} & 0.7\tablefootmark{c} & 0.7\tablefootmark{c} & 0.7\tablefootmark{c} & 0.7\tablefootmark{c} 
                                                      & 2.5\tablefootmark{c} & 0.7\tablefootmark{c} & 0.7\tablefootmark{c} & 0.7\tablefootmark{c} & 0.7\tablefootmark{c} 
                                                      & 0.7\tablefootmark{d} & 0.8\tablefootmark{d} & 0.4\tablefootmark{d} & 0.2\tablefootmark{d} & 0.7\tablefootmark{d} & 0.3\tablefootmark{d} \\ 
    \hline
    $\sigma(\zp)$\tablefootmark{b} (mmag) & 100\tablefootmark{g} & 15\tablefootmark{c} & 15\tablefootmark{c} & 15\tablefootmark{c} & 15\tablefootmark{c}  
                                          & 31\tablefootmark{f} & 11\tablefootmark{c} & 7\tablefootmark{c} & 25\tablefootmark{f} & 7\tablefootmark{c} 
                                          & 70\tablefootmark{g} & 11\tablefootmark{c} & 7\tablefootmark{c} & 7\tablefootmark{c} & 20\tablefootmark{c} 
                                          & 23\tablefootmark{e} & 9\tablefootmark{e} & 8\tablefootmark{e} & 7\tablefootmark{e} & 8\tablefootmark{e} & 8\tablefootmark{e}\\ 
    \hline
  \end{tabular*}
  \tablefoot{
    \tablefoottext{a}{Uncertainty in the mean filter wavelength.}
    \tablefoottext{b}{Internal sources of systematic uncertainty in the calibration. It includes the uncertainty in the calibration transfer between secondary and tertiary standards and systematic uncertainties in the SN photometry. For measurements reported in the Landolt system, it also includes an uncertainty associated to the color transformation of supernovae from the observer's  system.}
    \tablefoottext{c}{From \citetalias{2011ApJS..192....1C}.}
    \tablefoottext{d}{Includes the uncertainties on the new measurement of SWOPE transmission curves presented in \citet{2011AJ....142..156S} plus the effect of a 0.25 airmass change on the atmospheric extinction curve.}
    \tablefoottext{e}{From \citet{2012AJ....144...17M}.}
    \tablefoottext{f}{See Appendix~\ref{sec:consistency-low-z}.}
    \tablefoottext{g}{See Appendix~\ref{sec:syst-induc-color}.}
    \tablefoottext{h}{KEPLERCAM and 4Shooter are the two main photometric instruments used in the CfAIII survey. SWOPE is the photometric instrument of the CSP survey. We refer to the Landolt photometric system in which the historical measurements are color transformed as the ``standard'' instrument.}
  }
\end{table*}

\begin{table}
  \centering
  \caption{Uncertainties 
    in calibration parameters.} 
  \label{tab:sig}

\begin{tabular}{lrr}
\hline\hline      
& $\sigma(\zp)$ & $\sigma(\lambda^{eff})$\\
& $(mmag)$ & $(nm)$ \\
\hline

\hline 
MEGACAM (SNLS)&&\\
\hline
$g$&$3$ & $0.3$\\$r$&$6$ & $3.7$\\$i$&$4$ & $3.1$\\$z$&$8$ & $0.6$\\
\hline 
SDSS&&\\
\hline
$u$&$8$ & $0.6$\\$g$&$4$ & $0.6$\\$r$&$2$ & $0.6$\\$i$&$3$ & $0.6$\\$z$&$5$ & $0.6$\\
\hline 
STANDARD &&\\
\hline
$U$&$100$ & $2.5$\\$B$&$15$ & $1.2$\\$V$&$15$ & $1.2$\\$R$&$15$ & $2.5$\\$I$&$15$ & $2.5$\\
\hline 
4SHOOTER (CfAIII)&&\\
\hline
$Us$&$70$ & $2.5$\\$B$&$11$ & $0.7$\\$V$&$7$ & $0.7$\\$R$&$8$ & $0.7$\\$I$&$20$ & $0.7$\\
\hline 
KEPLERCAM (CfAIII)&&\\
\hline
$Us$&$31$ & $2.5$\\$B$&$11$ & $0.7$\\$V$&$7$ & $0.7$\\$r$&$25$ & $0.7$\\$i$&$8$ & $0.7$\\
\hline 
SWOPE (CSP)&&\\
\hline
$u$&$23$ & $0.7$\\$g$&$9$ & $0.8$\\$r$&$8$ & $0.4$\\$i$&$8$ & $0.2$\\$B$&$8$ & $0.7$\\$V$&$8$ & $0.3$\\
\hline 
NICMOS (HST)&&\\
\hline
$F110W$&$24$ & $0.0$\\$F160W$&$62$ & $0.0$\\
\hline 
ACSWF (HST)&&\\
\hline
$F606W$&$10$ & $0.0$\\$F625W$&$10$ & $0.0$\\$F775W$&$20$ & $0.0$\\$F814W$&$20$ & $0.0$\\$F850LP$&$20$ & $0.0$\\
\hline
\end{tabular}
  \tablefoot{A provision for the filter uncertainty in the NICMOS and ACS instruments is already included in the zero-point uncertainty available from the literature.}
\end{table}

\section{Joint training of the light-curve model}
\label{sec:joint-training-light}

\subsection{Supernova models and distance estimates}
\label{sec:supern-model-dist}

Distance estimation with SNe Ia is based on the empirical observation
that these events form a homogeneous class whose remaining variability
is reasonably well captured by two parameters (see, \emph{e.g.},
\citealt{1998A&A...331..815T}). One parameter describes the time stretching
of the light-curve (\xun in what follows), and the other describes
the supernova color at maximum brightness (\col in what follows).

Specifically, the distance estimator used in this analysis (and
in most similar cosmological analyses) assumes that supernovae with
identical color, shape and galactic environment have on average the
same intrinsic luminosity for all redshifts. This hypothesis is
quantified by a linear model, yielding a standardized distance
modulus $\mu = 5\log_{10}(d_L/10{\rm pc})$:
\begin{equation}
 \mu = 
 \mstar - 
 \left( M_B   - \alpha \times X_1 + \beta \times C \right)
  \label{eq:4}
\end{equation}
where $\mstar$ corresponds to the observed peak magnitude in
rest-frame \emph{B} band and $\alpha$, $\beta$ and $M_B$ are nuisance
parameters in the distance estimate. Both the absolute magnitude $M_B$
and $\beta$ parameter were found to depend on host galaxy properties
\citep{2011ApJ...737..102S,2012arXiv1211.1386J} although the mechanism
is not fully understood. We use the C11 procedure to approximately
correct for these effects assuming that the absolute magnitude is
related to the host stellar mass ($M_\text{stellar}$) by a simple step
function:\footnote{We do not consider an additional dependency of
  $\beta$ because it does not have a significant impact on the
  cosmology. }
\begin{equation}
  \label{eq:mabs}
  M_B = \left\lbrace
      \begin{array}{ll} 
        M^1_B &\quad \text{if}\quad  M_\text{stellar} < 10^{10}~M_{\odot}\,,\\
        M^1_B + \Delta_M & \quad \text{otherwise.}
      \end{array}
    \right.
\end{equation}

The light-curve parameters $(\mstar, \xun, \col)$ result from the fit
of a model of the SN Ia spectral sequence to the photometric
data. 
Light-curve fitting techniques have a long history, and the potential
biases introduced by specific model choices have raised some concerns
(see, \emph{e.g.}, \citealt{kessler_first-year_2009}). The estimate of
model systematics in the C11 analysis was based on the comparison of
light-curve parameters reconstructed from the same data by two
different models (SALT2 and SiFTO,
\citealt{2008ApJ...681..482C}). Such a scheme is only
moderately satisfying as both methods could share similar biases,
leading to underestimated errors, or one model could have
substantially larger errors than the other.

By using extensive Monte Carlo simulations, the analysis from
\citetalias{mosher} provides a significant improvement in the determination
of light-curve model biases. Varying the underlying
supernova model in the range currently allowed by data, it
demonstrates that the data-driven SALT2 method, trained on samples
comparable to the G10 sample, recovers the input distances
without introducing a significant bias between low and high-redshift distances (see
Sect.~\ref{sec:light-curve-model}). Therefore, we adopt the SALT2
method for the present analysis, and base our systematic estimate on the
\citetalias{mosher} results.

\subsection{The SALT2 model}
\label{sec:salt}

The SALT2 model is a first order description of the time-spectral
sequence of SNe Ia, multiplied by a time independent color-law. At
phase $p$ and wavelength $\lambda$, the flux density model for a given
supernova is:
\begin{equation}
  \label{eq:13}
  S_\text{SN}(p, \lambda) = \xzero \left(\mathcal{M}^0(p, \lambda) + \xun \mathcal{M}^1(p, \lambda)\right) \exp(\col \times \mathcal{C}_L(\lambda))\,,
\end{equation}
where the normalization, shape and color parameters \xzero, \xun and
\col, respectively, are evaluated for each SN.\footnote{For a given SN, \mstar can
  be readily computed from the adjusted model, and we equivalently use
  (\mstar, \xun, \col) instead of (\xzero, \xun, \col) as parameters
  in the cosmology fit.} The mean spectral sequence $\mathcal{M}^0$,
the first order deviation around the mean sequence $\mathcal{M}^1$ and
the phase-independent color-law $\mathcal{C}_L$ are trained on a
photometric and spectroscopic sample of spectroscopically identified
SNe Ia (see below Sect.~\ref{sec:light-curve-model-2}).

The model does not capture all the variability of observed
supernovae. The remaining deviations to the model, sometimes referred
to as the ``intrinsic scatter'', have to be accounted for in some
way. In SALT2, the remaining scatter $d_{b,p}$ affecting a measurement
point at phase $p$ in band $b$ is modeled as:
\begin{equation}
  \label{eq:14}
  d_{b,p} = \epsilon_{b,p} + \kappa_b m_{b,p} 
\end{equation}
where $m_{b,p}$ is the magnitude prediction from Eq.~(\ref{eq:13}) and
$\epsilon_{b,p}$ and $\kappa_b$ are assumed to be independent, 
centered Gaussian random variables. 
The term $\epsilon_{b,p}$ describes phase dependent variations in the
magnitude around the predicted light-curve in band $b$. The set of the
$\epsilon_{b,p}$ are referred to as the ``error snake''. The term
$\kappa_b$ describes variations in the relative amplitude of the
multiband light curves around the predicted color-law. The $\kappa_b$
are referred to as the ``$k$-correction error''. The variances of
$\epsilon_{b,p}$ and $\kappa_b$ are functions fitted as part of the
training process, so that the final model describes the observed
variability.

The training of $\mathcal{M}^0$, $\mathcal{M}^1$, $\mathcal{C}_L$ and
 the associated error model are done iteratively. A detailed
description of the model and training procedure is given in
\citetalias{mosher}.

\subsection{Training SALT2 on the JLA sample}
\label{sec:light-curve-model-2}

The previous public release of the SALT2 model (v2.2) was described in
\citetalias{2010A&A...523A...7G}. It has been trained on photometric
and spectroscopic data from the literature (\emph{e.g.},
\citealt{1994ApJ...434L..19S,1996MNRAS.278..111P,2002AJ....124..417H,2004MNRAS.348..261B,matheson_optical_2008})
and high-$z$ supernovae data from the SNLS
\citep{2008ApJ...674...51E,2009A&A...507...85B}. The training sample
also includes high quality data from nearby SNe Ia that are not in the
Hubble flow. Biasing of the model could occur due to contamination by
peculiar events, use of poorly sampled light curves with ill-defined
maximum date, or biased selection of brightest events.  To prevent
such biases, cuts in redshift and quality were applied to form the G10
sample.

Here we add SDSS-II photometric data to the
\citetalias{2010A&A...523A...7G} training sample, and we refer to this
extended G10+SDSS-II sample as the ``JLA'' training sample.  Our
selection of SDSS-II data follows the procedure of G10. 
We use only the spectroscopically identified sample, and we estimate
the initial fit parameters from a fit with the G10 SALT2 model.  As
selection biases become significant in the SDSS-II sample at $z>0.25$,
we discard SNe above this limit. We evaluated the dependence of the
cosmology on the redshift cut by using alternative cuts at $z=0.2$ and
$z=0.3$ and found that the dependence was weak enough to be
ignored. We require the date of maximum $t_0$ and the light-curve
shape parameter $X_1$ to be well constrained, which efficiently
selects light curves with good sampling. We discard SNe whose fitted
color and shape parameters lie well outside the range of model
validity. Finally, we select SNe with limited extinction by dust in
the Milky-Way. The cuts are listed in Table~\ref{tab:cuts} along with
the number of supernovae discarded at each step. The remaining
supernova light curves were visually inspected, as detailed in
appendix~\ref{sec:visual-inspection-1}. A total of 24 problematic
light curves were discarded, mostly because of apparent problems in
the sampling or photometry.
\begin{table}
  \centering
  \caption{Number of supernovae discarded by the successive cuts applied to the SDSS-II sample before inclusion in the training sample.}
  \label{tab:cuts}
  \begin{tabular}{lrr}
    
\hline
\hline
 & Discarded & Remaining \\
\hline
Initial & - & 507 \\
\hline
$z<0.25$ & 170 & 337 \\
$\sigma(t_0) < 0.5$ & 85 & 252 \\
$\sigma(X_1) < 0.5$ & 14 & 238 \\
$-0.3 < C < 0.3$ & 9 & 229 \\
$-3 < X_1 < 3$ & 1 & 228 \\
$ E(B-V)_\text{mw}< 0.15$ & 1 & 227 \\
Other\tablefootmark{a} & 24 & 203 \\
\hline
  \end{tabular}
  \tablefoot{
  \tablefoottext{a}{See Appendix~\ref{sec:visual-inspection-1}.}}
\end{table}

\begin{figure*}
  \centering
  \igraph{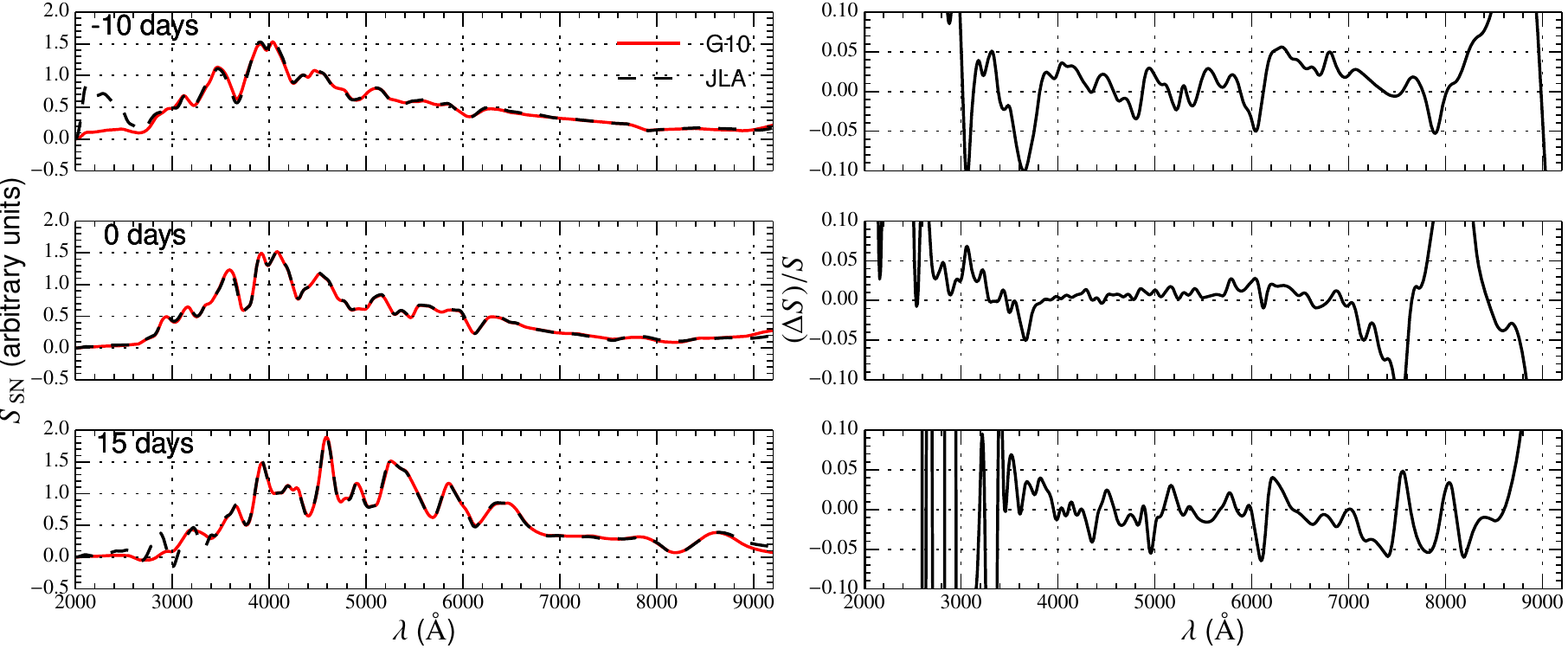}
  \caption{Comparison of $\mathcal{M}_0$ templates between the
    previous release of the SALT2 model
    \citepalias{2010A&A...523A...7G} and the present release trained
    on the JLA sample. \emph{Left:} The present model is shown as a
    black dashed line at three different phases: early (-10 days),
    close to maximum (0 day), and late (+15 days). The
    \citetalias{2010A&A...523A...7G} model is shown as the red solid
    line. \emph{Right:} Relative differences in the two models
    (JLA/G10 - 1) at the three selected phases.}
  \label{fig:M0}
\end{figure*}

\begin{figure}
  \centering
  \igraph{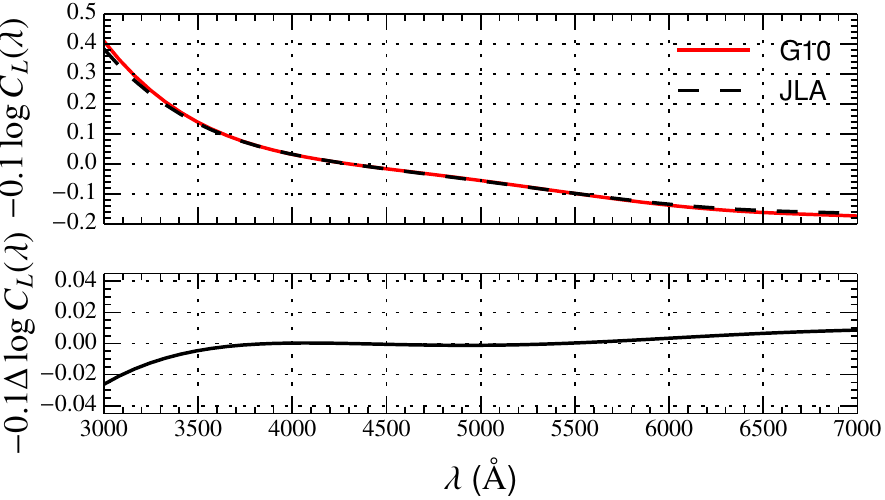}
  \caption{\emph{Top:} Comparison of the reconstructed color law $C_L$
    for two trainings of the SALT2 model (see Eq.~(\ref{eq:13}): the
    present release trained on the JLA sample (back dashed
    line) and the previous release of the SALT2 model from
    \citetalias{2010A&A...523A...7G} (red solid line).  We display
    $- 0.1 \log C_L$, which is approximately the rms magnitude variation from
    color variation. \emph{Bottom:} Difference in the two color laws.}
  \label{fig:clc}
\end{figure}

The resulting sample of 207 new SDSS-II SNe was added to the
\citetalias{2010A&A...523A...7G} training sample. A single SN in the
original sample was removed: SNLS 03D4gl, which does not have post-max
data. We retrained SALT2 on this extended and recalibrated training
sample. The resulting model functions are compared to those of the G10
model in Figs.~\ref{fig:M0} and~\ref{fig:clc}. The larger changes
occur in the UV and infrared region where the model is now better
constrained by the newly added data. The recalibrated and retrained
version of the SALT2 model can be obtained from the SALT2 web page
(see Appendix~\ref{sec:data-release}).


\subsection{Light-curve model uncertainties}
\label{sec:light-curve-model}

The SALT2 training procedure described above was evaluated with Monte
Carlo (MC) simulations in \citetalias{mosher}. The basic principle of
this analysis is to perform an end-to-end SALT2 analysis on realistic
simulations, where the analysis starts with training and ends with
fitting for $w$.  For a given reference supernova model, realistic
SALT2 training sets including light curves and spectra are generated
using the SNANA package~\citep{2009PASP..121.1028K}. They have the
same cadence, wavelength coverage and noise as the SALT2 training
sample used in the present study; in that sense, they are realistic.
For each of those training samples, the SALT2 model is trained, and
then used to fit the light curves of a statistically independent
sample of supernovae generated using the same reference model. The
light-curve parameters of those test supernovae are used to estimate
distances $\mu$ (Eq.~\ref{eq:4}). The distances are corrected for
selection effects using a dedicated simulation, similar to the one
described in Section~\ref{sec:select-bias-corr}. Finally, the set of
corrected distance moduli, $\mu_\text{corr}$, are compared to the
input cosmology to test for potential biases.

\subsubsection{Regularization in the training of SALT2}
\label{sec:regul-train-salt2}

The current training sample lacks reliable spectroscopic data at early
phases in the UV. Constrained only by photometric data, the training
of the spectroscopic sequences in these regions is an unstable
deconvolution process that requires some amount of regularization. A
high level of regularization, however, distorts the spectral model by
smoothing spectral features. 
The simulation tests in M14 have shown that our choice of
regularization results in a small bias in the distance modulus of
about 0.005~mag in the redshift range $0.2 < z < 1$.

\subsubsection{Limitations in the parameterization}
\label{sec:param-surf}

For extreme values of the shape parameter, the first order expansion
from Eq.~(\ref{eq:13}) cannot describe the actual shape of SN Ia
light curves at early and late phases. To test this limitation, we
evaluated the bias from input models based on the time stretching of
the spectral sequence established by \citet{Hsiao2007}, which cannot
be fully reproduced by SALT2. The simulations in \citetalias{mosher}
indicate that the distance modulus bias from this limitation of the
SALT2 model are less than 3~mmag.

\subsubsection{Residual scatter model}
\label{sec:resid-scatt-model}

The current SALT2 model compresses the multiwavelength information
available on SNe Ia into the two parameters $(\mstar, \col)$. This
compression is trained to describe the mean behavior of the SN Ia
population, but does not encompass all their diversity as already
explained in Sect.~\ref{sec:salt}.  Whatever the physical phenomenon
causing the remaining scatter (differences between extrinsic and
intrinsic extinction, unmodeled variability of spectral features...),
it is treated as noise independent from one broad-band to
another. Modeling the scatter restores the capacity to predict
distances when the SN Ia population is biased from selection effects
(see the discussion on selection biases in
Sect.~\ref{sec:select-bias-corr}) and enables proper error propagation
to distances. However, the accuracy of bias corrections depends on the
accuracy of the underlying assumptions about the scatter model. Thus,
the exact impact on cosmology depends on subtle interplay between the
SN model training, the light-curve fitting to determine distances, and the bias corrections.

The SALT2 model assumes there is no correlation of the scatter between
different photometric bands. As part of the JLA analysis, the impact
of this assumption was tested recently in \citet[hereafter
\ktreize]{2013ApJ...764...48K}, on simulated samples drawn from models
with correlations. In particular, models of the intrinsic scatter were
built from measurements of spectral scatter in SNFactory data
\citep{Chotard2011}. This analysis, however, did not include training
the SALT2 model on the simulated sample. Based on the same models of
intrinsic scatter, \citetalias{mosher} is the first analysis to
evaluate the effect using the entire analysis chain, including the
training of the SALT2 model and the bias correction. The
\citetalias{mosher} results, therefore, supersede those in
\citetalias{2013ApJ...764...48K}. They show that, in the most adverse
case, the bias on reconstructed distance moduli is less than 0.03 mag
at very high redshifts that are most sensitive to the rest-frame UV
region.

\citetalias{mosher} does not test explicitly for variations of
spectral features, in particular the strong variations of Calcium
features in the near UV. However, by introducing broadband magnitude
scatter trained on measured spectral scatter as given in
\citet{Chotard2011}, the simulation implicitly includes the impact of
spectral features on broadband magnitudes.

\begin{figure}
  \centering
  \igraph{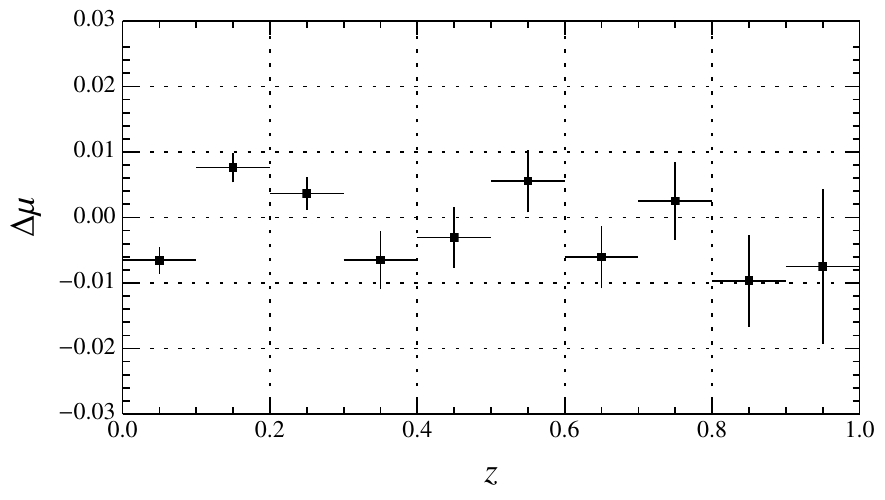}
  \caption{Bias in reconstructed distance modulus as a function of
    redshift. Simulations follow the color variation model described
    in \citet[Sect. 3.1]{2013arXiv1306.4050S}. The simulated sample
    includes low-$z$, SDSS-II and SNLS SNe Ia and is representative of
    our JLA sample described in
    Sect.~\ref{sec:sn-cosmology-sample}. The analysis of the simulated
    sample includes the bias correction described in
    Sect.~\ref{sec:select-bias-corr} (computed under the baseline
    SALT2 assumptions). The SALT2 model was not retrained on the
    simulated sample, similarly to what is done in
    \citet{2013ApJ...764...48K}.}
  \label{fig:scolnic}
\end{figure}

As noted recently by \citet[hereafter S14]{2013arXiv1306.4050S}, the
intrinsic scatter models from \citetalias{2013ApJ...764...48K} that
are used in \citetalias{mosher} do not vary the assumptions regarding
the distribution of extrinsic color. S14 propose an alternative model
where extinction is attributed entirely to interstellar dust with
properties similar to dust in the Milky Way (\emph{i.e.}, $\beta
\approx 4.1$). This reddening is then smeared by intrinsic color
variations uncorrelated with brightness, explaining both the observed
color distribution and the low recovered $\beta$ value. This model is
similar to the model labeled ``H-C11''\footnote{H-C11 definition: H
  refers to the \citet{Hsiao2007} spectral sequence that is modified
  with time stretching and the G10 color law; C11 refers to the
  intrinsic scatter model based on \citet{Chotard2011}, and it
  includes much more color variation compared to the G10 scatter
  model.} in \citetalias{mosher}. The differences are the value of
$\beta$, set to 4.1 in S14 instead of 3.1 in \citetalias{mosher}, and the
underlying distribution of extrinsic color, which in the S14 model
is 
strongly asymmetric with $c > -0.1$, instead of only slightly
asymmetric in \citetalias{mosher}. Biasing of distances arises if the
simulated bias corrections are based on an incorrect model of
intrinsic scatter. As the assumptions about intrinsic dispersion
remain the same in both cases, the changes proposed in S14 are not
expected to significantly alter the conclusion from
\citetalias{mosher}.
We check this expectation by modifying the simulations from
\citetalias{2013ApJ...764...48K} to use the S14 model. After applying
the full cosmology analysis, including bias corrections based on the
G10 intrinsic scatter model, the distance-modulus bias as a function
of redshift is shown in Fig.~\ref{fig:scolnic}. The resulting bias is
contained within 0.02~mag and is compatible with the bias obtained in
\citetalias{mosher} for the H-C11 model.

\subsubsection{SALT2 model uncertainties}
\label{sec:model-systematic}

The Hubble diagram biases for realistic simulated samples (mimicking
the JLA sample) and plausible assumptions about the SN Ia dispersion
are given in \citetalias[Fig.~16]{mosher}. They are shown
to be within 0.03 mag in all cases over the entire redshift range.

As part of the systematic uncertainty in the cosmology analysis, we
adopt a light-curve model uncertainty based on these results. We
consider the input model in \citetalias{mosher} that leads to the
largest Hubble diagram bias (labeled G10$^{\prime}$-C11 in
\citetalias[Table~7]{mosher}). We conservatively use the recovered bias as our
model uncertainty. In addition, statistical uncertainties from the
finite training sample size are propagated to distances using the
approximate error propagation described in \citetalias[Appendix
A]{2010A&A...523A...7G}.

\section{The JLA Hubble diagram}
\label{sec:hubble-diagram-its}

In this section, we present distance estimates (and associated
uncertainties) for our JLA SN Ia compilation. Those distance
estimates are based on light-curve parameters fitted with the
retrained SALT2 light-curve model from
Sect.~\ref{sec:joint-training-light}. The distance estimates also
depend on other ingredients; in particular, we rely on models for the:
\begin{itemize}
\item instrument responses (see
  Sect.~\ref{sec:supern-surv-sampl});
\item relation between the SN Ia luminosity and the host
  environment properties (in our case the host stellar mass-luminosity
  relation from Eq.~\ref{eq:mabs});
\item survey selection biases;
\item peculiar velocities of nearby supernovae;
\item extinction by dust in the Milky Way;
\item contamination of the sample by mis-classified non-Ia events.
\end{itemize}
While the parameters of these models are held fixed when fitting for
light-curve parameters and distances, we propagate the uncertainties
associated with these parameters to the cosmological fit, resulting in correlations between the distances of individual SN.

We built upon the C11 analysis that
modeled each item listed above and quantified the associated
systematic uncertainty. We revised only what was necessary to use the extended
dataset and the JLA work on the dominant systematics \citepalias{B12,
  mosher}. With respect to the
C11 analysis, we made the following changes:
\begin{enumerate}
\item The SDSS-II and SNLS light curves have been
  recalibrated.\label{item:18}
\item We use only the SALT2 light-curve fitting method, validated by
  simulation results from \citetalias{mosher}. The amount of
  regularization applied in the training has been validated on the
  simulations, and we include a model uncertainty derived from
  \citetalias{mosher} results.\label{item:12}
\item SALT2 was retrained with SDSS-II SN Ia light curves added to the
  G10 sample as discussed above, to maximize the benefit of the
  improved calibration and extended photometric coverage in both
  wavelength and epoch.\label{item:19}
\item A selection of photometric data from the full SDSS-II
  spectroscopic sample has been added to the cosmology
  sample.\label{item:20}
\item The impact of calibration uncertainties on the
  light-curve model and parameters has been recomputed.\label{item:15}
\item We gathered self-consistent host mass estimates for the full
  cosmology sample including SDSS-II SNe. We also revised some of the
  host mass estimates for the supernovae in the C11
  compilation.\label{item:13}
\item We recomputed selection bias corrections and associated
  uncertainties accounting for refined models of the intrinsic
  dispersion of SNe Ia \citepalias{2013ApJ...764...48K}.\label{item:14}
\item We doubled the systematic uncertainty in the Milky Way dust
  extinction correction described in
  \citetalias[Sect. 5.6]{2011ApJS..192....1C}, to encompass concerns raised by
  \citet{2011ApJ...737..103S} on the accuracy of the 
  \citet{1998ApJ...500..525S} extinction map .\label{item:1}
\item We used observer frame UV photometry from the low-$z$ and SDSS
  surveys. We determined a consistent calibration uncertainty for
  those as described in Appendix~\ref{sec:calibr-syst}.\label{item:16}
\item Based on considerations described in
  Appendix~\ref{sec:calibr-syst}, we revise uncertainties for low-z
  measurements that were {\it not} reported in the natural photometric
  system of their instrument.\label{item:17}
\end{enumerate}
The peculiar velocity model for low-$z$ supernovae, the correction for
Milky Way dust extinction and the estimated contamination of the
sample by non-Ia events are left unchanged with respect to the C11
analysis. We refer the reader to C11 for a description of these
components and the estimate of the associated
uncertainties. Items~\ref{item:18} and~\ref{item:12} have been
extensively treated in Sect.~\ref{sec:supern-surv-sampl}
and~\ref{sec:joint-training-light} respectively. In the rest of this
section, we describe the selection of SDSS-II events for the
JLA cosmology sample (item~\ref{item:20}). We then detail items~\ref{item:15},
\ref{item:13}, and~\ref{item:14} and discuss corresponding
uncertainties. Items~\ref{item:16} and~\ref{item:17} are detailed in 
Appendix~\ref{sec:calibr-syst}.

A substantial effort has been conducted by several authors over the
last few years to investigate potential redshift-dependent biases of
the supernova distances inferred from the simple stretch and color
corrections to supernova magnitudes.  Of concern are i) a correlation
of the intrinsic supernova color (especially UV) with metallicity (a
non exhaustive list of recent works:
\citealt{2012ApJ...748..127F,2012AJ....143..113F,2012ApJ...749..126W,2013ApJ...769L...1F,2012MNRAS.427..103W,2012MNRAS.426.2359M}),
ii) the existence of several subclasses of SNe~Ia with different
intrinsic colors as identified by their correlation with spectral
indicators
\citep{2009ApJ...699L.139W,2011ApJ...734...42N,2011ApJ...742...89F,2013Sci...340..170W},
iii) a potential evolution of the effective dust extinction law due to
a mixture of intrinsic color variation, circumstellar dust, and
extinction in the interstellar medium of the SNe host galaxies
\citep{2011ApJ...735...20A,2013ApJS..207....3S,2013MNRAS.436..222M,2013MNRAS.431L..43J,2013ApJ...779...38P}. In
the context of the present study, because the data do not allow
reliable identification of SN subclasses, we consider a single SN~Ia
population and work with the assumption that any redshift evolution of
SN properties (change of demographics, metallicity, dust properties)
should be also be imprinted in the properties of their host
galaxies. 

Within this hypothesis, redshift evolution is corrected for, on
average, when host galaxy properties are taken into account in the
distance estimate (see Sect.~\ref{sec:host-mass-estimates}). We have
not found evidence of any residual redshift-dependent evolution. For
instance, the average SN color as a function of redshift is well
described by selection effects, and we do not see a significant
evolution of the color-magnitude relation with redshift (more details
can be found in \citetalias{mosher}).

Diffuse intergalactic dust extinction could, however, be undetected in
our analysis and lead to biased distances. \citet{2010MNRAS.406.1815M}
have quantified the bias on $w$ to be as large 0.03 when combining
supernovae, CMB and BAO. Their study is based on a diffuse
intergalactic extinction model constrained by an observed correlation
between the color of distant quasars and the positions of foreground
galaxies. The impact of this potential systematic error has been
re-evaluated in~\citet{2010ApJ...716..712A}, who have obtained a much
smaller systematic uncertainty on $w$ of 0.012. In a recent study,
\citet{2012MNRAS.426.3360J} have further constrained the intergalactic
dust models using both quasar colors and the soft X-ray
background. Because it has a minor contribution, we have not
propagated this source of systematic uncertainty to our JLA analysis,
limiting the differences between our analysis and C11 analysis to the
more crucial points.

As in C11, our estimates of systematic uncertainties are summarized
into contributions to the covariance matrix of the light-curve
parameters.  At the end of this section we describe the statistical
and systematic contributions of the full distance modulus covariance
matrix to be used in the cosmological fits.

\subsection{The JLA cosmology sample}
\label{sec:sn-cosmology-sample}
We add SDSS-II SNe to the cosmology sample with cuts similar to those
imposed for the C11 compilation. The cosmology sample requires SNe in
the Hubble flow ($z>0.01$) but can accommodate less stringent
selection criteria than the training sample. In particular, we do not
impose an upper redshift cut to the sample, since selection bias can
be corrected for as described in Sect.~\ref{sec:select-bias-corr}. We
require the fitted color and shape parameters to lie in the range of
validity of the SALT2 model, and we discard supernovae affected by
strong Milky Way extinction. The cuts on the uncertainty on $t_0$ and
$X_1$, discarding poorly sampled light curves, are loosened since the
resulting uncertainty is accurately propagated to the cosmology fits.
In addition, 25 SNe have been discarded because they are either known
to be peculiar events, or because they have apparent problems in the
lightcurve sampling or photometry. The detailed list of SNe is given
in Appendix~\ref{sec:visual-inspection-1}. The selection requirements
are summarized in Table~\ref{tab:cutscosmo}, and result in \nsdssc
spectroscopically confirmed SDSS-II SNe Ia included in the JLA cosmology sample.
\begin{table}
  \centering
  \caption{Number of SDSS-II supernovae discarded by the successive cuts applied before inclusion in the cosmology sample.}
  \label{tab:cutscosmo}
  \begin{tabular}{lrr}
  
\hline
\hline
 & Discarded & Remaining \\
\hline
Initial & - & 507 \\
\hline
$-3 < X_1 < 3$ & 20 & 487 \\
$-0.3 < C < 0.3$ & 11 & 476 \\
$ E(B-V)_\text{mw} < 0.15$ & 6 & 470 \\
$\sigma(t_0) < 2$ & 19 & 451 \\
$\sigma(X_1) < 1$ & 52 & 399 \\
Other\tablefootmark{a} & 25 & 374 \\
\hline
\end{tabular}
\tablefoot{\tablefoottext{a}{See Appendix~\ref{sec:visual-inspection-1}.}}
\end{table}

For the rest of the sample, including low-$z$, SNLS and HST SNe, we
started from the original C11 selection, which was typically slightly
stricter. We did not allow for any SNe to reenter this part of the
sample, however we discarded a few problematic SNe: three SNLS SNe, 03D4gl
which does not have post-max data, 03D1bk and 04D3cp which are
extremely blue; 1 CfAIII SN the fast declining SN 2001da; and the 5
\emph{HST} SNe named Borg, Ferguson, Greenberg, Sasquatch and
Strolger, whose sampling is not sufficient to give a proper constraint
on the date of maximum.

The resulting cosmology sample includes \ntotc SNe Ia whose
best-fit light-curve parameters according to the retrained SALT2 model
are given in Table~\ref{table:lcfit}. 
The contributions of the different surveys to the final sample is
summarized in Table~\ref{tab:snnumbers}. The corresponding Hubble
diagram is shown in Fig.~\ref{fig:hd}.

\begin{table}
  \centering
  \caption{Contribution of the different surveys to the cosmology sample.}
  \label{tab:snnumbers}
  \begin{tabular}{lr}
    \hline
    \hline
    Source & Number \\
    \hline
    C\'alan/Tololo & \ncalantololoc\\
    CfAI & \ncfaic\\
    CfAII & \ncfaiic\\
    CfAIII\tablefootmark{a} & \ncfaiiic\\
    CSP\tablefootmark{a} & \ncspc\\
    Other low-$z$ & \nlowzc\\
    SDSS\tablefootmark{a} & \nsdssc\\
    SNLS & \nsnlsc\\
    HST & \nriesshstc\\
    \hline
    Total & \ntotc\\
    \hline
  \end{tabular}
  \tablefoot{
  \tablefoottext{a}{Supernovae followed by several surveys are counted only once.}}
\end{table}

\subsection{Host mass estimates for the extended sample}
\label{sec:host-mass-estimates}
\citet{Kelly2010}, \citet{Lampeitl2010} and \citet{Sullivan2010} have
found, in independent samples, a correlation between the shape and
color corrected luminosities of type Ia supernovae and the stellar
mass of their host galaxies (respectively for low-$z$, SDSS-II, and
SNLS SNe). The corrected luminosities are brighter for
supernovae in massive host galaxies:  the average
difference of Hubble residuals between SNe with host stellar masses
that are lower and larger than $10^{10}$~M$_{\odot}$ is found to be
$0.08 \pm 0.02$~mag (from \citealt{Sullivan2010}, Table~5). We 
use this value as a reference for error propagation in the following,
with the label $\Delta_M^\text{ref}$.  This correlation, known as the
``mass step'', could be a consequence of a correlation of both the SN
luminosities and the host masses with the metallicity of the host
galaxy (\emph{e.g.}, \citealt{2009Natur.460..869K}) or with the age of its
stellar population (\emph{e.g.}, \citealt{2010ApJ...719L...5K}).

\citet{Gupta2011} improve the stellar mass estimate of the SDSS-II
sample using UV and near-IR photometry in addition to the SDSS visible
photometry, and they confirm the SN-luminosity correlation with host
mass. \citet{DAndrea2011} (see also \citealt{Konishi2011}) use
host-galaxy spectra of star-forming galaxies to obtain gas-phase
metallicities and star-formation rates. They find that Hubble
residuals are correlated with both quantities, consistent with the
correlations of stellar mass to those same
quantities. \citet{Childress2013} and \citet{2013arXiv1311.6344P} find
similar results at low-$z$ from analyses of the Nearby Supernova
Factory and Palomar Transient Factory data samples. They also find a
correlation between SN Ia intrinsic color and host metallicity. Using
the same data and local measurements of the H$_\alpha$ emission in the
SN neighborhood, \citet{2013arXiv1309.1182R} recently reported a
$3.1\sigma$ difference in shape and color-corrected luminosity between
SNe Ia from H$_\alpha$ emitting regions (SNe Ia$\alpha$) and SNe Ia
from neutral environments ($\Delta M_B^\text{corr}(Ia\epsilon- Ia\alpha) =
-0.094\pm0.031$~mag). They show that invoking a subclass of SNe Ia
specific to passive environments (as traced by the absence of local
H$_\alpha$ emission) and intrinsically brighter by about 0.2~mag,
could explain both the observed difference and the mass step.

Those correlations indicate that the distance estimator of
Equation~\ref{eq:4}, which is based on SN light-curve width and color
corrections to the luminosity, does not capture an important remaining
source of variation in luminosity. Understanding the variation and
optimizing the technique to correct for the variation using broad-band
light curves or spectroscopic data (or both) is currently a subject of
active research. In this paper we use the approach in C11 and fit for
two different absolute magnitude parameters depending on the SN host
stellar mass, with a split at $10^{10}$~M$_{\odot}$. This method
corrects for the induced redshift-dependent bias on distance moduli,
at least approximately. We discuss an alternative model in
Sect.~\ref{sec:host-galaxy-relation}.

We use the host stellar mass estimates of C11 for all SNe except for
those in the SDSS-II sample. For the latter, we use a similar (but
independent) estimate of the masses, also based on the PEGASE spectral
synthesis code \citep{Fioc1997}. The SDSS-II host galaxy masses,
computed according to the technique of \citet{2012ApJ...755...61S},
are listed in \citet{SDSSRELEASE}. We obtain good agreement with the
host masses of the $\sim100$ SDSS-II SNe that were in the C11 sample,
with a dispersion of 0.2 dex, no significant offset ($0.03 \pm 0.03$),
and few outliers.\footnote{This agreement is better than what is
  obtained for the host stellar mass estimates from
  \cite{Gupta2011}. For consistency, we thus use the
  \citet{2012ApJ...755...61S} estimates. However the agreement does
  not exclude possible differences in mass estimates obtained from
  different photometry, and a fully consistent analysis of the host
  galaxy is desirable.}  

The C11 compilation is missing estimates of the host galaxy mass for
61 nearby SNe (mostly because of missing photometry for the host), 36
SDSS-II supernovae and 16 SNLS supernovae for which a host galaxy has
not been clearly identified. These SNe were assigned to the high mass
bin in the C11 analysis but were also assigned a large magnitude error
to account for a potentially incorrect assignment. In our analysis, we
recover estimates for 57 of the 61 missing galaxy mass values, as
described in Appendix~\ref{sec:host-mass-estimates-1}. According to
those new estimates 14 of the 61 SNe-Ia were reassigned to the
low-mass bin. The SDSS-II and SNLS SNe without identified host masses
are assigned to the low mass bin, with an uncertainty on distance
moduli of $\Delta_M^\text{ref}$ added in quadrature to the other sources of
uncertainty.

In the cosmology fit, we use as free parameters $M_B^1$, the absolute
magnitude of supernovae in hosts with $M_\text{stellar}<10^{10}~M_{\odot}$
(following the notation of \citealt{2011ApJ...737..102S}), and
$\Delta_M$, the magnitude offset of supernovae in more massive hosts
(see Equation~\ref{eq:mabs}). The results obtained with this
parameterization are further discussed in
Sect.~\ref{sec:host-galaxy-relation}. As in C11, we also consider the
systematic uncertainty of this correction. Because the stellar mass is
only a proxy for an uncertain environmental property that alters
supernova luminosities, the value of the mass cut is quite
arbitrary. We therefore consider the effects of choosing cuts at
$10^9$ and $10^{11}~M_{\odot}$. We assign correlated uncertainties of
$\Delta_M^\text{ref}$ to those supernovae that change from the low to high mass range
by adding a term to the covariance of the peak brightness:
\begin{equation}
  \tens C_\text{host} = \left( \Delta_M^\text{ref} \right)^2 \vec H_\text{low} \vec H_\text{low}^{\dag} + \left( \Delta_M^\text{ref} \right)^2 \vec H_\text{high} \vec H_\text{high}^{\dag} + \diag(\sigma_\text{host}^2) \label{eq:covmat_hosts}
\end{equation}
where, $^\dag$ denotes the matrix transposition, and for a supernova $i$,
\begin{equation}
  \label{eq:8}
  (\vec H_\text{low})_i = \left\lbrace
      \begin{array}{ll} 
        1 &\quad \text{if}\quad  10^9~M_{\odot} < M_{\text{stellar},i} < 10^{10}~M_{\odot}\,,\\
        0 & \quad \text{otherwise.}
      \end{array}
    \right. \nonumber
\end{equation}
\begin{equation}
  (\vec H_\text{high})_i = \left\lbrace
      \begin{array}{ll} 
        1 &\quad \text{if}\quad  10^{10}~M_{\odot} < M_{\text{stellar},i} < 10^{11}~M_{\odot}\,,\\
        0 & \quad \text{otherwise.}
      \end{array}
    \right. \nonumber
\end{equation}
and $(\sigma_\text{host})_i = \Delta_M^\text{ref}$ if, given the
uncertainty on the host mass, the supernova $i$ may be assigned to one
mass bin or the other. If not, $(\sigma_\text{host})_i = 0$.

\subsection{Bias correction}
\label{sec:select-bias-corr}

Flux-limited surveys are affected by selection biases which impact
the reconstructed distance moduli $\mu$ in a rather complex
manner. A detailed discussion of the biasing mechanisms can be found in
\citetalias[Sect.~6.2]{mosher}.
We determine a correction for $\mu$ in our analysis from simulations
using the SNANA simulation package \citep{2009PASP..121.1028K}. The
bias is computed in redshift bins as
follows: 
\begin{equation}
  \label{eq:11}
   \delta \mu_b(z) = \langle \mu_\text{FIT} - \mu_\text{SIM}\rangle\,,
\end{equation}
where $\mu_\text{SIM}$ is the input distance modulus in the simulation
and $\mu_\text{FIT}$ is the distance modulus reconstructed
  using the SALT2 fit parameters along with $\alpha$ and $\beta$ determined
  from a combined fit to the low-$z$+SDSS+SNLS simulated sample. It is argued in
C11 (\S2.7.2) that the \emph{HST} sample included in our analysis is
essentially free from selection bias. In addition, given the smallness
of the sample, its relative weight in the analysis is small. For these
reasons, we exclude the \emph{HST} sample from the simulations.

The uncertainty of the reconstructed bias is relatively large for two
reasons. First, the effective selection function of each survey is
the result of a combination of effects that are difficult to model
accurately (see \citealt{2008ApJ...682..262D,2010AJ....140..518P} for
evaluations of the SDSS and SNLS selection functions). The selection
of spectroscopic targets, in particular, involves human decisions
influenced by complex operational factors that cannot be simulated
from first principles. The uncertainty in the selection function is
the primary uncertainty in determining the bias. Second, the relation
between the selection function and the distance modulus bias depends
on the details of the underlying SNe Ia model, which are themselves
uncertain.

For the spectroscopic selection function, we computed the data/MC
ratio as a function of peak magnitude after all known selection
requirements are applied (see \citealt{2013ApJ...764...48K} for
details). The cuts are applied to the SDSS and SNLS \footnote{For the
  SNLS sample, some of the \citet{2011ApJS..192....1C} analysis cuts
  were left out of the MC analysis, resulting in 8\% too many
  simulated SNe; we ran additional simulations using these different
  cuts and found that the change in the best fit value of $\Omega_m$
  is negligible ($5\cdot10^{-4}$).} samples, for both data and MC.
For the nearby sample the analysis cuts are rather complicated for the
various subsamples and we therefore define the spectroscopic
selection function to include both the analysis and survey selections.
The SNLS spectroscopic selection function is evaluated as a function
of peak $i$ band magnitude. Modeling of the SDSS spectroscopic
selection requires a function of the peak $g-r$ color in addition
to $r$ band peak magnitudes, as noted in
\citet[Sect. 3]{2013ApJ...764...48K}. A possible explanation for this
color dependence
is that the selection of SDSS spectroscopic targets favored
intrinsically bluer events. Given the finite number of SNe in each
sample, the parameters of the selection functions are subject to
statistical uncertainty, which was estimated by applying our procedure
to 20 simulated random data samples.

The selection functions of the low-$z$ samples are more
uncertain. Most of the low-$z$ sample comes from galaxy targeted
searches where the discovery is not expected to be magnitude
limited. However, the color distribution of the low-$z$ sample
slightly trends to the blue when the redshift increases, indicating
that our sample likely suffer from a slight selection bias. We thus
consider two extreme cases 1) the search is free from selection bias
and 2) the search is entirely magnitude limited so that the selection
function can be determined from data/MC comparisons can be
  determined from data/MC comparisons of the peak B band
  magnitude. We use 2) as the most realistic case as it is better at
reproducing observed distributions. We use the difference between 2)
and 1) as a conservative estimate of the uncertainty on our
choice. This is an acceptable solution given the limited impact of
low-$z$ selection bias on cosmological parameters. In a \LCDM fit,
using one or the other solution shifts the recovered value of
$\Omega_m$ by only 0.004 which is one order of magnitude smaller than
the global uncertainty on this parameter.

\begin{figure}
  \centering
  \igraph{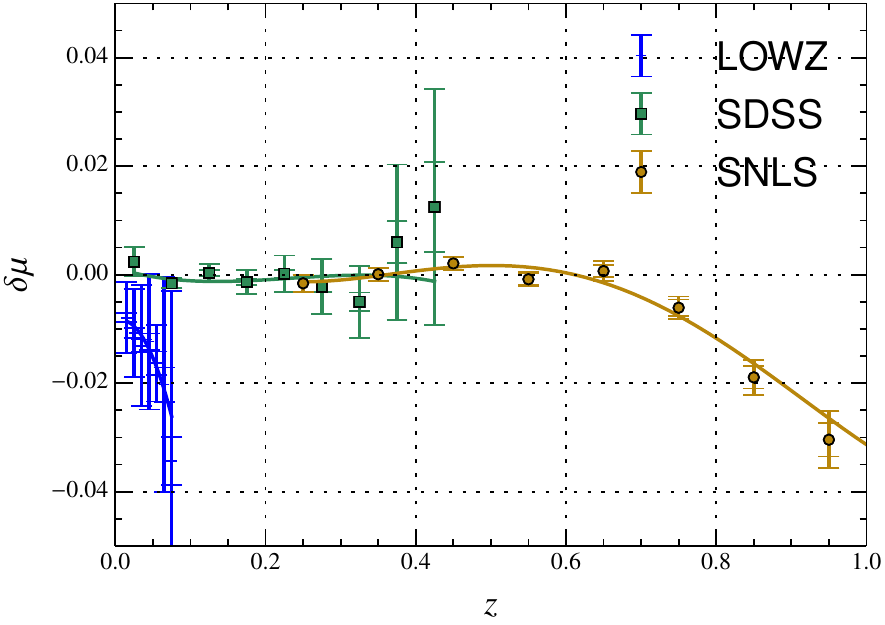}
  \caption{Bias corrections computed from Monte Carlo simulations of
    the cosmological analysis (see Eq.~\ref{eq:11}). Error bars show the statistical
    uncertainty of the correction due to MC noise and uncertainty in
    the selection function. The smallest error bars show the
    contribution from Monte Carlo noise alone.}
  \label{fig:bias}
\end{figure}

The baseline bias correction is shown in Fig.~\ref{fig:bias}. The
simulated statistics are about $400$ times greater than the data
statistics, leading to a small statistical uncertainty in the MC. The
dominant statistical uncertainty is in the determination of the
selection functions. The Monte Carlo results for the low-$z$, SDSS-II
and SNLS samples are fit with smooth polynomial functions,
shown as solid lines in Fig.~\ref{fig:bias}. The apparent brightness
parameter \mstar of each SN Ia is corrected for the value taken by the
corresponding function at the redshift of the supernova. 

The mu-bias from the nearby and SNLS subsamples decreases with
redshift as expected from a selection bias giving brighter SNe Ia with
increasing redshift. The SDSS-II sample, however, has a mu-bias that
is essentially flat with redshift. This flat (or slightly positive)
bias at the high-redshift end of the SDSS-II sample occurs when
intrinsic color scatter is considered. It arises from a compensation
of the selection of positive brightness fluctuations by the selection
of blue color fluctuations. The correction itself is strongly
uncertain (in particular above $z>0.3$) because the color dependency
of the SDSS selection function is itself uncertain. The relatively
flat bias correction for the SDSS data is a feature of the
spectroscopic selection efficiency; the photometric selection used in
\citet{2013ApJ...763...88C} results in a significant increase in bias
with redshift.

The uncertainty in the bias correction is computed by propagating the
statistical uncertainty in the polynominal coefficients to the
uncertainty on \mstar, forming an additional covariance matrix $\tens
C_\text{bias}$. We also estimate systematic uncertainties on the bias
correction by varying uncertain parameters in the simulation, such as
the intrinsic brightness of SNe Ia, the evolution of SN Ia rate and
the underlying cosmology, over the ranges allowed by current data. In
each case, we derive error estimates from the difference between the
baseline analysis and the alternative, and add them to the $\tens
C_\text{bias}$ matrix. We do not consider the uncertainty on the
intrinsic scatter model as a source of systematic error here because
it is already accounted for as part of the light-curve model
uncertainty (Sect.~\ref{sec:light-curve-model}).

\subsection{Propagation of the photometric calibration uncertainties}
\label{sec:prop-phot-calibr}

The systematic uncertainty induced by calibration uncertainties on
light-curve parameters is described by the covariance matrix:
\begin{equation}
  \label{eq:3}
  \tens C_\text{cal} = \tens J \tens C_\calpar \tens J^\dag
\end{equation}
where $\tens C_\calpar$ is the covariance matrix of $\calpar$
described in Sect.~\ref{sec:covar-matr-calibr} and $\tens J$ is the
Jacobian matrix of light-curve parameters with respect to the
calibration uncertainties. Defining the vector of light-curve parameters $\lcpar =
({\mstar}_{,i}, {\xun}_{,i}, \col_{,i})$ where i runs over all the
SNe Ia in the sample, the matrix $\tens J$ can be written as $\tens J =
\left(\partial \lcpar/\partial \calpar\right)$.

\begin{figure}
  \centering
  \igraph{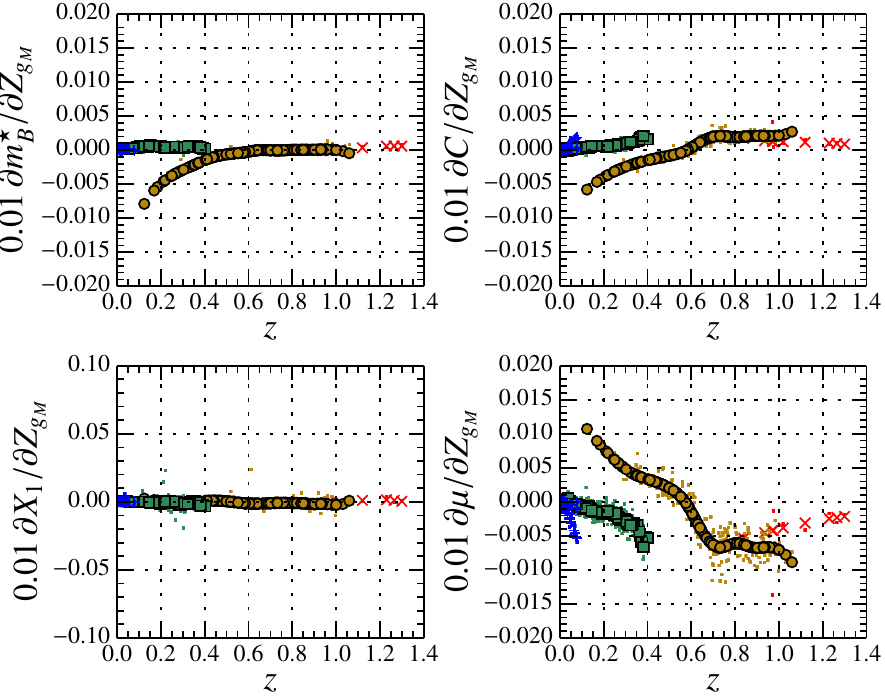}
  \caption{The effect of changing the zero-point in the $g$ MegaCam filter
    by 10~mmag on light-curve parameters and distance modulus as a
    function of redshift for SNe in the low-$z$ (blue crosses),
      SDSS (green squares), SNLS (orange circles) and HST (red x)
      samples. The effect on the distance modulus is computed using
    the fiducial values $\alpha = 0.14$ and $\beta=3.15$. The dots
    show the numerical values of the derivatives (including numerical noise) 
    before smoothing.\label{fig:derivatives}}
\end{figure}

We set up a pipeline to compute $\tens J$ numerically. The
derivatives are obtained by shifting each calibration quantity by a
small amount from its fiducial value ($0.01$~mag for zero-points and
$1$~nm for central wavelengths) and determining both \lcpar and
the SALT2 model spectral surfaces again. Thus, each partial derivative
computation involves a full retraining of the SALT2 model. Neglecting the
impact of the calibration uncertainties on the SALT2 training would result in
a significant underestimate of their effect on the
cosmology. To limit numerical and statistical noise arising in the
computation, the derivatives are smoothed as a function of
  redshift.\footnote{Within each survey, a smoothing spline $s_k(z)$
    is fit to the derivatives. The smoothness of the spline is
    adjusted so that $\sum_n (J_{ik} - s(z_i))^2 = N \sigma^2$ where
    $\sigma$ is an estimate of the SN to SN dispersion computed in
    redshift bins of ten SNe. We also checked that the cosmology result
    was stable when varying the spline smoothness parameter.}  Figure
\ref{fig:derivatives} illustrates the derivatives of light-curve
parameters with respect to a zero-point shift in the MegaCam $g$
filter. All SNe in the sample (not only the SNLS SNe) are affected by
shifts of the SNLS calibration because any change to the training sample 
changes the SALT2 model, which affects all the SNe.

\subsection{The Hubble diagram covariance matrix}
\label{sec:hubble-diagr-covar}

Following the prescription of the previous sections, we assemble a
$3N_\text{SN}\times3N_\text{SN} = 2220\times2220$ covariance matrix for the
light-curve parameters including statistical and systematic
uncertainties.
\begin{equation}
\begin{aligned}
\label{eq:18}
\tens C_{\lcpar} = \tens C_\text{stat} &+ \left(\tens C_\text{cal} + \tens C_\text{model} + \tens C_\text{bias} + \tens C_\text{host} + \tens C_\text{dust}\right)_\text{reevaluated} \\ &+ (\tens C_\text{pecvel} + \tens C_\tens{nonIa})_\text{C11}
\end{aligned}
\end{equation}
$C_\text{stat}$ is obtained from error propagation of light-curve fit
uncertainties as described in \citetalias[Appendix B]
{2010A&A...523A...7G}. We consider a total of $7$ sources of
systematic uncertainty.  The computation of the systematic uncertainty
matrices associated with the calibration $\tens C_\text{cal}$, the
light-curve model uncertainty $\tens C_\text{model}$, the bias
correction uncertainty $\tens C_\text{bias}$, and the mass step
uncertainty $\tens C_\text{host}$ were described above in
Sect.~\ref{sec:prop-phot-calibr},~\ref{sec:light-curve-model},~\ref{sec:select-bias-corr}
and ~\ref{sec:host-mass-estimates}, respectively. The model for
systematic uncertainties in the peculiar velocity corrections and the
contamination of the Hubble diagram by non-Ia are directly taken from
\citetalias{2011ApJS..192....1C}. As done in
\citetalias{2011ApJS..192....1C}, we include Milky-Way extinction as
part of the light-curve fitting model. However, we consider a
correlated systematic uncertainty twice as large ($20$\%) in the
$E(B-V)$ values, to encompass the systematic deviations found in
\cite{2011ApJ...737..103S}. The net effect of this increased
uncertainty is to decrease the weight of extinguished SNe Ia in the
cosmology fit, which reduces the sensitivity of our analysis to
incorrect determinations for Milky Way dust extinction.

The distance estimate of Eq~(\ref{eq:4}) can be rewritten with matrix
notation by forming a matrix\footnote{For the ordering of light-curve
  parameters as $\vec{\eta} = \left((\mstar)_1, (\xun)_1, (\col)_1, \cdots,
    (\mstar)_n, (\xun)_n, (\col)_n\right)$, one would have $\A = \A_0
  + \alpha \A_1 - \beta \A_2$ with $(\A_k)_{i,j} = \delta_{3i,j+k}$.}
$\A$ such that:
\begin{equation} 
  \label{eq:10}
  \muv = \A \lcpar - \vec M_B\,,
\end{equation}
with the components of the intrinsic luminosity vector $\vec M_B$
given by Eq.~(\ref{eq:mabs}).  The covariance matrix of the vector of
distance modulus estimates $\muv$ is 
\begin{equation}
  \label{eq:9}
  \tens C = \A \tens C_{\lcpar} \A^\dag + \diag\left(\frac{5\sigma_z}{z\log 10}\right)^2 + \diag(\sigma_\text{lens}^2 ) + \diag(\sigmaint^2) \,.
\end{equation}
The last three terms account for the uncertainty in cosmological redshift
due to peculiar velocities, the variation of magnitudes caused by
gravitational lensing, and the intrinsic variation in SN magnitude not
described by the other terms.  Our approximation of redshift
uncertainties by magnitude uncertainties is a good approximation only
at low redshift, but this term is negligible at higher redshifts. We
follow C11 in using $c \sigma_z = \SI{150}{\kilo\meter\per\second}$,
as well as $\sigma_\text{lens} = 0.055 \times z$ as suggested in
\citet{2010MNRAS.405..535J}. We now discuss further the estimate of
$\sigmaint$.

In the analysis of C11 (their section 3.4), an intrinsic
variation term, $\sigma_\text{int}$ in that paper, was
determined for each supernova sample, requiring that the best fit
$\chi^2$ per degree of freedom in a specific cosmological fit with
$\Omega_m$ and $w$ as free parameters be equal to 1. It has been
mentioned by several authors, including C11, that this procedure
precludes statistical tests of the adequacy of
the cosmological model to describe the data.

One possibility to circumvent this problem is to introduce additional degrees of
freedom in the fit of the Hubble diagram so that the best fit $\chi^2$
value is dominated by the scatter of the Hubble residuals at
similar redshifts and is insensitive to the choice of the fiducial
cosmological model. 
A simple implementation splits the SNe into several redshift bins and fits an arbitrary 
 average offset for each of bin when determining
the best fit $\chi^2$. 

The restricted log-likelihood (the REML method, see, \emph{e.g.},~\citealt{Harville77})
\begin{equation}
REML = \sum_i w_i (\mu_i - \bar \mu)^2 - \sum_i \log w_i + \log \left(  \sum_i w_i \right) 
\end{equation}
is defined for each bin and is minimized to determine the offset $\bar
\mu$ and $\sigmaint$ for that bin.  The $w_i = \tens C_{ii}^{-1}$ are
the inverses of the diagonal elements of $\tens C$, which contains the
$\sigmaint^2$ terms (Eq.~\ref{eq:9}). The minimum of the REML provides
an unbiased estimate of $\sigmaint^2$.

\begin{figure}
  \centering
  \igraph{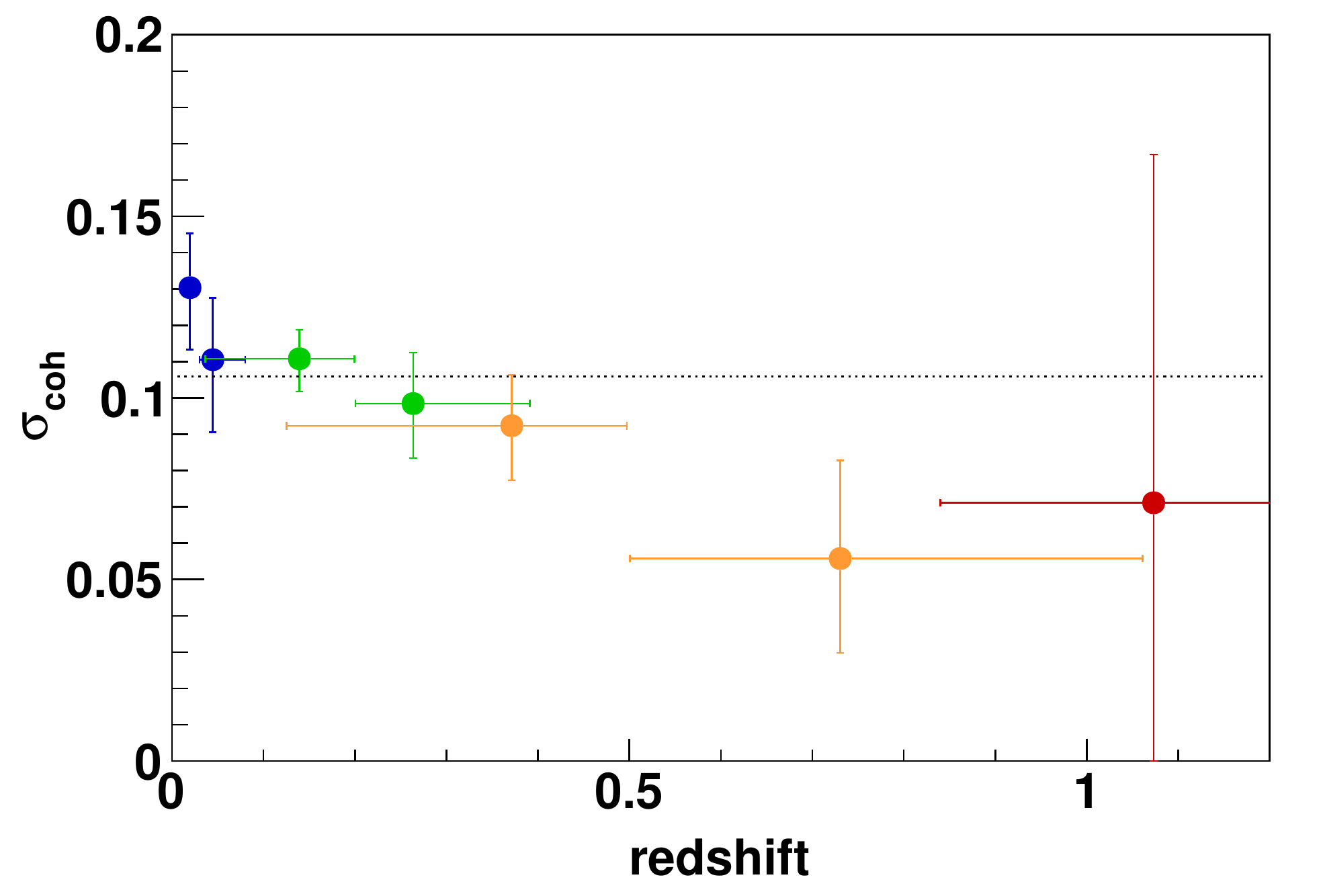}
  \caption{Values of $\sigmaint$ determined for seven subsamples of the Hubble residuals: low-$z$ $z<0.03$ and $z>0.03$ (blue), SDSS $z<0.2$ and $z>0.2$ (green), SNLS $z<0.5$ and $z>0.5$ (orange), and \emph{HST} (red).\label{fig:sigma_int}}
\end{figure}

We compute $\sigmaint$ for only seven subsamples in
order to have sufficient statistical precision: two bins for the low-$z$
sample split at the average redshift of $z=0.03$, two bins for SDSS-II
SNe split at $z=0.2$, two bins for SNLS split at $z=0.5$, and a single
bin for the few \emph{HST} supernovae. The results are shown in
Figure~\ref{fig:sigma_int}. The error bars represent 68\% confidence
levels based on the values of the REML. This likelihood is almost
Gaussian for the subsamples considered here (except for the 
small \emph{HST} sample).

The values of $\sigmaint$ are compatible with a constant value of
$0.106 \pm 0.006$ (with $\chi^2=7.2$ for six degrees of freedom) despite
an apparent trend with redshift.\footnote{One expects a small decrease
  of $\sigmaint$ with redshift because of Malmquist bias: a decrease
  of about 0.01 mag in the high-$z$ SNLS bin has been estimated with
  Monte Carlo simulations.} However, other factors may affect our results including 
survey-dependent errors in estimating the measurement uncertainty, survey dependent errors in calibration, and a redshift dependent tension in the SALT2 model which might arise because different redshifts sample different wavelength ranges of the model.
In addition, the fit value of $\sigmaint$ in the first redshift bin
depends on the assumed value of the peculiar velocity dispersion (here
$150 {\rm km \cdot s^{-1}}$) which is somewhat uncertain.

We follow the approach of C11 which is to use one value of $\sigmaint$
per survey. We consider the weighted mean per survey of the values
shown in Figure~\ref{fig:sigma_int}. Those values are listed in
Table~\ref{table:sigma_coh} and are consistent with previous analysis
based on the SALT2 method
\citep{2011ApJS..192....1C,2013ApJ...763...88C}.

\begin{table}
\centering
\begin{tabular}{lc}
\hline
\hline
sample & $\sigmaint$\\
\hline
low-$z$ & 0.12\\
SDSS-II & 0.11\\
SNLS & 0.08\\
HST & 0.11\\
\hline
\end{tabular}
\caption{Values of $\sigmaint$ used in the cosmological fits. Those values correspond to the weighted mean per survey of the values shown in Figure~\ref{fig:sigma_int}, except for \emph{HST} sample for which we use the average value of all samples. They do not depend on a specific choice of cosmological model (see the discussion in \S\ref{sec:hubble-diagr-covar}).\label{table:sigma_coh}}
\end{table}

\section{\LCDM constraints from SNe Ia alone}
\label{sec:lcdm-constr-from}

The SN Ia sample presented in this paper covers the redshift range
$0.01<z<1.2$. 
This lever-arm is sufficient to provide a stringent constraint on a
single parameter driving the evolution of the expansion rate. In
particular, in a flat universe with a cosmological constant (hereafter
\LCDM), SNe Ia alone provide an accurate measurement of the reduced
matter density $\Omega_m$.  However, SNe alone can only measure ratios
of distances, which are independent of the value of the Hubble
constant today ($H_0 = 100 h\,
\si{\kilo\meter\per\second\per\mega\parsec}$). In this section we
discuss \LCDM parameter constraints from SNe Ia alone. We also detail
the relative influence of each incremental change relative to the C11
analysis.

\subsection{\LCDM fit of the Hubble diagram}
\label{sec:lcdm-fit-hubble}

\begin{figure}
  \centering
  \igraph{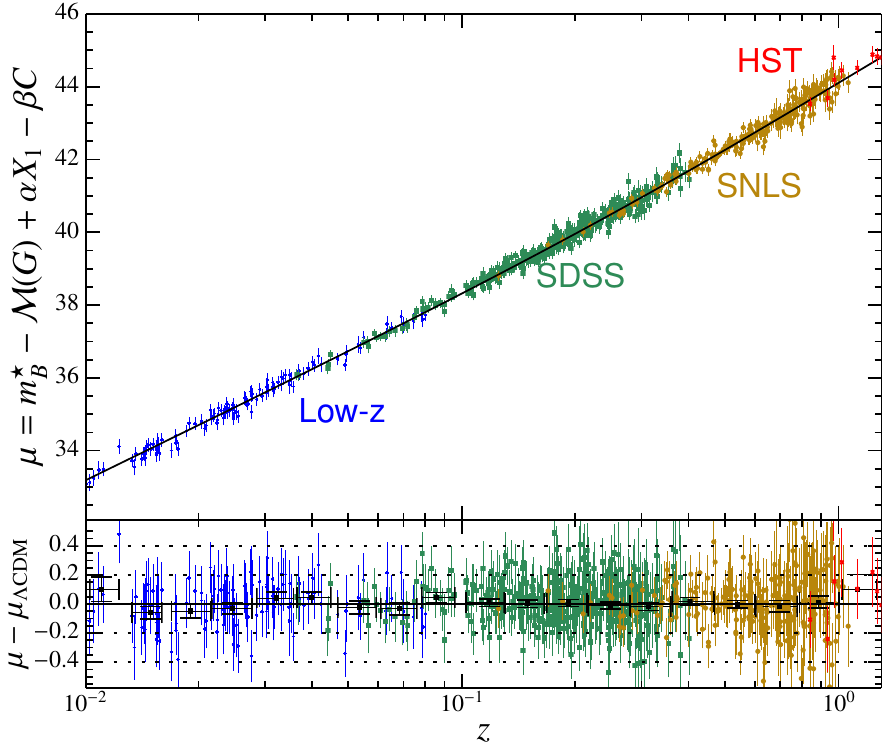}
  \caption{\emph{Top:} Hubble diagram of the combined
    sample. The distance modulus redshift relation of the
    best-fit \LCDM cosmology for a fixed $H_0 =
    \SI{70}{\kilo\meter\per\second\per\mega\parsec}$ is shown as the
    black line. \emph{Bottom:} Residuals from the best-fit \LCDM
    cosmology as a function of redshift. The weighted average of the residuals
     in logarithmic redshift bins of width $\Delta z / z \sim 0.24$ are
    shown as black dots. 
  }
  \label{fig:hd}
\end{figure}

Using the distance estimator given in Eq.~(\ref{eq:4}), we fit a \LCDM
cosmology to supernovae measurements by minimizing the following
function:
\begin{equation}
  \label{eq:7}
  \chi^2 = (\hat{ \vec \mu} - \vec \mu_{\LCDM}(z; \Omega_m))^\dag \tens C^{-1} (\hat{ \vec \mu} - \vec \mu_{\LCDM}(z; \Omega_m))
\end{equation}
with $\tens C$ the covariance matrix of $\hat{\vec \mu}$ described in
Sect.~\ref{sec:hubble-diagr-covar} and $\mu_{\LCDM}(z; \Omega_m) = 5
\logdec(d_L(z; \Omega_m)/10 {\rm pc})$ computed for a fixed fiducial
value of $H_0 =
\SI{70}{\kilo\meter\per\second\per\mega\parsec}$,\footnote{This value
  is assumed purely for convenience and using another value would not
  affect the cosmological fit (beyond changing accordingly the
  recovered value of $M_B^1$).} assuming an unperturbed
Friedmann-Lemaître-Robertson-Walker geometry, which is an acceptable
approximation \citep{2013JCAP...06..002B}. The free parameters in the
fit are $\Omega_m$ and the four nuisance parameters $\alpha$, $\beta$,
$M_B^1$ and $\Delta_M$ from Eq.~(\ref{eq:4}). The Hubble
diagram for the JLA sample and the \LCDM fit are shown in
Fig.~\ref{fig:hd}. We find a best fit value for $\Omega_m$ of
\bfomegamlcdm. The fit parameters are given in the first row of
Table~\ref{tab:lcdm}.

\begin{figure}
  \centering
  \igraph{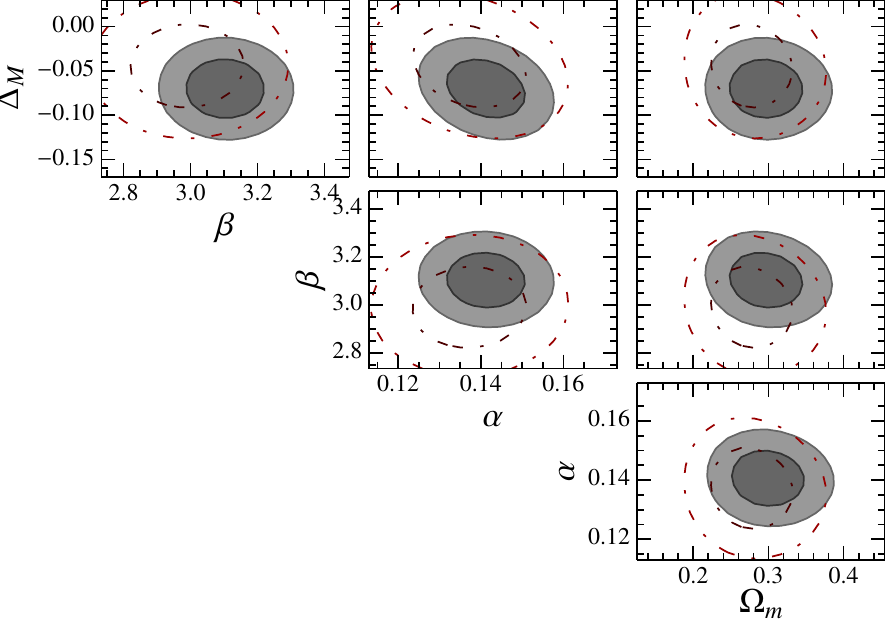}
  \caption{68\% and 95\% confidence contours for the \LCDM fit
    parameters. Filled gray contours result from the fit of the full
    JLA sample; red dashed contours from the fit of a subsample
    excluding SDSS-II data (lowz+SNLS).}
  \label{fig:nuisance}
\end{figure}

For consistency checks, we fit our full sample excluding systematic
uncertainties and we fit subsamples labeled according to the data
included: SDSS+SNLS, lowz+SDSS and lowz+SNLS. Confidence contours for
$\Omega_m$ and the nuisance parameters $\alpha$, $\beta$ and
$\Delta_M$ are given in Fig.~\ref{fig:nuisance} for the JLA and the
lowz+SNLS sample fits. The correlation between $\Omega_m$ and any of
the nuisance parameters is less than 10\% for the JLA sample.
\begin{table*}
  \centering
  \caption{Best-fit \LCDM parameters for SNe Ia alone.}
  \label{tab:lcdm}
  \begin{tabular}{l*{6}{c}}
    
\hline
\hline
 & $\Omega_m$ & $\alpha$ & $\beta$ & $M_B^1$ & $\Delta_M$ & $\chi^2/{\rm d.o.f.}$\\
\hline
JLA (stat+sys) & $0.295 \pm 0.034$ & $0.141 \pm 0.006$ & $3.101 \pm 0.075$ & $-19.05 \pm 0.02$ & $-0.070 \pm 0.023$ & $682.9/735$\\
JLA (stat) & $0.289 \pm 0.018$ & $0.140 \pm 0.006$ & $3.139 \pm 0.072$ & $-19.04 \pm 0.01$ & $-0.060 \pm 0.012$ & $717.3/735$\\
\hline
SDSS+SNLS (stat+sys) & $0.311 \pm 0.042$ & $0.140 \pm 0.007$ & $3.140 \pm 0.082$ & $-19.04 \pm 0.03$ & $-0.072 \pm 0.025$ & $577.9/608$\\
SDSS+SNLS (stat) & $0.305 \pm 0.022$ & $0.139 \pm 0.007$ & $3.178 \pm 0.079$ & $-19.03 \pm 0.01$ & $-0.062 \pm 0.013$ & $599.7/608$\\
lowz+SDSS (stat+sys) & $0.337 \pm 0.072$ & $0.145 \pm 0.007$ & $3.059 \pm 0.093$ & $-19.02 \pm 0.03$ & $-0.088 \pm 0.028$ & $445.4/487$\\
lowz+SDSS (stat) & $0.298 \pm 0.052$ & $0.144 \pm 0.007$ & $3.096 \pm 0.090$ & $-19.04 \pm 0.02$ & $-0.059 \pm 0.015$ & $471.9/487$\\
lowz+SNLS (stat+sys) & $0.281 \pm 0.043$ & $0.138 \pm 0.009$ & $3.024 \pm 0.107$ & $-19.08 \pm 0.03$ & $-0.045 \pm 0.033$ & $315.0/352$\\
lowz+SNLS (stat) & $0.282 \pm 0.023$ & $0.139 \pm 0.009$ & $3.074 \pm 0.104$ & $-19.05 \pm 0.02$ & $-0.060 \pm 0.018$ & $336.0/352$\\
\hline
  \end{tabular}
\end{table*}

The \LCDM model is already well constrained by the SNLS and low-$z$
data thanks to their large redshift lever-arm. However, the addition
of the numerous and well-calibrated SDSS-II data to the C11 sample
is interesting in several respects. Most importantly, cross-calibrated
accurately with the SNLS, the SDSS-II data provide an alternative
low-$z$ anchor to the Hubble diagram, with better understood
systematic uncertainties.  This redundant anchor adds some weight in
the global \LCDM fit, thanks to high statistics, and helps in the
determination of $\Omega_m$ with a 25\% reduction in the total
uncertainty.

\begin{figure}
  \centering
  \igraph{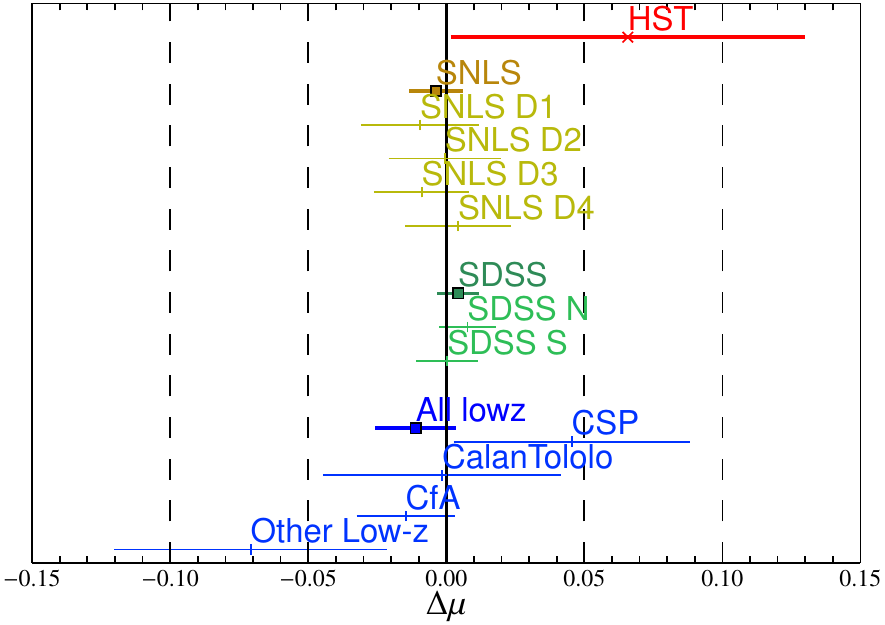}
  \caption{Residuals from the \LCDM fit of the JLA Hubble diagram by
    survey. The weighted bin average and error bars are computed
    without systematic uncertainties. We further distinguish the
    different low-$z$ surveys, the four non-contiguous SNLS fields, and
    the nothern and southern part of the SDSS stripe.}
  \label{fig:consistency}
\end{figure}

The complete redshift coverage makes it possible to assess
the overall consistency of the SN data with the \LCDM model. Residuals
from the \LCDM fit can be seen for the entire redshift range in the
bottom panel of Fig.~\ref{fig:hd}. The consistency is better assessed
when residuals are binned by survey as in
Fig.~\ref{fig:consistency}. A notable feature of this plot is the high
statistical precision of the SDSS data, constraining the mean relative
distance at $z\sim0.16$ with an accuracy of $0.007$~mag (statistical
error only).

The large number of SNe from the SDSS-II also improves constraints on
the nuisance parameters as can be seen in Fig.~\ref{fig:nuisance}. In
particular, the mass step parameter is measured more accurately with
SDSS-II data (Table~\ref{tab:lcdm}, row 3 vs. 5). 

\begin{figure}
  \centering
  \igraph{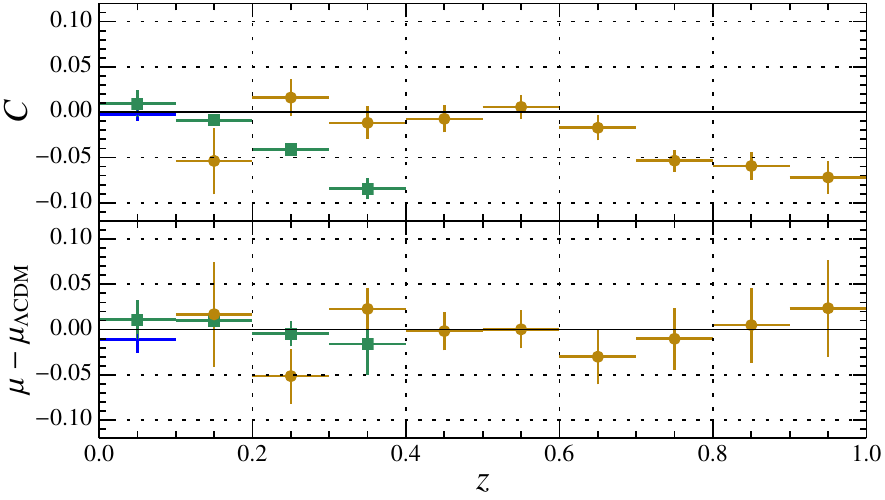}
  \caption{\emph{Top:} Average color of the samples in redshift
    bins. The low-$z$ (blue crosses), the SDSS-II (green squares) and
    the SNLS (orange circles) samples are binned separately. The plot
    provides a visual assessment of the selection bias affecting each
    survey. \emph{Bottom:} Residuals from the \LCDM fit of the JLA
    Hubble diagram as a function of redshift. The different
    surveys are shown separately.}\label{fig:hdres}
\end{figure}

Finally, the region of overlap between the SDSS-II and SNLS in the
redshift range $0.2 < z < 0.4$ provides an opportunity to assess the
accuracy of the bias corrections
(Sect.~\ref{sec:select-bias-corr}). In this redshift range, we can
compare distance estimates from the complete SNLS sample to the
distance estimates from the incomplete SDSS-II sample. The upper panel
in Fig.~\ref{fig:hdres} presents the evolution with redshift of the
mean color of the SNLS and SDSS-II samples. The residuals from the
Hubble diagram are shown on the lower panel in the same plot. The
shift in mean color induced by selection bias becomes about
$-0.08$~mag at $z\sim0.35$ for the SDSS-II sample (a value comparable
to the bias in the SNLS sample at $z\sim0.9$). Nevertheless, distances
measured from the SDSS-II and SNLS SNe at $z\sim0.35$ are in good
agreement, giving confidence in the bias corrections. For example, a
$\beta$ value error of $-1$ would appear in this plot as a bias of
$\sim-0.08$ in SDSS-II distances at $z\sim0.35$, a possibility which
is excluded by the data.

\subsection{The relative importance of the sources of uncertainty}
\label{sec:relat-import-syst}

Sect.~\ref{sec:hubble-diagr-covar} presents our composite model
(Eq.~\ref{eq:18}) of the measurement error. To gain
insight into the relative importance of each component, we decompose the
variance $V$ of the fit parameter $\Omega_m$. Close to the
likelihood maximum, 
the fit parameters $\vec{\theta}$ are determined from the measurements by:
\begin{equation}
\vec{\theta} = (\tens J^\dag \tens C^{-1} \tens J)^{-1} \tens J^\dag \tens C^{-1} \A \lcpar \label{eq:22}
\end{equation}
where $\tens J$ is the Jacobian matrix at the maximum 
likelihood. Defining $\tens W = (\tens J^\dag \tens C^{-1} \tens
J)^{-1} \tens J^\dag \tens C^{-1} \A$, we evaluate the contribution
$V_x$ of each component $x$ from Eq.~(\ref{eq:18}) using
\begin{equation}
  \label{eq:19}
  \tens V_x = \tens W \tens C_x \tens W^\dag\,,
\end{equation}
We report the diagonal entries of $V_x$ for the $\Omega_m$ parameter 
(denoted $\sigma^2_x(\Omega_m)$) in Table~\ref{tab:sensitivity}. As
an aid to interpretation, we also report in Table~\ref{tab:sensitivity}
$\sigma^2_x(\Omega_m)/\sigma^2(\Omega_m)$ as a percentage of the total
variance. These values are not the result of a proper sensitivity analysis 
because the weights are held fixed, but they
provide a useful, qualitative overview of the relative importance of the uncertainties.

\begin{table}
  \centering
  \caption{Contribution of various source of measurement uncertainties to the uncertainty in $\Omega_m$.}
  \begin{tabular}{lcr}
  
\hline
\hline
Uncertainty sources& $ \sigma_x(\Omega_m)$ & \% of $\sigma^2(\Omega_m)$\\
\hline
Calibration & 0.0203 & 36.7\\
Milky Way extinction & 0.0072 & 4.6\\
Light-curve model & 0.0069 & 4.3\\
Bias corrections & 0.0040 & 1.4\\
Host relation\tablefootmark{a} & 0.0038 & 1.3\\
Contamination & 0.0008 & 0.1\\
Peculiar velocity & 0.0007 & 0.0\\
\hline
Stat & 0.0241 & 51.6\\
\hline
  \end{tabular}
  \tablefoot{For the computation of $\sigma_\text{stat}(\Omega_m)$, we include the diagonal terms of Eq.~(\ref{eq:9}) in $\tens C_\text{stat}$.\tablefoottext{a}{We discuss an alternative model for the environmental dependence of the SN luminosity in Sect.~\ref{sec:host-galaxy-relation}.}}
  \label{tab:sensitivity}
\end{table}
 
Calibration uncertainties still stand out as the dominant systematic,
but the improvement in the accuracy of the calibration, made possible
by the joint calibration analysis, results in an uncertainty that is
smaller than the statistical uncertainty. And fitting our sample using
the calibration uncertainties from C11 would have
produced 
a $15$\% increase in the uncertainty, with the contribution from
calibration uncertainty dominating all the other sources. On the other
hand, in spite of a conservative estimate, the uncertainty on the bias
correction does not significantly affect the overall accuracy of the
$\Omega_m$ estimate.

Uncertainties associated with the SALT2 model and host relation are
still subdominant assuming that the standardization model of
Eq.~(\ref{eq:4}) holds and, in particular, that the
host-mass-luminosity relation of Eq.~(\ref{eq:mabs}) captures the full
effect of the environmental dependence. As already mentioned, the
subject is an open question, and we discuss it further below.

\subsection{Assessment of the mass step correction}
\label{sec:host-galaxy-relation}

Recent analyses of large samples of type Ia supernovae have produced
evidence for a remaining environmental dependence of the SN Ia shape
and color-corrected luminosities. Correlations were found (see
Sect.~\ref{sec:host-mass-estimates}) between the Hubble residuals and
several characteristics of host galaxies (stellar mass, star formation
rate, inferred stellar age, metallicity) which evolve with redshift
and are therefore likely to cause 
 a bias if not corrected. Unfortunately, no correction for these effects based on 
measured SN Ia light-curve properties is known.

The most significant empirical correlation is with the host
mass. Therefore, a correction for this effect was adopted in the C11
analysis, which we also use in the present analysis. It takes the form
given in Eq.~(\ref{eq:mabs}), namely a step function of the host mass,
which is the functional form suggested by current data (see,
\emph{e.g.}, \citealt{Childress2013,2012arXiv1211.1386J}).

\begin{figure}
  \centering
  \igraph{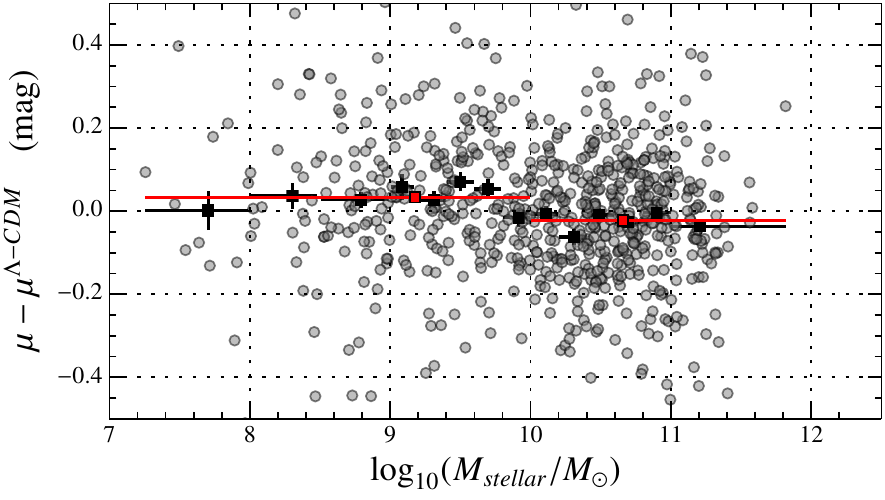}
  \caption{Residuals from the \LCDM fit of the JLA hubble diagram as a
    function of the host galaxy mass. The fit does not include the
    mass step correction. Binned residuals are shown as black
    squares. The red line shows the mass step correction for a step at
    $M_{\rm stellar} = 10^{10} M_\odot$.}
  \label{fig:hostmass}
\end{figure}

We confirm the measurement of a non-zero mass dependent step in Hubble
residuals at $5\sigma$ in our sample. In the framework of \LCDM, we
determine $\Delta_M = $\deltam for the full JLA sample, including all
systematic uncertainties except the uncertainty from the mass step
correction itself (Eq.~\ref{eq:covmat_hosts}). The Hubble residuals of
the JLA sample as a function of the host galaxy stellar mass are shown
in Fig.~\ref{fig:hostmass}.

Since there is no clear understanding of the underlying phenomena, it
is important to explore possible models for this apparent mass step
effect. \citet[Sect. 6.1.2]{2013arXiv1309.1182R} propose an
alternative explanation for the mass step origin that involves a
subclass of SNe Ia, peculiar to passive environments, that are about
0.26 mag brighter than the bulk of the population after
standardization. In this model, the mean intrinsic magnitude of SNe Ia
in passive and active environment differs by a quantity denoted
$\Delta_\alpha$ due to this subclass. The subclass is also
subdominant in low-mass host galaxies, explaining the observed mass
step. Assuming that the proportion of SNe Ia from active environments
follows the specific star formation rate, this model predicts that an
evolution of the induced mass step with redshift is possible, in which
case a redshift-independent mass step correction is incorrect.

\begin{figure}
  \centering
  \igraph{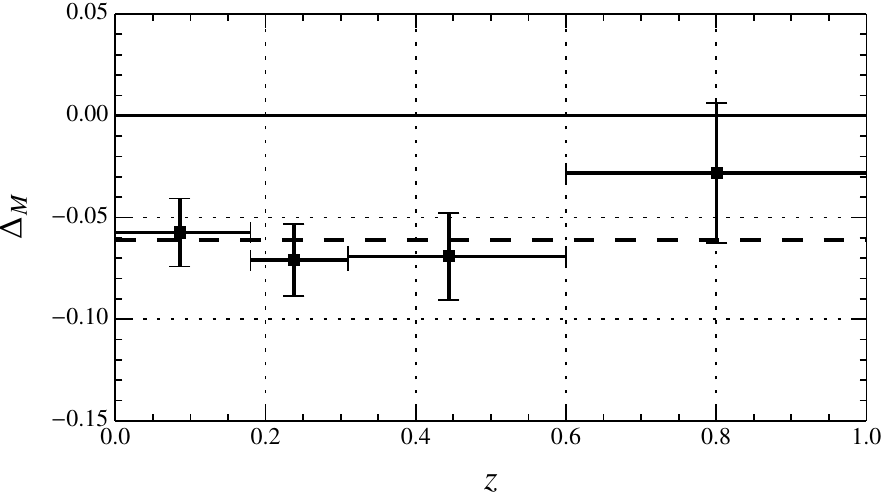}
  \caption{Measurements of the mass step in four redshift bins: $z
      < 0.18$, $0.18 \leq z < 0.31$, $0.31 \leq z < 0.6$, $z >
      0.6$. The dashed line shows the measured value for a
      redshift-independent mass step.}
  \label{fig:msbins}
\end{figure}

In this model, the predicted bias on cosmology can be computed and is
directly related to the evolution of the mass
step. Fig.~\ref{fig:msbins} shows the mass steps measured as a
function of redshifts for the JLA sample. Our data does not show any
significant evolution of the mass step with redshift and therefore
allowing for an evolution of the mass step in the cosmology fit has
little effect on the result, shifting $\Omega_m$ by only
$\omshiftvarm$ for example. Further splitting Hubble residuals between
globally passive and globally star-forming hosts in the SNLS and SDSS
subsamples does not show measurable difference after correcting for
the mass-step. A significant remaining environmental bias unrelated to
the mass effect is therefore unlikely.

Alternatively, \citet{2013arXiv1312.1688S} address the issue of
  environmental dependence by introducing three independent pairs of
$M$ parameters in three redshift bins. This approach suppresses the
cosmological information on SN distance ratios across those bins. In
our sample, it is roughly equivalent to fitting only the low redshift
slice ($z<0.5$) which is the one with the largest weight. Again the
shift in cosmological parameter is small ($\Delta \Omega_m = 0.007$)
and compatible with the statistical fluctuation expected from the
change in the model.\footnote{The expected rms of this change,
  evaluated on simulations including calibration uncertainties, is
  $0.02$.} The recovered $M$ parameters are also compatible across the
three bins.

Overall, our sample does not provide evidence for a significant
evolution of the mass step in the covered redshift
range.\footnote{Similar conclusions hold for the $w$-CDM model fit in
  combination with \emph{Planck}.}  Because the evidence for this
model is weak and the uncertainty of our baseline model is already
consistent with the results from alternative models, we have
not increased the systematic error associated with variations in the
SN environment, leaving our analysis of this effect the same as that
of C11.

\subsection{Comparison with the C11 analysis}
\label{sec:comp-with-snls3}

\begin{table*}
  \centering
  \caption{Drift in the parameters with respect to the C11 analysis.}
  \label{tab:lcdmdrift}
  \begin{tabular}{l*{6}{c}}
    
\hline
\hline
 & $\Omega_m$ & $\alpha$ & $\beta$ & $M_B^1$ & $\Delta_M$ & $\chi^2/{\rm d.o.f.}$\\
\hline
C11 Combined  (stat+sys) & $0.228 \pm 0.038$ & $1.434 \pm 0.093$ & $3.272 \pm 0.100$ & $-19.16 \pm 0.03$ & $-0.047 \pm 0.023$ & $428.8/467$\\
C11 SALT2  (stat+sys) & $0.249 \pm 0.043$ & $1.708 \pm 0.156$ & $3.306 \pm 0.109$ & $-19.15 \pm 0.03$ & $-0.044 \pm 0.024$ & $395.1/468$\\
C11 SiFTO  (stat+sys) & $0.225 \pm 0.038$ & $1.360 \pm 0.072$ & $3.401 \pm 0.111$ & $-19.15 \pm 0.03$ & $-0.047 \pm 0.022$ & $439.1/463$\\
C11 SALT2  (stat) & $0.246 \pm 0.018$ & $1.367 \pm 0.071$ & $3.133 \pm 0.087$ & $-19.15 \pm 0.02$ & $-0.065 \pm 0.015$ & $484.9/468$\\
C11 SiFTO  (stat) & $0.272 \pm 0.016$ & $1.366 \pm 0.059$ & $3.049 \pm 0.078$ & $-19.12 \pm 0.01$ & $-0.064 \pm 0.013$ & $509.8/463$\\
\hline
C11-reanalized (stat only) & $0.230 \pm 0.018$ & $0.140 \pm 0.008$ & $2.771 \pm 0.085$ & $-19.06 \pm 0.02$ & $-0.053 \pm 0.016$ & $427.3/453$\\
C11-recalibrated (stat only) & $0.291 \pm 0.022$ & $0.136 \pm 0.009$ & $2.907 \pm 0.095$ & $-19.02 \pm 0.02$ & $-0.061 \pm 0.017$ & $407.8/453$\\
\hline
JLA (stat) & $0.289 \pm 0.018$ & $0.140 \pm 0.006$ & $3.139 \pm 0.072$ & $-19.04 \pm 0.01$ & $-0.060 \pm 0.012$ & $717.3/735$\\
\hline
  \end{tabular}
  \tablefoot{The difference in $\alpha$ between the C11 and JLA samples is due to a different parameterization of light-curve shapes: while for the C11 sample, a stretch parameter $s$ is reported, we use the SALT2 \xun parameter which is roughly $10\times (s-1)$. The high value of $\alpha$ in the ``SALT2 (stat)'' case was due to a convergence problem in the computation of a covariance matrix. We confirmed that fixing this problem did not change the recovered values of $\Omega_m$.}
\end{table*}

Our best fit value for $\Omega_m$ differs from the value published in
C11 ($\Omega_m=$\omtrois) by $\diffomsigma\sigma$ (stat+sys). This
discrepancy is not simply a statistical fluctuation because a large
part of the data sample remains the same. The C11 value was dependent
on the light-curve model and we discuss that difference below.

\subsubsection{SALT2/SiFTO differences in the C11 analysis}
\label{sec:salt2s-diff-snls3}

The upper part of Table~\ref{tab:lcdmdrift} provides the best-fit
\LCDM parameters for the C11 sample. The ``C11 combined'' analysis
combines light-curve parameters derived from the SALT2 and SiFTO
light-curve models. We also report separately the results obtained
from the two models.

As previously noted in \citet[Sect. 5.4]{2013arXiv1303.5076P} the
light-curve parameters and covariance matrices obtained for the SiFTO
light-curve model lead to an $\Omega_m$ value significantly lower than
that obtained from the SALT2 analysis. 
Interestingly, the comparison of the SiFTO and SALT2 analyses obtained
when the systematic uncertainties are not taken into account (``stat''
rows in Table~\ref{tab:lcdmdrift}) produces a difference with the
opposite sign, with the SiFTO value increasing by $0.047$ ($>2\sigma$
stat). Applying SiFTO systematic uncertainties to the SALT2 Hubble
diagram results in a smaller but still significant shift ($0.027$ in
$\Omega_m$).

The inclusion of systematic uncertainties in the fit, particularly
calibration uncertainties, alters the relative weight of each survey
in the global fit. The shift in $\Omega_m$ when the weighting is
changed is evidence of tensions between the C11 Hubble diagram and the
\LCDM model.  The present analysis does not exhibit the same
behavior. The best-fit parameters do not change significantly when
including systematic uncertainties in the fit, as witnessed by the two
first rows in Table~\ref{tab:lcdm}.\footnote{This statement continues to be true even when
we artificially vary the weights. As an illustration, we can
  fit \LCDM to the subsample of the JLA supernovae that are part of
  the ``C11 SiFTO'' sample, but using the weights resulting from the
  systematics of the C11 SiFTO analysis.
  We then recover an $\Omega_m$ value that differs by only $0.013$
  ($\sim0.5\sigma$) from the stat-only value.} In particular the difference in
$\Omega_m$ is only 0.006.

We conclude that the SiFTO/SALT2 differences observed in the C11
analysis are related to tensions between the C11 datasets (before the
recalibration of the present study) and the \LCDM model, and that the
tensions are reduced by the recalibration. Those tensions result in
differences between SALT2 and SiFTO analyses because of different
weights assigned to SNe on the Hubble diagram rather than differences
in the models.

\subsubsection{Differences with the C11 SALT2 analysis}
\label{sec:diff-with-snls3}

To further understand the relative impact of the changes introduced by
the present analysis, we decompose the list of changes given at
the beginning of Sect.~\ref{sec:hubble-diagram-its}, taking C11
SALT2 (stat) as a starting point. For the purposes of this discussion
we will distinguish three steps. For the first step, we employ the
changes in analysis relative to C11, namely the changes in the SALT2
training procedure (item~\ref{item:12}), the revision of host-mass
estimates (item~\ref{item:13}), the change in the computation of bias
corrections (item~\ref{item:14}) and the revision of low-$z$
measurement uncertainties (items~\ref{item:16}
and~\ref{item:17}). Together, these changes constitute a fully
consistent \emph{reanalysis} of the C11 sample that we label
``C11-reanalyzed''. In a second step we apply the B13 recalibration
to the SNLS and first-year SDSS light curves
(item~\ref{item:18}). This constitutes the
\emph{recalibrated}-reanalysis of the C11 sample that we label
``C11-recalibrated''. Finally, we include the full, recalibrated,
SDSS-II spectroscopic data set in both the training and the cosmology
sample (item~\ref{item:19} and~\ref{item:20}), which is the final JLA
result.

We performed a \LCDM fit for each of these three steps; the results
are reported in the bottom part of Table~\ref{tab:lcdmdrift}. We
consider only the statistical uncertainties in these fits, so that
changes in parameters cannot be attributed to different
interpretations of systematic uncertainties. The full reanalysis of
the C11 sample (C11-reanalyzed) is consistent with the C11 SALT2
results. The recalibration is the most important effect, shifting
$\Omega_m$ by $0.06$ (\emph{i.e.}, $3\sigma$ of the statistical
uncertainty). The most important calibration changes are the revision
of the MegaCam zero-points in the $g$ band (by $0.012$ or
$\sim3\sigma$) and $z$ band (by $0.018$ or $\sim1\sigma$), and the
corrections to the MegaCam $r$ and $i$ filter bandpasses ($\sim3$~nm
on the central wavelength). The revision of SNLS zero-points included
the correction of a sign error (described in
\citetalias[Sect.~10.4]{B12}) and the addition of more calibration
data, including SDSS calibration data and direct observations of
\emph{HST} standards (see Sect.~\ref{sec:supern-surv-sampl}). We note
that, after recalibration, the \LCDM model is a better fit to the data
with a $\chi^2$ decrease of $\sim20$. Adding the full SDSS-II sample to
both the training and cosmology sample leaves the $\Omega_m$ value
virtually unchanged but further reduces the uncertainty.

The $\beta$ values in Table~\ref{tab:lcdmdrift} vary by as much as 0.6, much larger than
the $\sim0.1$ uncertainty. As discussed in M14, the color definition,
training procedure, and selection biases can introduce a bias on beta
that is comparable to the variations seen in Table~\ref{tab:lcdmdrift}. Since $\beta$ is a
nuissance parameter, we do not attempt to report bias-corrected
values. However, its impact on the cosmology analysis has been
included in M14 and in the systematic uncertainties reported here.

\subsection{Differences with the SDSS first year data analysis}
\label{sec:diff-with-sdss}

The first-season SDSS-II analysis \citep{kessler_first-year_2009}
reported two different sets of
distances\footnote{\url{http://das.sdss.org/va/SNcosmology/sncosm09_fits.tar.gz}}
computed using the SALT2 model and the MLCS2k2 model
\citep{2007ApJ...659..122J}. 
Fitting the \LCDM model to the nearby+SDSS sample gives respectively
$\Omega_m = 0.340\pm0.083_\text{stat}$ for the SALT2 model, and
$\Omega_m = 0.278\pm0.084_\text{stat}$ for the MLCS2k2
model.\footnote{Since K09 reports $\Omega_m$ with BAO and CMB priors,
  we have fit their published distances without these priors.} The
difference in these two results was traced to the difference between
light-curve models, particularly in the rest-frame ultraviolet region,
and to the contribution of the observed SN color to the standardized
SN magnitude.  In addition to the obvious improvements in sample size
and calibration uncertainty, this JLA result substantially reduces the
uncertainties in the earlier SDSS-II analysis.

The SALT2 model has been retrained with the full SDSS-II sample,
which, like the lower redshift SN data, contains significant data in
the observer-frame ultraviolet wavelength range. In addition, we have
examined the accuracy and consistency of the other low redshift data
as described in detail in Appendix B.  These studies have resulted in
an improved SALT2 model and model errors that are consistent with the
data. We did not attempt to retrain MLCS2k2, partly because of the
significant effort that would be required, but primarily because we
favor use of SALT2, which models details of the SN Ia spectrum.

Another result of the \citet{kessler_first-year_2009} SALT2 analysis
was evidence for a variation in the effective value of the
standardized magnitude-color correlation parameter $\beta$ with
redshift.  This potential systematic was addressed in G10 and found to
be an artifact of poorly determined model uncertainties. Here we
perform extensive end-to-end simulations including the interrelated
problems of intrinsic scatter, correlations between color and
standardized magnitude, and selection effects (particularly Malmquist
bias).  The basic technique is to measure the sensitivity to different
spectral variations that are consistent with the SALT2 light-curve
residuals. These simulations were not available in previous analyses.
The studies of intrinsic scatter \citepalias{2013ApJ...764...48K} and
systematic uncertainties from the SALT2 method \citepalias{mosher}
show that the SN light-curve data are consistent with a redshift
independent value of $\beta$.

\subsection{Comparison with other measurements of  $\Omega_m$.}
\label{sec:comp-with-other}

\begin{figure}
  \centering
  \igraph{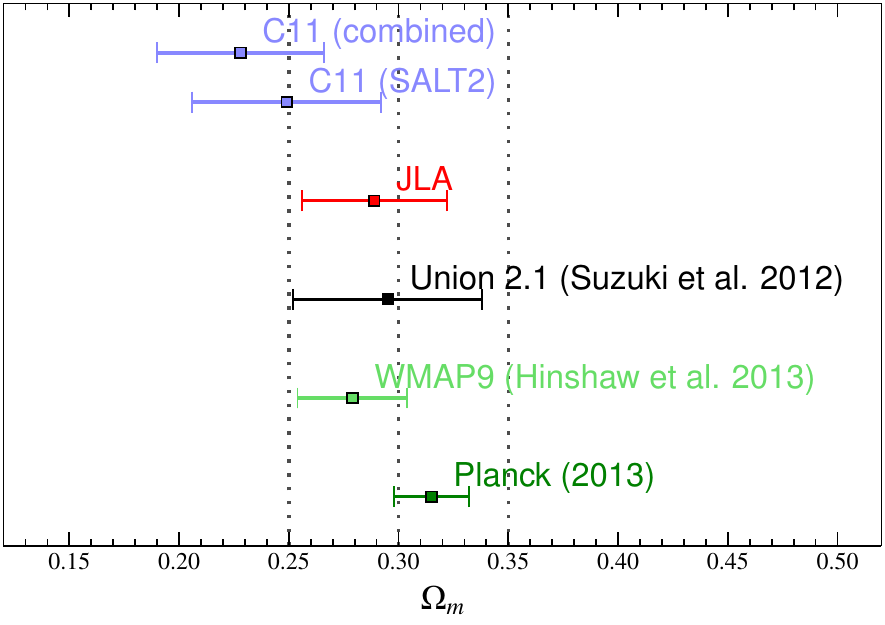}
  \caption{Comparison of various measurements of $\Omega_m$ for a \LCDM
    cosmology.}
\label{fig:frise}
\end{figure}

The comparison of our \LCDM constraints with other analyses and data
sets is summarized in Fig.~\ref{fig:frise}. Our value is in good
agreement with the CMB value from \emph{Planck}
\citep{2013arXiv1303.5076P}, eliminating the previously noted
$\sim2\sigma$ discrepancy between \emph{Planck} and C11. As discussed
in Sect.~\ref{sec:diff-with-snls3}, this change is primarily a result
of the recalibration of the SDSS-II and SNLS light curves. The
recalibration analysis elucidated and corrected an unanticipated
systematic effect (the aging of MegaCam $r$ and $i$ band filters, see
Sect.~\ref{sec:flux-interpr-phot}), and is further bolstered by more
precise and redundant calibration observations. We conclude,
therefore, that the previously found discrepancy should be attributed
to systematic errors in the supernova measurements and that, with our
new analysis, the two probes yield consistent measurements of
$\Omega_m$ in the \LCDM model.
Our value is also compatible with the $\Omega_m$ \LCDM measurement
from \emph{WMAP9} \citep{2013ApJS..208...19H}. The CMB measurement of
$\Omega_m$ with \emph{Planck} and our SN measurement would have
comparable precision, if systematic uncertainties in the SN analysis
were neglected, showing that, despite notable improvements,
systematic measurement uncertainties remain a crucial issue.

Our measurement is also in agreement with the SN Ia measurement from
the Union 2.1 sample \citep{2012ApJ...746...85S}. This is not,
however, a fully independent confirmation as both analyses share part
of the dataset and methodology. There are nonetheless notable
differences between the two SN samples: the second and third years of
the SDSS-II and SNLS surveys, which constitute the large majority of
our sample, are not part of \citet{2012ApJ...746...85S} while the
ESSENCE survey \citep{miknaitis_essence_2007}, most of the high-$z$
\emph{HST} supernovae, as well as some older samples are included
in the Union 2.1 sample.

\section{Dark energy constraints from the combination of supernovae
  and complementary probes}
\label{sec:dark-energy-constr}

The redshift lever arm of our SN Ia sample is insufficient to constrain
all the parameters in more general dark energy models. In this section
we combine SNe Ia with other probes to test extensions of the \LCDM
model. We do not seek to be comprehensive, but restrict our study to
combining our SNe constraints with the most recent measurements of the
CMB fluctuations and of the BAO scale.

\subsection{Complementary data}
\label{sec:complementary-data}

\subsubsection{Power spectrum of the Cosmic Microwave Background
  fluctuations}
\label{sec:planck}
The most recent measurement of the CMB temperature fluctuations has
been provided by the 2013 release of the \emph{Planck} experiment
results \citep{2013arXiv1303.5062P}. This release is based on data
gathered in the first 15.5 months of satellite operation. It delivered
maps of temperature fluctuations over the entire sky in nine frequency
bands (30-857~GHz). The analysis of \emph{Planck} data exploits the
multiwavelength coverage to determine the CMB temperature fluctuation
power spectrum after removing the foreground emissions
\citep{2013arXiv1303.5072P,2013arXiv1303.5075P}. Their results are
summarized by a likelihood function for the CMB spectrum given the
\emph{Planck} data \citep{2013arXiv1303.5075P}.

The CMB temperature power spectrum is directly sensitive to matter
densities and measures precisely the angular diameter distance at the
last-scattering surface ($z\approx1090$). This precise measurement of the
early universe complements very well the SN Ia distance
measurements in the late Universe. The combination produces
constraints on dark-energy models that cannot be obtained from the
CMB alone because of the geometric degeneracy.

For this analysis, we use the \emph{Planck} measurement of the
CMB temperature fluctuations and the \emph{WMAP} measurement of the
large-scale fluctuations of the CMB polarization
\citep{2013ApJS..208...20B}. This combination of CMB data is denoted
``Planck+WP'' to follow the nomenclature used by
\citet{2013arXiv1303.5076P}. We summarize the geometrical constraints
inferred from those measurements by a Gaussian prior on the value of
the baryon density today $\omega_b = \Omega_b h^2$, the cold dark 
matter density today $\omega_c=\Omega_c h^2$, and
$\theta_\text{MC}$ the CosmoMC approximation of the sound horizon angular
size computed from the \citet{1996ApJ...471..542H} fitting
formulae. This combination of parameters is well constrained by the
temperature power spectrum and is independent of any assumptions about
dark energy (for the range of models considered in this paper). The
\emph{WMAP} polarization information slightly improves the \emph{Planck}
constraints by reducing degeneracies, which involve the damping of
small scale fluctuations by reionization and are unresolved by the
temperature spectrum alone.  Our prior has the form:\footnote{Those
  numbers correspond to the best-fit parameters and covariance for the
  exploration of the \emph{Planck} temperature and \emph{WMAP}
  polarization likelihood (Planck+WP in \citet{2013arXiv1303.5076P}
  terminology) to a flat $w$-CDM cosmology as retrieved from the
  Planck Legacy Archive
  \url{http://pla.esac.esa.int/pla/aio/planckProducts.html}}
\begin{equation}
\chi^2_\text{cmb} = (v - v_\text{cmb})^\dag \tens C_\text{cmb}^{-1}
(v - v_\text{cmb})\label{eq:25}
\end{equation}
where:
\begin{equation}
  v_\text{cmb}=(\omega_b, \omega_c, 100\theta_\text{MC})_\text{cmb}
  = (0.022065, 0.1199, 1.041)\label{eq:16}
\end{equation}
and $\tens C_\text{cmb}$ is the best fit covariance matrix for $v$
(marginalized over all other parameters):
\begin{equation}
\tens C_\text{cmb} = 10^{-7}\left(
\begin{array}{ccc}
 0.79039 &  -4.0042 &  0.80608 \\
 -4.0042 & 66.950 & -6.9243 \\
 0.80608 & -6.9243 & 3.9712 \\
\end{array}
\right)\,.
\label{eq:17}
\end{equation}

The use of a distance prior is only an approximate summary of CMB
constraints for dark energy. In particular, the sensitivity of the CMB to the 
late-time growth of structure is neglected. 
However, these effects are small, and
our approximation is known to adequately represent more sensitive
combinations such as CMB+SNe Ia and CMB+BAO (see, \emph{e.g.}, the
discussions in \citealt[Sect. 5.5]{2011ApJS..192...18K} and references
therein). Our approach has the advantage of being purely geometrical
and easy to calculate. We provide a comparison of our results with the full
 \emph{Planck} likelihood
\citep{2013arXiv1303.5075P} in Appendix~\ref{sec:accur-cmb-dist}: in
the case of a flat universe model with a constant equation of state, the
difference in best fit values for $w$ is less than 0.3$\sigma$ and
the uncertainties are the same. 
We provide the tools to use our data in
investigations of more general dark energy models in which the above
approximation is not valid (see Appendix~\ref{sec:data-release}).

\emph{Planck} also provides a reconstruction of the CMB weak-lensing
potential \citep{2013arXiv1303.5077P} that breaks part of the
geometric degeneracy that arises from the CMB temperature spectrum alone. Better
constraints on the foreground contamination of the temperature spectrum
can also be obtained from higher resolution experiments, such as the
Atacama Cosmology Telescope \citep{2013arXiv1301.1037D} and the South
Pole Telescope \citep{2012ApJ...755...70R}. Exhaustive investigations
of constraints provided by the various combinations of CMB data are
conducted in \citet{2013arXiv1303.5076P}. These results suggest little
difference from the additional lensing and high-$\ell$ likelihoods in
dark energy studies when used in combination with later distance
measurements such as SNe Ia and BAOs. Therefore, we do not consider
their use in the present study.

We also present constraints obtained in combination with \emph{WMAP}
for comparison (labeled WMAP9). For this purpose, we use the distance
prior given in \citet[Sect. 4.6.1]{2013ApJS..208...19H}.

\subsubsection{Baryon Acoustic Oscillations}
\label{sec:bao}

The detection of the characteristic scale of the baryon acoustic
oscillations (BAO) in the correlation function of different matter
distribution tracers provides a powerful standard ruler to probe the
angular-diameter-distance versus redshift relation and Hubble
parameter evolution. The BAO scale has now been detected in the
correlation function of various galaxy surveys
\citep{2005ApJ...633..560E,2011MNRAS.416.3017B,2011MNRAS.418.1707B,2012MNRAS.427.3435A},
as well as in the Ly$\alpha$ forest of distant quasars
\citep{2013A&A...552A..96B,2013JCAP...04..026S}. Large-scale surveys
also probe the horizon size at matter-radiation equality. However,
this latter measurement appears to be more affected by systematic
uncertainties than the robust BAO scale measurement.

BAO analyses usually perform a spherical average of their scale
measurement constraining a combination of the angular scale and
redshift separation:
\begin{equation}
  \label{eq:21}
  d_z = \frac{r_s(z_\text{drag})}{D_v(z)}
\end{equation}
with:
\begin{equation}
  \label{eq:20}
  D_v(z) = \left((1+z)^2D_A^2 \frac{cz}{H(z)}\right)^{1/3}
\end{equation}
For this work, we follow \citet{2013arXiv1303.5076P} in using the
measurement of the BAO scale at $z=0.106, 0.35,$ and $0.57$ from
\citet{2011MNRAS.416.3017B,2012MNRAS.427.2132P,2012MNRAS.427.3435A}, respectively. We
consider a BAO prior of the form:
\begin{equation}
\chi^2_\text{bao} = (d_z - d_z^\text{bao})^\dag C_\text{bao}^{-1}
(d_z - d_z^\text{bao})\label{eq:15}
\end{equation}
with $z_\text{drag}$ computed from the \citet{1998ApJ...496..605E}
fitting formulae, $d_z^\text{bao} = (0.336, 0.1126, 0.07315)$ and
$C_\text{bao}^{-1}=\diag(4444, 215156, 721487)$.

\subsection{Constraints on cosmological parameters for various dark
  energy models}
\label{sec:main-results}

We consider three alternatives to the base \LCDM model:
\begin{itemize}
\item the one-parameter extension allowing for non-zero spatial
  curvature $\Omega_k$, labeled o-\LCDM.
\item the one-parameter extension allowing for dark energy in a
  spatially flat universe with an arbitrary constant equation of state
  parameter $w$, labeled $w$-CDM.
\item the two-parameter extension allowing for dark energy in a
  spatially flat universe with a time varying equation of state
  parameter parameterized as $w(a) = w_0 + w_a (1 - a)$ with
  $a=1/(1+z)$ \citep{2003PhRvL..90i1301L} and labeled $w_z$-CDM.
\end{itemize}
We follow the assumptions of \citet{2013arXiv1303.5076P} to achieve
consistency with our prior. In particular we assume massive neutrinos
can be approximated as a single massive eigenstate with $m_\nu =
0.06$~eV and an effective energy density when relativistic:
\begin{equation}
  \label{eq:26}
  \rho_\nu = N_\text{eff} \frac78 \left(\frac4{11}\right)^{4/3} \rho_\gamma
\end{equation}
with $\rho_\gamma$ the radiation energy density and $N_\text{eff} =
3.046$. We use $T_\text{cmb} = 2.7255$~K for the CMB temperature
today.

\begin{figure}
  \centering
  \igraph{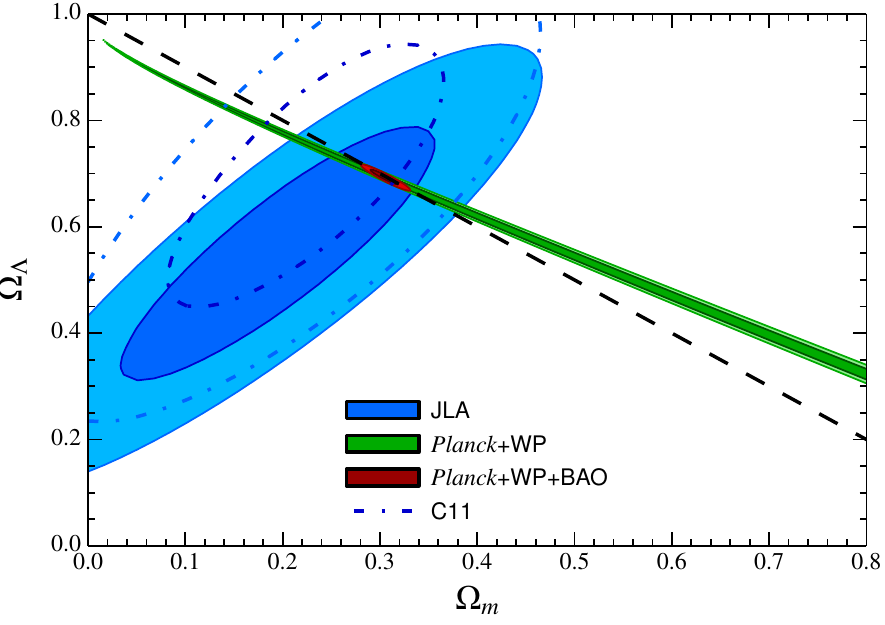}
  \caption{ 68\% and 95\% confidence contours (including systematic
    uncertainty) for the $\Omega_m$ and $\Omega_\Lambda$ cosmological
    parameters for the o-\LCDM model. Labels for the various data sets
    correspond to the present SN Ia compilation (JLA), the
    \citet{2011ApJS..192....1C} SN Ia compilation (C11), the
    combination of \emph{Planck} temperature and \emph{WMAP}
    polarization measurements of the CMB fluctuation (\emph{Planck}+WP), and
    a combination of measurements of the BAO scale (BAO). See
    Sect.~\ref{sec:complementary-data} for details. The black dashed
    line corresponds to a flat universe.}
  \label{fig:omegamomegal}
\end{figure}

\begin{figure}
  \centering
  \igraph{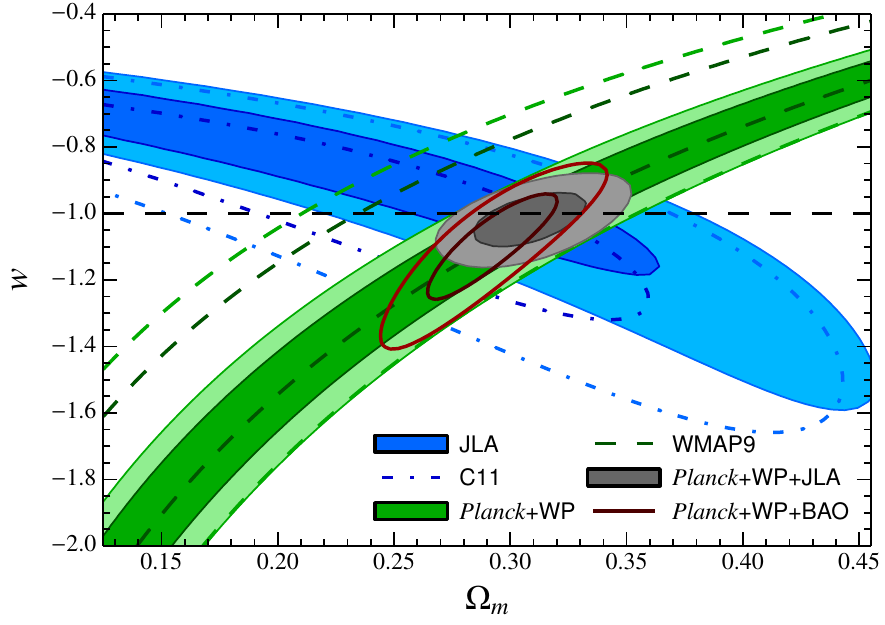}
  \caption{Confidence contours at 68\% and 95\% (including systematic
    uncertainty) for the $\Omega_m$ and $w$ cosmological parameters
    for the flat $w$-\LCDM model. The black dashed line corresponds to
    the cosmological constant hypothesis.}
  \label{fig:omegamw}
\end{figure}

\begin{figure}
  \centering
  \igraph{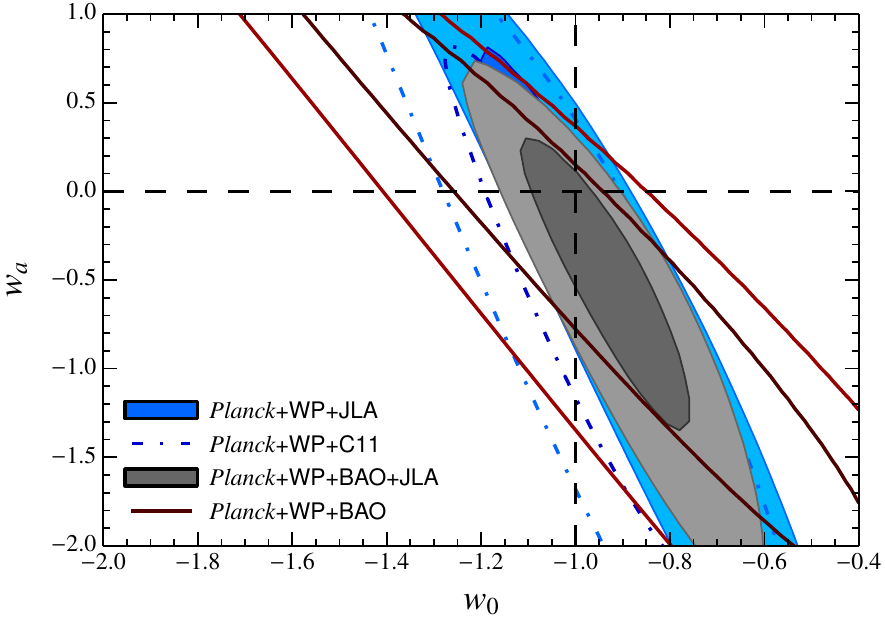}
  \caption{Confidence contours at 68\% and 95\% (including systematic uncertainty) for the
    $w$ and $w_a$ cosmological parameters for the
    flat $w$-\LCDM model.}
  \label{fig:wwa}
\end{figure}

Best-fit parameters for different probe combinations are given in
Tables~\ref{tab:oLCDM},~\ref{tab:FwCDM} and~\ref{tab:FwwaCDM}. Errors
quoted in the tables are 1-$\sigma$ Cramér-Rao lower bounds from the
approximate Fisher Information Matrix. Confidence contours
corresponding to $\Delta \chi^2 =2.28\; (68\%)$ and $\Delta \chi^2 = 6\;
(95\%)$ are shown in Figs.~\ref{fig:omegamomegal},~\ref{fig:omegamw}
and~\ref{fig:wwa}. For all studies involving SNe Ia, we used 
likelihood functions similar to Eq.~(\ref{eq:7}), with both
statistical and systematic uncertainties included in the computation
of $\tens C$. We also performed fits involving the SNLS+SDSS subsample
and the C11 ``SALT2'' sample for comparison (see
Sect.~\ref{sec:lcdm-constr-from}).

\begin{table*}
  \centering
  {\tiny
  \caption{Best fit parameters for the o-\LCDM cosmological model.}
  \label{tab:oLCDM}
  \begin{tabular*}{\linewidth}{@{\extracolsep{\fill}}l*{9}{@{\extracolsep{\fill}}c}@{\extracolsep{\fill}}}
    
\hline
\hline
 & $\Omega_m$ & $\Omega_k$ & $H_0$ & $\Omega_bh^2$ & $\alpha$ & $\beta$ & $M_B^1$ & $\Delta_M$ & $\chi^2/{\rm d.o.f.}$\\
\hline
\emph{Planck}+WP+BAO+JLA & $0.305 \pm 0.010$ & $0.002 \pm 0.003$ & $68.34 \pm 1.03$ & $0.0221 \pm 0.0003$ & $0.141 \pm 0.006$ & $3.099 \pm 0.074$ & $-19.10 \pm 0.03$ & $-0.070 \pm 0.023$ & $684.1/738$\\
\hline
\emph{Planck}+WP+BAO & $0.306 \pm 0.010$ & $0.002 \pm 0.003$ & $68.25 \pm 1.06$ & $0.0221 \pm 0.0003$ & \\
\emph{Planck}+WP+SDSS & $0.397 \pm 0.108$ & $-0.019 \pm 0.026$ & $59.93 \pm 8.17$ & $0.0221 \pm 0.0003$ & $0.145 \pm 0.008$ & $3.115 \pm 0.108$ & $-19.34 \pm 0.27$ & $-0.091 \pm 0.031$ & $350.7/369$\\
\emph{Planck}+WP+SDSS+SNLS & $0.309 \pm 0.046$ & $0.001 \pm 0.011$ & $67.94 \pm 5.15$ & $0.0221 \pm 0.0003$ & $0.140 \pm 0.007$ & $3.141 \pm 0.082$ & $-19.10 \pm 0.15$ & $-0.072 \pm 0.025$ & $577.9/608$\\
\emph{Planck}+WP+JLA & $0.292 \pm 0.037$ & $0.005 \pm 0.009$ & $69.85 \pm 4.44$ & $0.0221 \pm 0.0003$ & $0.141 \pm 0.006$ & $3.102 \pm 0.075$ & $-19.05 \pm 0.12$ & $-0.070 \pm 0.023$ & $682.9/735$\\
\hline
\emph{Planck}+WP+C11 & $0.244 \pm 0.047$ & $0.015 \pm 0.010$ & $76.48 \pm 7.36$ & $0.0221 \pm 0.0003$ & $1.708 \pm 0.156$ & $3.306 \pm 0.109$ & $-18.96 \pm 0.19$ & $-0.045 \pm 0.024$ & $395.1/468$\\
\hline
  \end{tabular*}
}
\end{table*}

\begin{table*}
  \centering
  {\tiny
    \caption{Best fit parameters for the flat $w$-CDM cosmological model.}
  \label{tab:FwCDM}
  \begin{tabular*}{\linewidth}{@{\extracolsep{\fill}}l*{9}{@{\extracolsep{\fill}}c}@{\extracolsep{\fill}}}
    
\hline
\hline
 & $\Omega_m$ & $w$ & $H_0$ & $\Omega_bh^2$ & $\alpha$ & $\beta$ & $M_B^1$ & $\Delta_M$ & $\chi^2/{\rm d.o.f.}$\\
\hline
\emph{Planck}+WP+BAO+JLA & $0.303 \pm 0.012$ & $-1.027 \pm 0.055$ & $68.50 \pm 1.27$ & $0.0221 \pm 0.0003$ & $0.141 \pm 0.006$ & $3.102 \pm 0.075$ & $-19.10 \pm 0.03$ & $-0.070 \pm 0.023$ & $684.1/738$\\
\hline
\emph{Planck}+WP+BAO & $0.295 \pm 0.020$ & $-1.075 \pm 0.109$ & $69.57 \pm 2.54$ & $0.0220 \pm 0.0003$ & \\
\emph{Planck}+WP+SDSS & $0.341 \pm 0.039$ & $-0.906 \pm 0.123$ & $64.68 \pm 3.56$ & $0.0221 \pm 0.0003$ & $0.145 \pm 0.008$ & $3.116 \pm 0.108$ & $-19.17 \pm 0.10$ & $-0.091 \pm 0.031$ & $350.7/369$\\
\emph{Planck}+WP+SDSS+SNLS & $0.314 \pm 0.020$ & $-0.994 \pm 0.069$ & $67.32 \pm 1.98$ & $0.0221 \pm 0.0003$ & $0.140 \pm 0.007$ & $3.139 \pm 0.082$ & $-19.12 \pm 0.05$ & $-0.072 \pm 0.025$ & $577.9/608$\\
\emph{Planck}+WP+JLA & $0.307 \pm 0.017$ & $-1.018 \pm 0.057$ & $68.07 \pm 1.63$ & $0.0221 \pm 0.0003$ & $0.141 \pm 0.006$ & $3.100 \pm 0.075$ & $-19.11 \pm 0.04$ & $-0.070 \pm 0.023$ & $683.0/735$\\
\hline
WMAP9+JLA+BAO & $0.296 \pm 0.012$ & $-0.979 \pm 0.063$ & $68.19 \pm 1.33$ & $0.0224 \pm 0.0005$ & $0.141 \pm 0.006$ & $3.099 \pm 0.075$ & $-19.10 \pm 0.03$ & $-0.070 \pm 0.023$ & $684.4/738$\\
\emph{Planck}+WP+C11 & $0.288 \pm 0.021$ & $-1.093 \pm 0.078$ & $70.33 \pm 2.34$ & $0.0221 \pm 0.0003$ & $1.707 \pm 0.156$ & $3.306 \pm 0.109$ & $-19.15 \pm 0.05$ & $-0.043 \pm 0.024$ & $395.4/468$\\
\hline
  \end{tabular*}
}
\end{table*}

\begin{table*}
  \centering
  {
    \fontsize{7pt}{1em}
  \caption{Best fit parameters for the flat $w_z$-CDM cosmological model. The point $(w_0, w_a) = (-1, 0)$ corresponds to the cosmological constant hypothesis.}
  \label{tab:FwwaCDM}
  \begin{tabular*}{\linewidth}{@{\extracolsep{\fill}}l*{10}{@{\extracolsep{\fill}}c}@{\extracolsep{\fill}}}
    
\hline
\hline
 & $\Omega_m$ & $w_0$ & $w_a$ & $H_0$ & $\Omega_bh^2$ & $\alpha$ & $\beta$ & $M_B^1$ & $\Delta_M$ & $\chi^2/{\rm d.o.f.}$\\
\hline
${\rm \emph{Planck}+WP+BAO+JLA}$ & $0.304 \pm 0.012$ & $-0.957 \pm 0.124$ & $-0.336 \pm 0.552$ & $68.59 \pm 1.27$ & $0.0220 \pm 0.0003$ & $0.141 \pm 0.006$ & $3.099 \pm 0.075$ & $-19.09 \pm 0.04$ & $-0.070 \pm 0.023$ & $683.7/737$\\
\hline
${\rm \emph{Planck}+WP+BAO}$ & $0.291 \pm 0.042$ & $-1.134 \pm 0.490$ & $0.167 \pm 1.318$ & $70.09 \pm 5.05$ & $0.0221 \pm 0.0003$ & \\
${\rm \emph{Planck}+WP+BAO+SDSS}$ & $0.315 \pm 0.019$ & $-0.848 \pm 0.200$ & $-0.582 \pm 0.702$ & $67.31 \pm 2.04$ & $0.0220 \pm 0.0003$ & $0.145 \pm 0.008$ & $3.126 \pm 0.108$ & $-19.09 \pm 0.05$ & $-0.091 \pm 0.031$ & $352.0/371$\\
${\rm \emph{Planck}+WP+JLA}$ & $0.296 \pm 0.022$ & $-0.886 \pm 0.206$ & $-0.698 \pm 1.090$ & $69.36 \pm 2.40$ & $0.0221 \pm 0.0003$ & $0.141 \pm 0.006$ & $3.099 \pm 0.075$ & $-19.06 \pm 0.08$ & $-0.070 \pm 0.023$ & $682.6/734$\\
\hline
${\rm \emph{Planck}+WP+BAO+C11}$ & $0.293 \pm 0.014$ & $-1.073 \pm 0.146$ & $-0.066 \pm 0.563$ & $69.90 \pm 1.64$ & $0.0220 \pm 0.0003$ & $1.706 \pm 0.156$ & $3.307 \pm 0.109$ & $-19.15 \pm 0.04$ & $-0.044 \pm 0.025$ & $396.4/470$\\
\hline
  \end{tabular*}
}
\end{table*}

In all cases the combination of our supernova sample with the two
other probes is compatible with the cosmological constant solution in
a flat universe, which could have been anticipated from the agreement
between CMB and SN Ia measurements of \LCDM parameters (see
Sect.~\ref{sec:comp-with-other}). This concordance is the main result
of the present paper. We note that this conclusion still holds if we
use the \emph{WMAP} CMB temperature measurement in place of the
\emph{Planck} measurement (see Table~\ref{tab:FwCDM}).

For the $w$-CDM model, in combination with Planck, we measure
$w=$\bfw.  This represents a substantial improvement in uncertainty
($30\%$) over the combination \emph{Planck}+WP+C11 ($w=$\wtrois). The
$\sim1\sigma$ (stat+sys) change in $w$ is caused primarily by the
recalibration of the SNLS sample as discussed in detail in
Sect.~\ref{sec:lcdm-constr-from}. The improvement in errors is due to
the inclusion of the full SDSS-II spectroscopic sample and to the
reduction in systematic errors due to the joint recalibration of the
SDSS-II and SNLS surveys. As an illustration of the relative influence
of those two changes, using the C11 calibration uncertainties would
increase the uncertainty of $w$ to $6.5$\%.

Interestingly, the CMB+SNLS+SDSS combination delivers a competitive
measurement of $w$ with an accuracy of $\ews\%$, despite the absence
of the low-$z$ SNe Ia. This measurement is expected to be robust since
the dominant systematic uncertainty (photometric calibration error)
was the subject of careful review in the joint analysis of the SDSS-II
and SNLS surveys. This subsample is also likely to be less sensitive to
errors in the environmental dependence of the SN Ia luminosity as the
distribution of SNLS and SDSS host properties are closer than are the
distribution of SNLS and low-$z$ surveys. As an illustration, fitting
the $w$-CDM model to the CMB+SNLS+SDSS data, and imposing $\Delta_M = 0$,
provides $w=$\bfwksnlssdss, a small shift ($\delta w < 0.003$) with
respect to the value reported for the same sample and $\Delta_M =
-0.070\pm0.023$ in Table~\ref{tab:FwCDM}.

Combined with CMB and BAO, SNe Ia yields a $5.4$\% measurement of $w$
which represents significantly tighter constraint than what can be
obtained from CMB and BAO alone ($11.0$\%). The combination of CMB,
BAO and SNe Ia constrains models with a varying equation of state
$w=$\bfwwaw and $w_a=$\bfwwawa (see Table~\ref{tab:FwwaCDM}), yielding
a figure of merit as defined by the dark energy task force
\citep[DETF;][]{2006astro.ph..9591A} of $31.3$. This is a factor 2
improvement in the FoM with respect to the C11+DR7+WMAP7 combination
considered in \citet{2011ApJ...737..102S}. This gain is attributable,
for roughly equal parts, to our improvement in SN measurements and to
the improvement in CMB and BAO external constraints.

Finally, the combination of CMB, BAO and SN Ia data constrains the
value of the Hubble parameter $H_0$ at better than 2\% even in generic
dark energy models. Our result,
$H_0=$\hoall$\SI{}{\kilo\meter\per\second\per\mega\parsec}$, is
slightly lower ($1.9\sigma$) than the direct measurement of
$H_0=73.8\pm2.4 \SI{}{\kilo\meter\per\second\per\mega\parsec}$ given
in \citet{2011ApJ...730..119R}. A recalibration of the absolute
distance of NGC 4258, one of the three distance anchors involved in this
direct measurement, is given in \citet{2013arXiv1307.6031H}. They
report a slightly smaller value determined from this anchor $H_0 =
72.0 \pm 3.0 \SI{}{\kilo\meter\per\second\per\mega\parsec}$. In
addition, \citet{2013arXiv1311.3461E} suggests that possible biases
were introduced in the Cepheid period-luminosity relation by
subluminous low metallicity Cepheids and shows some sensitivity of
the results to outlier rejections. He finds, using a revised outlier
rejection algorithm, $H_0 = 70.6 \pm
3.3\SI{}{\kilo\meter\per\second\per\mega\parsec}$, using only the
recalibrated NGC 4258 distance anchor and $H_0 = 72.5 \pm
2.5\SI{}{\kilo\meter\per\second\per\mega\parsec}$ combining the three
anchors. In conclusion, the recalibrated direct measurement of $H_0$
improves agreement ($1.4\sigma$) with our indirect determination.

\section{Summary and perspectives}
\label{sec:conclusion}

We have reported improved cosmological constraints from the Hubble
diagram of type Ia supernovae, based on a joint analysis of the SNLS
and SDSS-II SN Ia samples. These results are based on combining the
SN Ia compilation assembled in \citet{2011ApJS..192....1C} by
SNLS with the full SDSS-II three-year SN Ia sample \citep{SDSSRELEASE}. We
have explicitly chosen not to include all newly available SN Ia data,
and instead focus on the control of systematic uncertainties.

The results obtained here benefit from joint SNLS/SDSS analyses
addressing dominant systematic issues.  The effects of the systematic
studies on the cosmological parameters were unknown until the
systematic studies were completed; in this sense, our analysis is a
``blind'' analysis.  The largest systematic error has been reduced by
the notable improvement in the accuracy of the SNLS and SDSS
photometric calibration that resulted from a joint analysis of the
calibration data of both surveys \citep{B12}.  The other major
improvement was the result of detailed investigations of systematic
uncertainties and biases associated with the model of the type-Ia
supernovae spectral evolution \cite{2013ApJ...764...48K,mosher}. In
particular, \citet{mosher} performs a thorough analysis of the
SALT2 light-curve model \citep{2007A&A...466...11G} used in the
present analysis. Thanks to these analyses we are able to derive
distances for \ntotc SNe Ia with improved and well-understood
measurement systematics. The data release is succinctly described in
Appendix~\ref{sec:data-release}.

In the \LCDM model, the JLA sample provides a measurement of the
reduced matter density parameter $\Omega_m= $\bfomegamlcdm, independent
of the CMB measurement. Our result is in good agreement with the
recent measurement from the \emph{Planck} satellite. We show that the
$\sim2\sigma$ disagreement previously observed with
\citet{2011ApJS..192....1C} is largely eliminated when using the joint
recalibration results reported in \citet{B12}. Combining our sample
with the \emph{Planck} CMB measurement, we find no evidence for
dynamical dark energy. Assuming a flat universe, we measure a constant
dark-energy equation of state parameter of $w=$\bfw, where both
statistical and systematic uncertainties are included. In all the
cases we considered, our results are compatible with the cosmological
constant hypothesis.

About half of the gain in precision obtained with respect to the
\citet{2011ApJS..192....1C} result arises from the improvement in the
calibration accuracy.  This demonstrates that substantial gain was
obtained by working on improving calibration systematics. In spite of
these improvements, the accuracy of the photometric calibration
remains (by far) the limiting systematic uncertainty. However, there
is no known reason why this situation can not be improved in future
surveys. Our result is based on a photometric calibration which is
limited at the $\sim0.5\%$ level by the accuracy of the primary
stellar calibration standards. Further improvements, either in the
accuracy of stellar spectrophotometric standards, or in the delivery
of laboratory-made calibration sources, should make it possible to
approach the current systematic limit of $\sim1$~mmag with CCD-based
photometric measurements \citep{A13}. The use of CCD detectors with
enhanced sensitivity in the near infrared will make it possible to
observe low and high-redshift supernovae in more nearly similar
rest-frame bands. Better wavelength coverage would alleviate the
partial degeneracy between the cosmology, the calibration and the SNe
Ia model, the degeneracy that is responsible for a large part of the
sensitivity of cosmology to calibration uncertainties. 
In particular, the Dark Energy Survey (DES) experiment, which has just
begun its high-redshift SNe survey \citep{2012ApJ...753..152B}, will
exploit improved sensitivity in the infrared to reduce the
model-calibration-cosmology degeneracy. Also DES, unlike SNLS, has a
shallow survey ($24{\rm deg}^2$) that will provide both high and
medium redshift SN in the same experiment. DES will also have a
continual, \emph{in situ}, modeling of the filter transmission.  All
three features should ease the calibration problem. LSST should be
even better than DES in both these respects. However, neither DES nor
LSST will be able to spectroscopically identify a large fraction of
their candidates because the required spectroscopic time would be
prohibitive. Each survey will hence have to contend with the problem
of contamination in their SN~Ia samples. Several analyses have already
addressed this issue (\emph{e.g.}, in SNLS and SDSS
\citealt{2011A&A...534A..43B,2013ApJ...763...88C}) and showed that
good control of the contamination ($<4\%$) can be reached.

According to our estimates, issues related to the
environmental dependence of SNe Ia standardized luminosity contribute
a subdominant part of the error budget. However, this conclusion is
less robust than our understanding of other uncertainties because the
phenomena are, at best, only partially understood. This topic is currently the
subject of active research and there is hope that firm conclusions can
be reached with increased statistics, complementary data (\emph{e.g.},
local properties of the SN environment, \citealt{2012A&A...545A..58S})
and possibly improved theoretical modeling. Increased statistics are
already available: about 150 additional spectroscopically confirmed
SNe Ia from the 5-year SNLS sample; extended samples from several
low-$z$ experiments \citep{2011AJ....142..156S,2012ApJS..200...12H};
and numerous spectral data samples
\citep{2012AJ....143..126B,2012MNRAS.425.1789S,2012MNRAS.426.2359M,2013A&A...554A..27P},
which should provide better constraints on the SNe Ia model.

\begin{acknowledgements}
  Funding for the SDSS and SDSS-II has been provided by the Alfred
  P. Sloan Foundation, the Participating Institutions, the National
  Science Foundation, the U.S. Department of Energy, the National
  Aeronautics and Space Administration, the Japanese Monbukagakusho,
  the Max Planck Society, and the Higher Education Funding Council for
  England. The SDSS Web Site is \url{http://www.sdss.org/}.

  The SDSS is managed by the Astrophysical Research Consortium for the
  Participating Institutions. The Participating Institutions are the American
  Museum of Natural History, Astrophysical Institute Potsdam, University of
  Basel, Cambridge University, Case Western Reserve University, University of
  Chicago, Drexel University, Fermilab, the Institute for Advanced Study, the
  Japan Participation Group, Johns Hopkins University, the Joint Institute for
  Nuclear Astrophysics, the Kavli Institute for Particle Astrophysics and
  Cosmology, the Korean Scientist Group, the Chinese Academy of Sciences
  ({\small LAMOST}), Los Alamos National Laboratory, the Max-Planck-Institute
  for Astronomy ({\small MPIA}), the Max-Planck-Institute for Astrophysics
  ({\small MPA}), New Mexico State University, Ohio State University,
  University of Pittsburgh, University of Portsmouth, Princeton University,
  the United States Naval Observatory, and the University of Washington.

  The Hobby-Eberly Telescope (HET) is a joint project of the
  University of Texas at Austin, the Pennsylvania State University,
  Stanford University, Ludwig-Maximillians-Universit\"at M\"unchen,
  and Georg-August-Universit\"at G\"ottingen.  The HET is named in
  honor of its principal benefactors, William P. Hobby and Robert
  E. Eberly.  The Marcario Low-Resolution Spectrograph is named for
  Mike Marcario of High Lonesome Optics, who fabricated several optics
  for the instrument but died before its completion; it is a joint
  project of the Hobby-Eberly Telescope partnership and the Instituto
  de Astronom\'{\i}a de la Universidad Nacional Aut\'onoma de
  M\'exico.  The Apache Point Observatory 3.5-meter telescope is owned
  and operated by the Astrophysical Research Consortium.  We thank the
  observatory director, Suzanne Hawley, and site manager, Bruce
  Gillespie, for their support of this project.  The Subaru Telescope
  is operated by the National Astronomical Observatory of Japan.  The
  William Herschel Telescope is operated by the Isaac Newton Group,
  and the Nordic Optical Telescope is operated jointly by Denmark,
  Finland, Iceland, Norway, and Sweden, both on the island of La Palma
  in the Spanish Observatorio del Roque de los Muchachos of the
  Instituto de Astrofisica de Canarias.  Observations at the {\small
    ESO} New Technology Telescope at La Silla Observatory were made
  under programme {\small ID}s 77.A-0437, 78.A-0325, and 79.A-0715.
  Kitt Peak National Observatory, National Optical Astronomy
  Observatory, is operated by the Association of Universities for
  Research in Astronomy, Inc. ({\small AURA}) under cooperative
  agreement with the National Science Foundation.  The {\small WIYN}
  Observatory is a joint facility of the University of
  Wisconsin-Madison, Indiana University, Yale University, and the
  National Optical Astronomy Observatories.  The W.M.\ Keck
  Observatory is operated as a scientific partnership among the
  California Institute of Technology, the University of California,
  and the National Aeronautics and Space Administration.  The
  Observatory was made possible by the generous financial support of
  the W.M.\ Keck Foundation.  The South African Large Telescope of the
  South African Astronomical Observatory is operated by a partnership
  between the National Research Foundation of South Africa, Nicolaus
  Copernicus Astronomical Center of the Polish Academy of Sciences,
  the Hobby-Eberly Telescope Board, Rutgers University,
  Georg-August-Universit\"at G\"ottingen, University of
  Wisconsin-Madison, University of Canterbury, University of North
  Carolina-Chapel Hill, Dartmough College, Carnegie Mellon University,
  and the United Kingdom SALT consortium.  The Telescopio Nazionale
  Galileo ({\small TNG}) is operated by the Fundaci\'on Galileo
  Galilei of the Italian {\small INAF} (Istituo Nazionale di
  Astrofisica) on the island of La Palma in the Spanish Observatorio
  del Roque de los Muchachos of the Instituto de Astrof\'{\i}sica de
  Canarias.  

  This paper is based in part on observations obtained with
  MegaPrime/MegaCam, a joint project of CFHT and CEA/IRFU, at the
  Canada-France-Hawaii Telescope (CFHT) which is operated by the
  National Research Council (NRC) of Canada, the Institut National des
  Sciences de l'Univers of the Centre National de la Recherche
  Scientifique (CNRS) of France, and the University of Hawaii.  Part
  of the results are derived from observations obtained with
  \emph{Planck} (\url{http://www.esa.int/Planck}), an ESA science
  mission with instruments and contributions directly funded by ESA
  Member States, NASA, and Canada. We also makes use of data products
  from the Two Micron All Sky Survey, which is a joint project of the
  University of Massachusetts and the Infrared Processing and Analysis
  Center/California Institute of Technology, funded by the National
  Aeronautics and Space Administration and the National Science
  Foundation. We acknowledge the use of the NASA/IPAC Extragalactic
  Database (NED) which is operated by the Jet Propulsion Laboratory,
  California Institute of Technology, under contract with the National
  Aeronautics and Space Administration. This work was completed in
  part with resources provided by the University of Chicago Research
  Computing Center.

  The French authors acknowledge support from CNRS/IN2P3, CNRS/INSU
  and CEA.  GL is supported by the Swedish Research Council through
  grant No. 623-2011-7117. DARK is funded by DNRF. J.F. and R.K. are
  grateful for the support of National Science Foundation grant
  1009457, a grant from “France and Chicago Collaborating in the
  Sciences” (FACCTS), and support from the Kavli Institute for
  Cosmological Physics at the University of Chicago.  A.V.F. has
  received generous financial assistance from the Christopher
  R. Redlich Fund, the TABASGO Foundation, and NSF grant
  AST-1211916. MSu acknowledges support from the Royal Society.
\end{acknowledgements}

\bibliography{cosmo}
\bibliographystyle{aa}

\appendix

\section{Visual inspection}
\label{sec:visual-inspection-1}

In addition to the software cuts, we performed a visual inspection of
the SN light-curve fits. We discarded the following SNe Ia, for which
the SALT2 fits were particularly poor:\footnote{Supernovae denoted
  with a star would not enter the training sample anyway because they
  fail other selection requirements (typically the redshift cut).}
\begin{enumerate}
\item Fit probability $< 0.01$ due to apparent problems in the
  photometry: SDSS739$^\star$, SDSS1316$^\star$, SDSS3256 (2005hn),
  SDSS6773 (2005iu), SDSS12780, SDSS12907, SDSS13327$^\star$,
  SDSS16287, SDSS16578$^\star$, SDSS16637$^\star$, SDSS17176$^\star$,
  SDSS18456, SDSS18643, SDSS19381 (2007nk), SDSS20376$^\star$,
  SDSS20528 (2007qr), SDSS21810$^\star$.
\item Poor fit, probable 1986G-like: SDSS17886 (sn2007jh) \citep{2011AJ....142..156S}.
\item Poor fit, 2002cx-like: SDSS20208 (sn2007qd) \citep{2010ApJ...720..704M,2013ApJ...767...57F}.
\item Pathological sampling leading to unstable fit results: SDSS17500$^\star$, SDSS16692$^\star$.
\end{enumerate}

We also discarded the following four events that are $>3\sigma$ outliers on the Hubble diagram:
\begin{enumerate}
\item  over-luminous: SDSS14782 (2006jp), SDSS15369 (2006ln).
\item subluminous: SDSS15459 (2006la), SDSS17568 (2007kb).
\end{enumerate}

Last, a proper and stable determination of the date of maximum is
necessary for SNe Ia entering in the training sample, because the date
of maximum is held fixed in the training. We looked for remaining
poorly sampled light curves in the training sample, and discarded the
following nine SNe (only from the training sample):
\begin{enumerate}
\item Too few observations after the epoch of peak brightness (despite a reported uncertainty on $t_0$
  passing the cuts): SDSS10434, SDSS19899, SDSS20470, SDSS21510.
\item Too few observations before the epoch of peak brightness: SDSS6780, SDSS12781, SDSS12853 (2006ey), SDSS13072, SDSS18768.
\end{enumerate}

\section{Details on calibration systematics}
\label{sec:calibr-syst}

\subsection{Consistency of the CfAIII and CSP photometric calibration}
\label{sec:consistency-low-z}

A few low-$z$ SNe Ia have been observed contemporaneously with several telescopes
 which provides a way to assess their relative
calibration. \citet{2012AJ....144...17M} studied nine
spectroscopically confirmed Type Ia supernova observed by both the CSP and the
SDSS-II surveys. The study provides us with stringent constraints on possible
differences between the CSP calibration and the SDSS/SNLS calibration
of \citetalias{B12}. The \citet{2012AJ....144...17M} results are reproduced in
Table~\ref{tab:lowzcaloffset}. 

We performed a similar study on SNe Ia observed by both the CfAIII and
CSP surveys. To increase the statistics available for this comparison,
we consider SNe Ia from both the first \citep{2010AJ....139..519C} and
second \citep{2011AJ....142..156S} CSP data release. The list of all
SNe Ia in common is given in Table~\ref{table:SNeCfaCsp}.

We use SALT2 to interpolate between measurements (in phase and
wavelength) as follows: for each SN Ia, we perform an initial fit
using all available data to determine its shape, color, and date of
maximum. Holding these parameters fixed, we redetermine the amplitude
parameter $x_0$ for each band independently. In a given band,
comparing the values of $-2.5\logdec (x_0)$ obtained for two different
instruments gives an estimate of the calibration difference between
them. This method is similar to the S-correction and spline
interpolation applied in \citet{2012AJ....144...17M}.  However,
instead of transforming the CfA data to bring them to the CSP native
system, both sets data are transformed in the same manner. Applied to
the same sample, the two methods deliver very similar results.

We exclude peculiar type-Ia supernovae from the comparison. Light
curves with aberrant photometric points were rejected: SN2005M $U$ and
$r$ band light curves, SN2005ir, SN2006ev and SN2005mc $r$
band. Finally, $B$, $V$ and $r'$ band data for 2006hb are too long
after maximum brightness to be reliably compared to CSP
measurements. The results are given in the second part of
Table~\ref{tab:lowzcaloffset}. Our analysis shows an excellent
agreement in the $B$, $V$ and $i'$ bands. The offset measured in $r'$
appears statistically significant, justifying the upward adjustment of
the $r'$ calibration uncertainty quoted in
\citetalias{2011ApJS..192....1C}. The $U$ band also shows surprisingly
good consistency considering the fact that CfAIII $U$ band
measurements are color-corrected to the Landolt system using a color
transformation determined using ordinary stars. However, given the
small number of SNe Ia in the $U$-band comparison, we are concerned
that the agreement may be fortuitous and do not revise the $0.07$~mag
uncertainty used by \citet{2009ApJ...700..331H}. This choice of a
relatively large $U$-band uncertainty is justified in
\S\ref{sec:syst-induc-color} where a SN $U$-band color-correction
error is evaluated.

\begin{table}
  \centering
  \caption{Calibration offsets}
  \label{tab:lowzcaloffset}

  \begin{tabular*}{\linewidth}{lr@{\extracolsep{0.5em}}lrr}
    \hline
    \hline
    Band & \multicolumn{2}{l}{Mean offset}  & Scatter & $N_\text{SN}$ \\
         & \multicolumn{2}{l}{\it(mag)} & {\it(mag)} &\\
    \hline
    \multicolumn{4}{l}{CSP - SDSS\tablefootmark{a}}\\
    \hline
    $u$ & $-0.008$ & $\pm 0.016$ (stat) $\pm 0.013$ (sys) & $0.038$ & 4 \\
    $g$ & $-0.002$ & $\pm 0.006$ (stat) $\pm 0.005$ (sys) & $0.028$ & 7 \\
    $r$ & $ 0.011$ & $\pm 0.005$ (stat) $\pm 0.005$ (sys) & $0.025$ & 6 \\
    $i$ & $-0.012$ & $\pm 0.005$ (stat) $\pm 0.002$ (sys) & $0.032$ & 7 \\
    \hline
    \multicolumn{4}{l}{CSP - KeplerCam}\\
    \hline
    $U$ & $ 0.021$ & $\pm 0.013$ (stat) & $0.071$ & 6  \\
    $B$ & $ 0.005$ & $\pm 0.004$ (stat) & $0.042$ & 17 \\
    $V$ & $-0.009$ & $\pm 0.003$ (stat) & $0.021$ & 17 \\
    $r$ & $ 0.024$ & $\pm 0.004$ (stat) & $0.039$ & 17 \\
    $i$ & $ 0.003$ & $\pm 0.012$ (stat) & $0.049$ & 18 \\
    \hline
  \end{tabular*}
  \tablefoot{\tablefoottext{a}{From \citet[Table 11]{2012AJ....144...17M}}. Systematic uncertainties are the combination of interpolation and S-correction uncertainties.}
\end{table}

\begin{table}[hh]
\centering
\caption{\small{Spectroscopically confirmed SNe Ia in common between CSP and CFA}\label{table:SNeCfaCsp}}
\begin{tabular}{ccccc}
\hline\hline
SN IAU Name & $z_\text{helio}$ & Peculiar  \\
\hline
2005M  & 0.022   & \\
2005hj & 0.0580  & \\
2005ir & 0.0764  & \\
2005mc & 0.0252  &  \\
2006bd & 0.0257  & 91bg-like \\
2006br & 0.0246  &   \\
2006bt & 0.0322  & Yes\tablefootmark{a} \\
2006ef & 0.0179  & \\
2006ej & 0.0205  & \\
2006et & 0.0226  & \\
2006ev & 0.0287  & \\
2006gj & 0.0284  & \\
2006hb & 0.0153  & 86G-like\\
2006is & 0.0314  & \\
2006kf & 0.0213  & \\
2006D  & 0.00852 & \\
2006os & 0.0328  & \\
2007N  & 0.0129  & 91bg-like \\
2007S  & 0.0139  & 91T-like\\
2007af & 0.0055  & \\
2007ai & 0.0317  & 91T-like \\
2007ax & 0.0069  & 91bg-like \\
2007ba & 0.0385  & 91bg-like \\
2007bc & 0.0208  & \\ 
2007bd & 0.0309  & \\
2007ca & 0.0062  & \\ 
\hline
\end{tabular}
\tablefoot{\tablefoottext{a}{\cite{2010ApJ...708.1748F}}}
\end{table}

\subsection{Errors induced by the color-transformation of nearby Supernova
  measurements}
\label{sec:syst-induc-color}

A substantial fraction of our low-$z$ sample is composed of SNe Ia
with photometry reported in the Landolt system, which means that flux
measurements in the natural system have been transformed to the
Landolt system using color transformations determined by ordinary 
stars. This procedure introduces errors because
SNe Ia have spectral properties different from those of
main sequence stars (see, \emph{e.g.}, the discussion in
\citealt[\S2.4, hereafter J06]{2006AJ....131..527J}). Here we seek
quantitative estimates for these errors.

J06 provides effective filter transmissions for several combinations
of $UBVRI$ filter sets and CCD cameras used for the SN
observations. Using these transmissions, along with an effective model of Landolt
filters\footnote{Landolt filters from \citealt{Bessel90}, with
  wavelength shifts of $-31$, $+8$, $+3$, $+22$, and
  $+\SI{11}{\angstrom}$ for the $UBVRI$ bands respectively; see the
  Appendix A of C11 for a detailed discussion}, we can compute
synthetic magnitudes of stars in both the natural and the Landolt
system.  We use the stellar libraries of \cite{1983ApJS...52..121G}
and \cite{1998PASP..110..863P}, selecting stars in a range of $U-B$
and $B-V$ colors matching that of the SN calibration stars. For SNe,
we use the SALT2 average spectral sequence ($\xun = \col = 0$).

Using those synthetic magnitudes, we compare the ''true'' (synthetic)
Landolt magnitude to the Landolt magnitude estimated with a color
transformation of the (synthetic) natural magnitudes. For these color
transformations, we use the color terms given in Table 3 of J06, and
define $\delta m \equiv m_\text{Landolt}^\text{true} -
m_\text{Landolt}^{\text{color}-\text{corr.}}$ to be the difference between those two values. The
calibration bias for SNe is given by the difference of $\delta m$ for
SNe and main sequence stars. Indeed the latter value sets the
normalization of SN magnitudes through the assignement of a zero-point
to the images. We label this difference $\Delta m \equiv \delta
m(\mathrm{SN}) - \delta m(\mathrm{stars})$.

\begin{figure}
  \centering
  \igraph{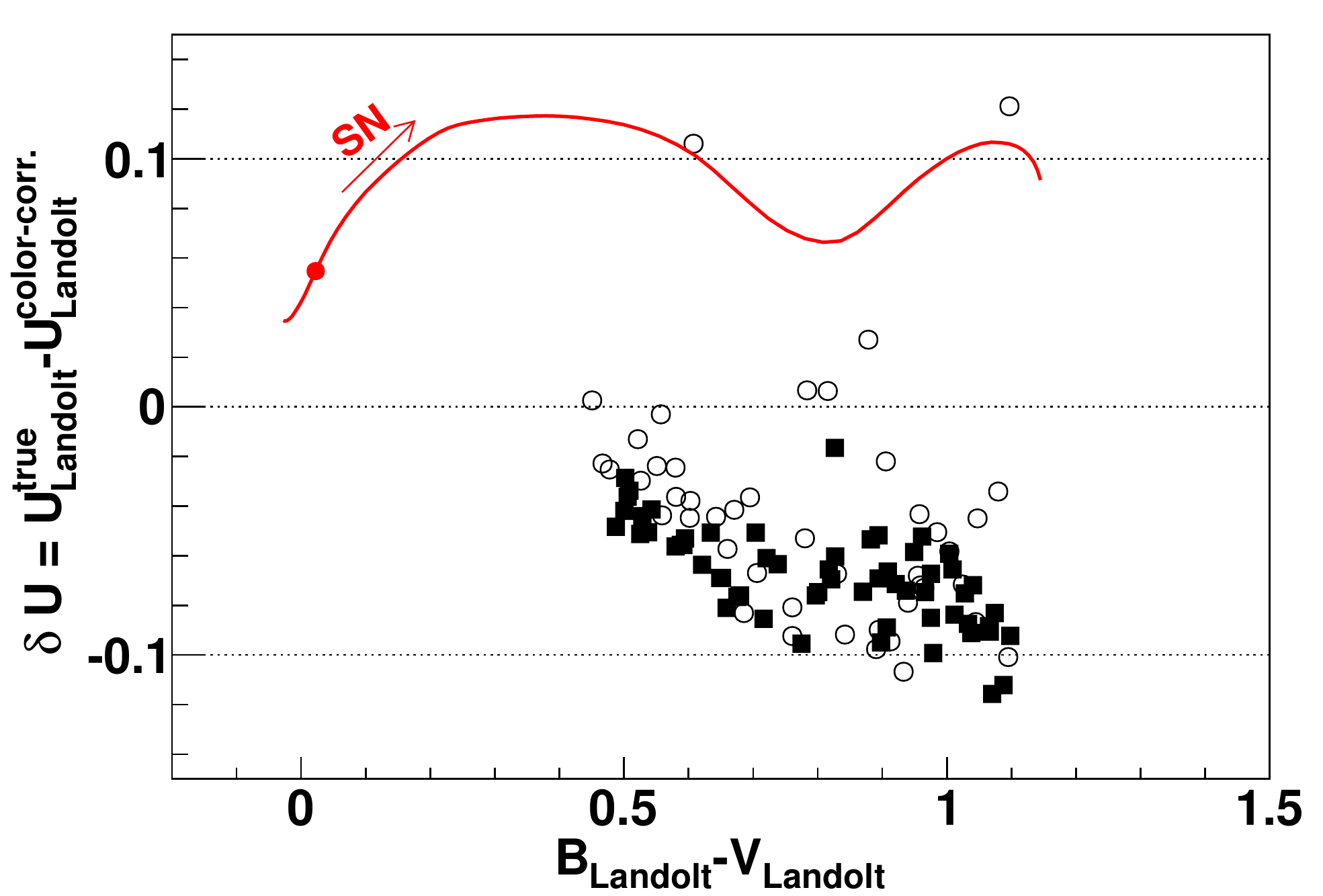}
  \caption{Synthetic values of $\delta U \equiv U_\text{Landolt}^\text{true} -
    U_\text{Landolt}^{\text{color}-\text{corr.}}$ as a function of $B-V$ color for stars
    from \cite{1983ApJS...52..121G} (open circles) and
    \cite{1998PASP..110..863P} (filled squares), and for an average SN
    at various epochs, from $-5$ to $+30$ days, based on the SALT2
    spectral sequence. The natural effective filter set is that of the
    4Shooter camera, chip 1, with SAO filters, given in
    \cite{2006AJ....131..527J}, Table 5. Only the difference between
    SN and stars are relevant here, not the absolute $\delta U$
    values.}
  \label{fig:colorcorrection}
\end{figure}

An uncertainty on the quantity $\Delta m$ can be estimated by varying
the SN model, the spectral library, or the filter transmissions. In
practice, the uncertainty on the filter transmissions is dominant. Figure~\ref{fig:colorcorrection} 
shows that $\delta m$ is a function
of the star color, which means that the filter model is inadequate.
By construction, $\delta m$ is color-independent for real
observations. One can adjust wavelength shifts of the filter
transmissions in order to obtain a color-independent value of $\delta
m$ for stars. This approach also results in a change of $\Delta m$
that we can subsequently use as an estimate of the uncertainty due to
approximate filter transmissions.

For the AndyCam CCD camera (CfA) with the Harris filter set
\citep{1981PASP...93..507H}, we have found $\Delta B = 0 \pm
0.015$~mag, $\Delta V = 0.03 \pm 0.01$~mag, and $\Delta R = 0.03 \pm
0.03$~mag. In other words, the color-correction does not significantly
bias the measurements for the $BVR$ bands.  The situation for the
$U$-band is, however, different. We have found a value as large as
$0.1$ for the 4Shooter camera (CfA), chip 1, with SAO filters. The values of
$\delta m $ for SNe and stars are represented in
Figure~\ref{fig:colorcorrection} for this latter instrumental
setup. One can also see on the figure that the residual color term is
quite important. A $U$-band shift of $\sim$3~nm is needed to obtain a
flat distribution of $\delta m$, and in that case one finds an even
larger value of $\Delta U = 0.15$.

A primary motivation for this study is the existence of significant
calibration offsets between observer-frame UV observations from
different instruments (see, \emph{e.g.}, \citealt{2013AJ....145...11K}
for a longer discussion of this effect) and with rest-frame UV
observations at higher redshift. \citet{kessler_first-year_2009} found
that this latter discrepancy was responsible for a large part of the
difference between the SALT2 and MLCS2k2 \citep{2007ApJ...659..122J}
models, MLCS2k2 being trained solely on low-$z$ SNe. This $U$-band
offset introduced by the application of a color-correction to SNe data
could explain some of the discrepancy. However, the $U$-band filter
transmissions are too uncertain to secure a good interpretation of
natural magnitudes. For this reason, we adopt the magnitudes that are
color-transformed to the Landolt system for the low-$z$ samples
(except for the CSP data and the CfA-III $BVri$ light curves where we
use the natural magnitudes and measured filter response functions),
but assign a coherent systematic uncertainty of 0.1 mag to the
amplitude of $U$-band light curves.

In all bands, the (phase dependent) error introduced by color
transformations is not included, so measurement errors are typically
underestimated. As a consequence, the uncertainties in the fit
light-curve parameters are underestimated. The training of SALT2 is
also affected by this problem. At present, we cannot afford discarding
the color-transformed low-$z$ and must deal with this issue. We
estimate the measurement errors again for color transformed
measurements in the low-$z$ sample as follows. Since the SDSS-II and
SNLS measurement errors are reliable, we trained a version of SALT2 as
described in Sect.~\ref{sec:joint-training-light}, but considering
only the SNLS and SDSS-II measurements in the computation of the
``error-snake''. We then use this version with reliable modeling of
the intrinsic dispersion to fit all the color transformed low-$z$
light curves. For each light-curve, we fit an ad-hoc two parameter
($\gamma_2$ and $\gamma_3$) correction of the measurement errors
$\sigma_i$ affecting the measurement $d_i$ by minimizing the following
residual likelihood:
\begin{equation}
REML = \sum _i w_i (d_i - \gamma_1 m_i)^2  - \sum_i \log w_i + \log \left(\sum_i w_i\right)
\end{equation}
with $w_i^{-1} = \sigma_i^2 + (\sigma_i^M)^2 + \gamma_2^2 m_i^2 +
\gamma_3^2$, where $m_i$ is the flux predicted by the best fit
light-curve model and $\sigma_i^M$ the model value of the intrinsic
dispersion. We simultaneously fit for $\gamma_1$, $\gamma_2$ and
$\gamma_3$. When the light-curve contains less than five points, we fix
the value of $\gamma_3$ to zero. We then alter the errors in the
light-curve accordingly to the fit values of $\gamma_2$ and
$\gamma_3$. We found a mean value of 0.007~mag for $\gamma_2$.

\section{Estimates of missing host stellar masses in the C11 sample}
\label{sec:host-mass-estimates-1}

The C11 compilation is missing estimates of the galaxy host mass for
61 nearby SNe (mostly because of missing photometry for the host). We
describe estimates obtained for 57 of the 61 missing galaxy mass
values.

For 49 of the nearby SN host galaxies, we derived an estimate based on
$Ks$ photometry \citep{2001ApJ...550..212B,2003ApJS..149..289B} from the
2003 2MASS All-Sky Data Release of the Two Micron All Sky Survey
\citep{2006AJ....131.1163S}. The photometric data are extracted from
the NASA/IPAC Extragalactic Database (NED) database. A linear model is
fit between the mass and the $Ks$ absolute magnitude on 51 objects with
stellar mass estimates from \citetalias{2011ApJS..192....1C}. This
linear model yields a residual of 0.15 dex and is used to provide
galaxy mass estimates. For 8 galaxies without 2MASS Ks magnitudes, we
rely on less precise models based on the total B band RC3 magnitude
\citep[three objects]{1991rc3..book.....D}, the $r$ C-Model magnitude (1
object from the SDSS DR6, \citealt{2008ApJS..175..297A}), the $B$
magnitude (three objects published in \citealt{2000AJ....120.1479H}), and
the $B$ magnitude in \citet{2002AJ....124.2905S} for the low-luminosity
host of SN 1999aw. The four remaining supernovae have no identified host
and were assigned to the low-mass bin with an uncertainty on distance
moduli of $\Delta_M^\text{ref}$ added in quadrature to the other sources of
uncertainty.

\section{Accuracy of the CMB distance prior}
\label{sec:accur-cmb-dist}

In Sect.~\ref{sec:dark-energy-constr}, we summarized the dark energy
constraints from the CMB in the form of a distance prior. A
computationally intensive, but more general, approach is to directly
compare the CMB data to theoretical predictions for the fluctuation
power spectra computed from a Boltzmann code. In this appendix, we
briefly compare the results from both approaches for a fit of the
$w$-CDM model to the combination of our SNe Ia JLA sample with CMB
constraints.

The \emph{Planck} collaboration \citep{2013arXiv1303.5075P} has
released code to compute the likelihood of theoretical models given
\emph{Planck} data.\footnote{We use the publicly available clik code
  to compute the high and low-$\ell$ \emph{Planck} and \emph{WMAP}
  low-$\ell$ polarization likelihood functions. Both the \emph{Planck}
  likelihood code v1.0 and corresponding data (CAMSPEC
  v6.2TN\_2013\_02\_26, commander v4.1\_lm49 and lowlike v222 are
  available from the \emph{Planck} Legacy archive
  \url{http://pla.esac.esa.int/pla/aio/planckResults.jsp?}}  This
enables the marginalization of several sources of systematic uncertainty in the
CMB spectra, such as errors in the instrumental beams and
contamination by astrophysical foregrounds. In our comparison we make
use of the full \emph{Planck} temperature likelihood complemented with
the \emph{WMAP} measurement of the large scale CMB polarization
\citep{2013ApJS..208...20B}.  We use the CAMB Boltzmann code
\citep[March 2013]{2000ApJ...538..473L} for our computation of CMB
spectra. We follow assumptions from \cite{2013arXiv1303.5076P},
fitting for the baryon density today $\omega_b = \Omega_b
h^2$, the cold dark matter density today $\omega_c=\Omega_c
h^2$, $\theta_\text{MC}$, the CosmoMC approximation of the sound horizon
angular size computed from the Hu\&Sugiyam (1996) fitting formulae,
$\tau$, the Thomson scattering optical depth due to reionization,
$\ln(10^{10}A_s)$, the log power of the primordial curvature
perturbations at the pivot scale $k_0 = 0.05 {\rm Mpc}^{-1}$, $n_s$,
the primordial spectrum index, and $w$, the dark energy equation of
state parameter.
\begin{figure}
  \centering
  \igraph{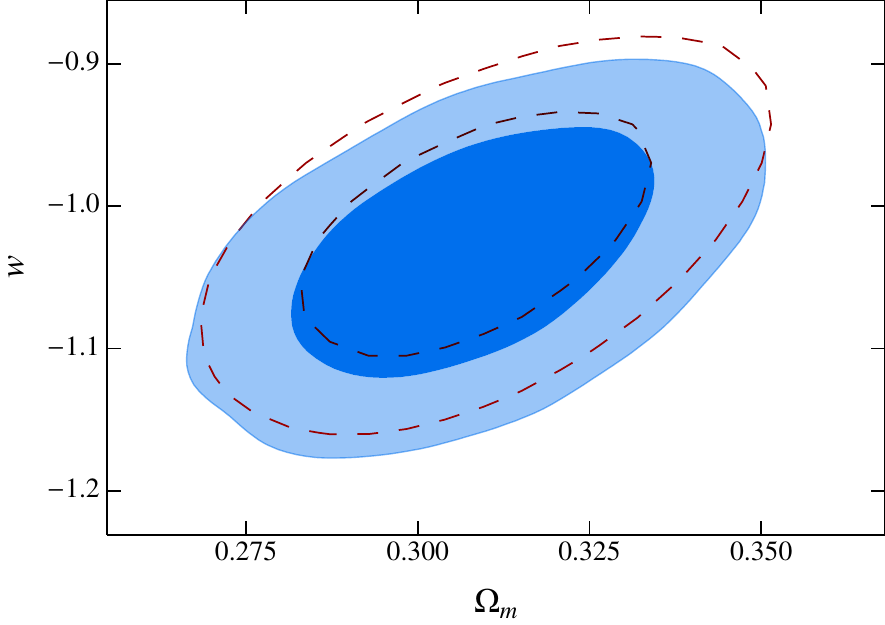}
  \caption{Comparison of two derivations of the $68$ and $95\%$
    confidence contours in the $\Omega_m$ and $w$ parameters for a
    flat $w$-CDM cosmology. In one case, constraints are derived from
    the exploration of the full Planck+WP+JLA likelihood (Blue). In
    the other case CMB constraints are summarized by the geometric
    distance prior described in Sect.~\ref{sec:complementary-data}
    (dashed red).}
  \label{fig:prior}
\end{figure}

We explored the Planck+WP+JLA likelihood with Markov Chain Monte Carlo
(MCMC) simulations of the posterior distribution assuming flat priors
for parameters as given in \citet[Table 1]{2013arXiv1303.5076P}.
Eight sample chains were drawn using CosmoMC
\citep{2002PhRvD..66j3511L,2013PhRvD..87j3529L}. Convergence of the
simulation is monitored using the \citet{1992} $R$
statistic.\footnote{We impose $R-1<0.01$ in the least converged
  orthogonalized parameter.}

The mean value of the posterior distribution and 68\% limits for the
fit parameters to the \emph{Planck}+WP+JLA likelihood are given in
Table~\ref{tab:bfcosmomc}. Best-fit parameters obtained using the
distance prior in Sect.~\ref{sec:main-results} are shown for
comparison. The 68\% and 95\% contours from these simulations are drawn
in Fig.~\ref{fig:prior}. Overplotted is the \emph{Planck}+WP+JLA contour from
Fig.~\ref{fig:omegamw}. The differences are small as expected from the
fact that the supplementary constraints brought by the complete CMB
power spectrum are weak compared to the supernova constraints.

\begin{table}
  \centering
  \begin{tabular}{lr@{}lr@{}l}
    
\hline\hline
Parameter & \multicolumn{2}{c}{\emph{Planck}+WP+JLA} & \multicolumn{2}{c}{DP+JLA}\\
\hline
$\Omega_bh^2$ & $0.02201$ & $^{+0.00028}_{-0.00028}$ & 
 $0.02208$ &$\pm 0.00028$ \\
$\Omega_m$ & $0.308$ & $^{+0.017}_{-0.017}$ & 
 $0.307$ &$\pm 0.017$ \\
$H_0$ & $68.1$ & $^{+1.6}_{-1.6}$ & 
 $68.1$ &$\pm 1.6$ \\
$\tau$ & $0.089$ & $^{+0.012}_{-0.015}$ & 
-- \\
$n_s$ & $0.9590$ & $^{+0.0071}_{-0.0071}$ & 
-- \\
${\rm{ln}}(10^{10}A_s)$ & $3.088$ & $^{+0.024}_{-0.027}$ & 
-- \\
$w$ & $-1.034$ & $^{+0.059}_{-0.055}$ & 
 $-1.018$ &$\pm 0.057$ \\
\hline
  \end{tabular}
  \caption{Best-fit parameters of the $w$-CDM fit for the full \emph{Planck}+WP+JLA likelihood, and for the distance prior (DP+JLA).}
  \label{tab:bfcosmomc}
\end{table}

\section{Compressed form of the JLA Likelihood}
\label{sec:appr-jla-likel}

Fig.~\ref{fig:nuisance} shows that the correlation between the
nuisance parameters ($\alpha$, $\beta$, $\Delta_M$) and the
cosmological parameter $\Omega_m$ is small as a result of the high
density of SNe in this Hubble diagram (especially in the SDSS sample
at intermediate redshifts). This suggests that, for a limited class of
models (those predicting isotropic luminosity distances evolving
smoothly with redshifts), the estimate of distances can be made
reasonably independent of the estimate of cosmological parameters. In
this appendix, we seek to provide the cosmological information of the
JLA Hubble diagram in a compressed form that is faster and easier to
evaluate and still remains accurate for the most common cases. Studies
investigating alternate cosmology or alternate standardization
hypotheses for SNe-Ia should continue to rely on the complete form.

\subsection{Binned distance estimates}
\label{sec:binn-dist-estim}

The distance modulus is typically well approximated by a piece-wise
linear function of $\log(z)$, defined on each segment $z_b \leq z
<z_{b+1}$ as:
\begin{equation}
\label{eq:28}
  \bar \mu(z) = (1 - \alpha) \, \mu_b + \alpha \, \mu_{b+1}
\end{equation}
with $\alpha = \log(z/z_b)/\log(z_{b+1}/z_{b})$ and $\mu_b$ the
distance modulus at $z_b$. As an example, for 31 log-spaced control
points $z_b$ in the redshift range $0.01 < z < 1.3$, the difference
between the $\Lambda$-CDM distance modulus and its linear interpolant
is everywhere smaller than 1~mmag.

\begin{figure}
  \centering
  \igraph{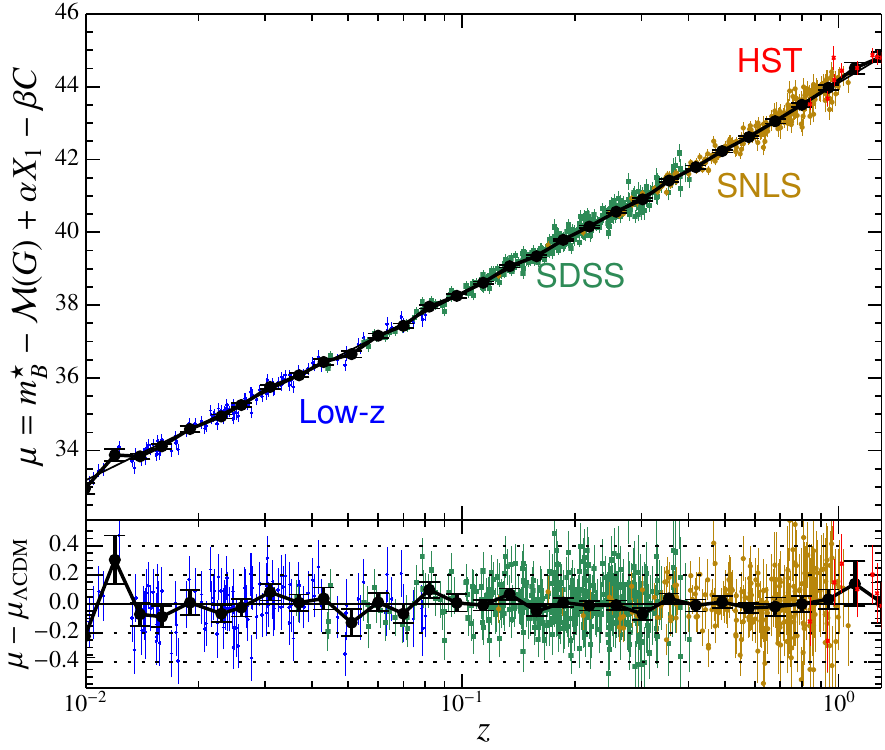}
  \caption{Binned version of the JLA Hubble diagram presented in
    Fig.~\ref{fig:hd}. The binned points are solid circles. Their are
    significant correlations between bins. The error bars are the
    square root of the diagonal of the covariance matrix given in
    Table~\ref{tab:cmub}.}
  \label{fig:hdb}
\end{figure}

\begin{figure}
  \centering
  \igraph{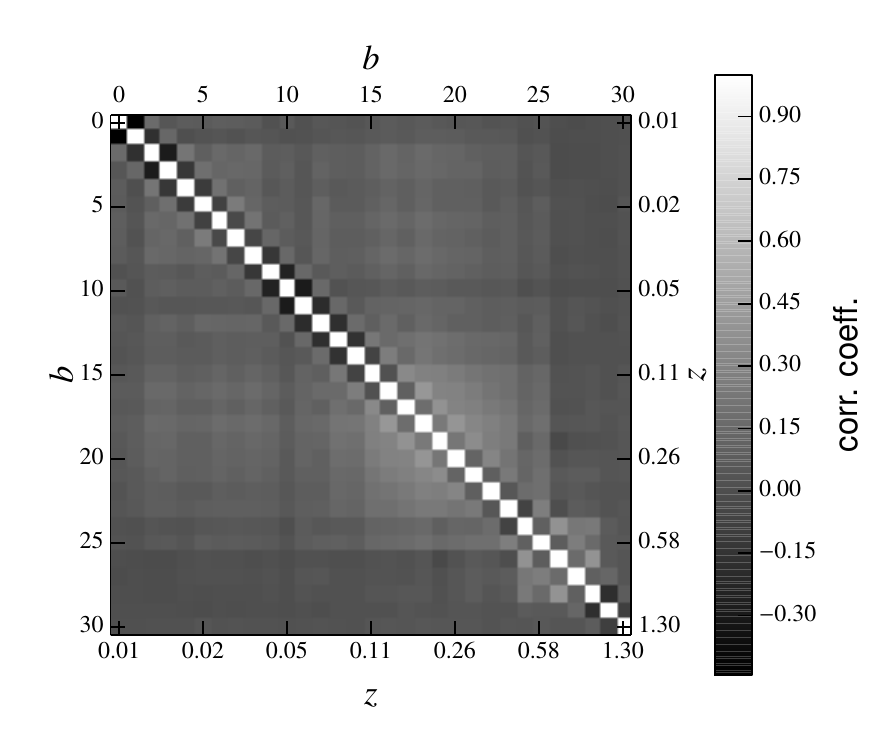}
  \caption{Correlation matrix of the binned distance modulus $\mu_b$.}
  \label{fig:corb}
\end{figure}

Such an interpolant can be fit to our measured Hubble diagram by
minimizing a likelihood function similar to the one proposed in
Eq. (\ref{eq:7}):
\begin{equation}
\label{eq:27}
  \chi^2 = (\hat{ \vec \mu} - \bar{\mu}(\vec z))^\dag \tens C^{-1} (\hat{ \vec \mu} - \bar{\mu}(\vec z))\,,
\end{equation}
The free parameters of the fit are $\alpha$, $\beta$, $\Delta_M$ and
$\mu_b$ at the chosen control points.  We use a fixed fiducial value
of $M^1_B = -19.05$ to provide uniquely determined $\mu_b$. Results
are compared to the best fit $\Lambda$-CDM cosmology in
Fig.~\ref{fig:hdb}. The structure of the correlation matrix of the
best-fit $\vec \mu_b$ is shown in Fig. \ref{fig:corb}. It displays
significant large scale correlation mostly due to systematic
uncertainties. The tri-diagonal structure arises from the linear
interpolation.

\subsection{Cosmology fit to the binned distances}
\label{sec:appr-likel}
\begin{figure}
  \centering
  \igraph{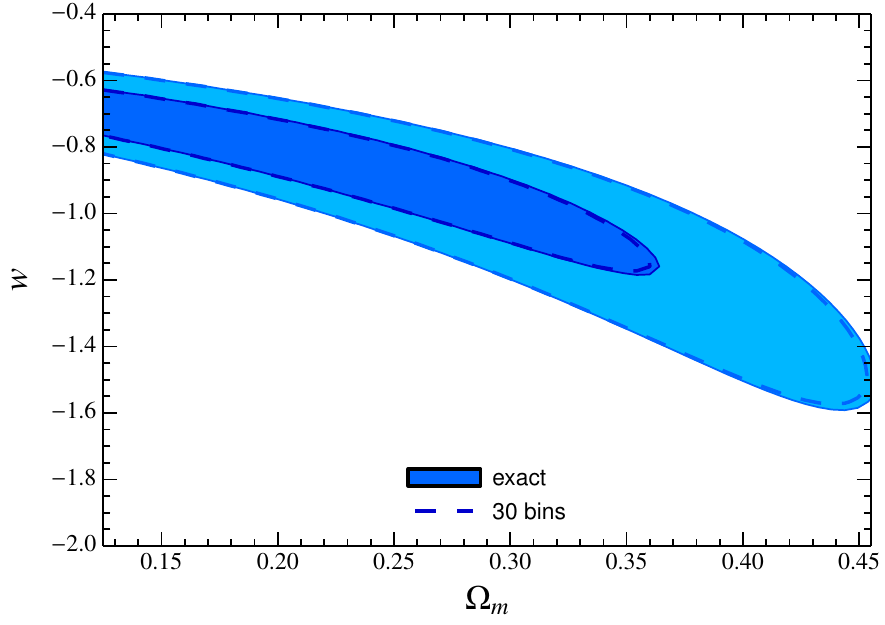}
  \caption{Comparison of the cosmological constraints obtained from
    the full JLA likelihood (filled contour) with approximate version
    derived by binning the JLA supernovae measurements in 20 bins
    (dashed blue contour) and 30 bins (continuous red
    contour).\label{fig:contour}}
\end{figure}

Cosmological models predicting isotropic luminosity distances
  evolving smoothly with redshifts can be fitted directly to the
  binned distance estimates. We denote $D_L(z; \theta)$ the luminosity
  distance predicted by a model dependent of a set of cosmological
  parameters $\theta$. A good approximation of the full JLA likelihood
  is generally given by the following likelihood function:
\begin{equation}
\label{eq:12}
  \chi^2(\theta, M) = \vec r^\dag \tens C_b^{-1} \vec r
\end{equation}
with:
\begin{equation}
\vec r=\vec \mu_b - M - 5\log_{10}D_L(\vec z_b; \theta)\,,\label{eq:23}
\end{equation}
$M$ a free normalization parameter, and $\tens C_b$ the covariance
matrix of $\vec \mu_b$ (see Table~\ref{tab:cmub}). As an illustration,
a comparison of the cosmological constraints obtained from the
approximate and full version of the JLA likelihood for the $w$-CDM
model is shown in Fig.~\ref{fig:contour}.
For the models evaluated in Sect.~\ref{sec:dark-energy-constr} in
combination with CMB and BAO constraints, the difference in best-fit
estimates between the approximate and full version is at most
$0.018\sigma$ and reported uncertainties differ by less than 0.3\%.

We warn that the normalization parameter $M$ must be left free in the
fit and marginalized over when deriving uncertainties. Not doing so
would be equivalent to introducing artificial constraints on the $H_0$
parameter and would result in underestimated errors.

\section{Data release}
\label{sec:data-release}

The light-curve fit parameters for the JLA sample are given in
Table~\ref{table:lcfit}. We provide the covariance matrices, described
in Sect.~\ref{sec:hubble-diagr-covar}, of statistical and systematic
uncertainties in light-curve parameters. These two products contain
all the information required to compute the likelihood function from
Eq.~(\ref{eq:7}) in a cosmological fit. We provide the necessary
computer code in two forms: a CosmoMC plugin and an independent C++
code.

Alternatively, we deliver estimates of binned distance modulus $\mu_b$
obtained, as described in appendix~\ref{sec:appr-jla-likel}, for 31
control points (30 bins) in Table~\ref{tab:mub} and the associated
covariance matrix in Table~\ref{tab:cmub}. These values can be used to
evaluate the approximate version of the JLA likelihood function
proposed in Eq.~(\ref{eq:12}).

In addition, we provide the retrained SALT2 model, the covariance
matrix of calibration parameters, and the SNLS recalibrated
light curves. The SDSS-II light curves can be obtained from the SDSS
SN data release \citep{SDSSRELEASE}. All data and software can be
retrieved from
\url{http://supernovae.in2p3.fr/sdss_snls_jla/ReadMe.html}.

\onecolumn
\begin{table}
  \centering
  \caption{Binned distance modulus fitted to the JLA sample.}
  \label{tab:mub}
    \begin{tabular}{*{2}{rr|}rr}
    \hline
    \hline
    $z_b$ & $\mu_b$ &$z_b$ & $\mu_b$ &$z_b$ & $\mu_b$ \\
\hline
0.010 & 32.9538  & 0.051 &  36.6511 &  0.257 &  40.5649\\
0.012 & 33.8790  & 0.060 &  37.1580 &  0.302 &  40.9052\\ 
0.014 & 33.8421  & 0.070 &  37.4301 &  0.355 &  41.4214\\
0.016 & 34.1185  & 0.082 &  37.9566 &  0.418 &  41.7909\\
0.019 & 34.5934  & 0.097 &  38.2532 &  0.491 &  42.2314\\
0.023 & 34.9390  & 0.114 &  38.6128 &  0.578 &  42.6170\\
0.026 & 35.2520  & 0.134 &  39.0678 &  0.679 &  43.0527\\
0.031 & 35.7485  & 0.158 &  39.3414 &  0.799 &  43.5041\\
0.037 & 36.0697  & 0.186 &  39.7921 &  0.940 &  43.9725\\
0.043 & 36.4345  & 0.218 &  40.1565 &  1.105 &  44.5140\\
&&&&                                        1.300 &44.8218\\
    \hline
  \end{tabular}
\tablefoot{An electronic version of this table, is available at the Centre de Données astronomiques de
  Strasbourg (CDS). It can also be downloaded at \url{http://supernovae.in2p3.fr/sdss_snls_jla/ReadMe.html}}
\end{table}
\begin{table}
  \centering
\caption{Covariance matrix of the binned distance modulus. }
  \label{tab:cmub}
  {\fontsize{6pt}{0.3em}
    \setlength{\arraycolsep}{1pt}
  
\begin{align*}10^{-6}\left(\begin{array}{*{31}{r}}
21282 & -10840 & 1918 & 451 & 946 & 614 & 785 & 686 & 581 & 233 & 881 & 133 & 475 & 295 & 277 & 282 & 412 & 293 & 337 & 278 & 219 & 297 & 156 & 235 & 133 & 179 & -25 & -106 & 0 & 137 & 168
\\
 & 28155 & -2217 & 1702 & 74 & 322 & 380 & 273 & 424 & 487 & 266 & 303 & 406 & 468 & 447 & 398 & 464 & 403 & 455 & 468 & 417 & 444 & 351 & 399 & 83 & 167 & -86 & 15 & -2 & 76 & 243
\\
 &  & 6162 & -1593 & 1463 & 419 & 715 & 580 & 664 & 465 & 613 & 268 & 570 & 376 & 405 & 352 & 456 & 340 & 412 & 355 & 317 & 341 & 242 & 289 & 119 & 152 & -69 & -33 & -44 & 37 & 209
\\
 &  &  & 5235 & -722 & 776 & 588 & 591 & 583 & 403 & 651 & 212 & 555 & 353 & 355 & 323 & 442 & 319 & 372 & 337 & 288 & 343 & 210 & 272 & 92 & 167 & -48 & -29 & -21 & 50 & 229
\\
 &  &  &  & 7303 & -508 & 1026 & 514 & 596 & 315 & 621 & 247 & 493 & 320 & 375 & 290 & 383 & 286 & 350 & 300 & 269 & 313 & 198 & 251 & 99 & 126 & 18 & 46 & 13 & 10 & 203
\\
 &  &  &  &  & 3150 & -249 & 800 & 431 & 358 & 414 & 173 & 514 & 231 & 248 & 221 & 293 & 187 & 245 & 198 & 175 & 231 & 126 & 210 & 103 & 170 & 51 & 66 & -8 & -51 & 308
\\
 &  &  &  &  &  & 3729 & -88 & 730 & 321 & 592 & 188 & 546 & 316 & 342 & 290 & 389 & 267 & 341 & 285 & 252 & 301 & 189 & 242 & 122 & 159 & 35 & 72 & 30 & 28 & 255
\\
 &  &  &  &  &  &  & 3222 & -143 & 568 & 421 & 203 & 491 & 257 & 280 & 240 & 301 & 221 & 275 & 227 & 210 & 249 & 148 & 220 & 123 & 160 & 43 & 69 & 27 & 7 & 253
\\
 &  &  &  &  &  &  &  & 3225 & -508 & 774 & 156 & 502 & 273 & 323 & 276 & 370 & 260 & 316 & 273 & 231 & 273 & 171 & 226 & 111 & 154 & 0 & 29 & 19 & 23 & 206
\\
 &  &  &  &  &  &  &  &  & 5646 & -1735 & 691 & 295 & 362 & 316 & 305 & 370 & 280 & 346 & 313 & 276 & 310 & 217 & 274 & 131 & 175 & 38 & 118 & 78 & 48 & 303
\\
 &  &  &  &  &  &  &  &  &  & 8630 & -1642 & 944 & 152 & 253 & 184 & 274 & 202 & 254 & 233 & 196 & 237 & 156 & 207 & 27 & 115 & -32 & 7 & -15 & 0 & 176
\\
 &  &  &  &  &  &  &  &  &  &  & 3855 & -754 & 502 & 225 & 278 & 294 & 274 & 285 & 253 & 239 & 255 & 173 & 229 & 181 & 177 & 93 & 124 & 132 & 108 & 227
\\
 &  &  &  &  &  &  &  &  &  &  &  & 4340 & -634 & 660 & 240 & 411 & 256 & 326 & 276 & 235 & 290 & 184 & 256 & 135 & 222 & 90 & 152 & 67 & 17 & 318
\\
 &  &  &  &  &  &  &  &  &  &  &  &  & 2986 & -514 & 479 & 340 & 363 & 377 & 362 & 315 & 343 & 265 & 311 & 144 & 198 & 17 & 62 & 86 & 147 & 226
\\
 &  &  &  &  &  &  &  &  &  &  &  &  &  & 3592 & -134 & 606 & 333 & 422 & 374 & 333 & 349 & 267 & 300 & 157 & 184 & 9 & 71 & 85 & 136 & 202
\\
 &  &  &  &  &  &  &  &  &  &  &  &  &  &  & 1401 & 22 & 431 & 343 & 349 & 302 & 322 & 245 & 284 & 171 & 186 & 70 & 70 & 93 & 142 & 202
\\
 &  &  &  &  &  &  &  &  &  &  &  &  &  &  &  & 1491 & 141 & 506 & 386 & 356 & 394 & 278 & 306 & 188 & 212 & 79 & 71 & 106 & 145 & 240
\\
 &  &  &  &  &  &  &  &  &  &  &  &  &  &  &  &  & 1203 & 200 & 435 & 331 & 379 & 281 & 311 & 184 & 209 & 49 & 51 & 110 & 197 & 181
\\
 &  &  &  &  &  &  &  &  &  &  &  &  &  &  &  &  &  & 1032 & 258 & 408 & 398 & 305 & 330 & 197 & 223 & 78 & 79 & 113 & 174 & 225
\\
 &  &  &  &  &  &  &  &  &  &  &  &  &  &  &  &  &  &  & 1086 & 232 & 453 & 298 & 328 & 120 & 189 & -48 & 22 & 42 & 142 & 204
\\
 &  &  &  &  &  &  &  &  &  &  &  &  &  &  &  &  &  &  &  & 1006 & 151 & 329 & 282 & 169 & 195 & 58 & 80 & 95 & 192 & 188
\\
 &  &  &  &  &  &  &  &  &  &  &  &  &  &  &  &  &  &  &  &  & 1541 & 124 & 400 & 199 & 261 & 150 & 166 & 202 & 251 & 251
\\
 &  &  &  &  &  &  &  &  &  &  &  &  &  &  &  &  &  &  &  &  &  & 1127 & 72 & 227 & 222 & 93 & 118 & 93 & 171 & 161
\\
 &  &  &  &  &  &  &  &  &  &  &  &  &  &  &  &  &  &  &  &  &  &  & 1723 & -105 & 406 & -3 & 180 & 190 & 198 & 247
\\
 &  &  &  &  &  &  &  &  &  &  &  &  &  &  &  &  &  &  &  &  &  &  &  & 1550 & 144 & 946 & 502 & 647 & 437 & 215
\\
 &  &  &  &  &  &  &  &  &  &  &  &  &  &  &  &  &  &  &  &  &  &  &  &  & 1292 & 187 & 524 & 393 & 387 & 284
\\
 &  &  &  &  &  &  &  &  &  &  &  &  &  &  &  &  &  &  &  &  &  &  &  &  &  & 3941 & 587 & 1657 & 641 & 346
\\
 &  &  &  &  &  &  &  &  &  &  &  &  &  &  &  &  &  &  &  &  &  &  &  &  &  &  & 2980 & 360 & 1124 & 305
\\
 &  &  &  &  &  &  &  &  &  &  &  &  &  &  &  &  &  &  &  &  &  &  &  &  &  &  &  & 4465 & -1891 & 713
\\
 &  &  &  &  &  &  &  &  &  &  &  &  &  &  &  &  &  &  &  &  &  &  &  &  &  &  &  &  & 23902 & -1826
\\
 &  &  &  &  &  &  &  &  &  &  &  &  &  &  &  &  &  &  &  &  &  &  &  &  &  &  &  &  &  & 19169
\\
\end{array}\right)
\end{align*}
}
\tablefoot{An electronic version of this table, is available at the Centre de Données astronomiques de
  Strasbourg (CDS). It can also be downloaded at \url{http://supernovae.in2p3.fr/sdss_snls_jla/ReadMe.html}}
\end{table}

\begin{longtable}{l|ccccc}
\caption{\label{table:lcfit}Parameters for the type Ia supernovae in the joint JLA cosmology sample.}\\
\hline\hline
Name & $z_{cmb}$ & $\mstar$ & $\xun$ & $\col$ & $M_{stellar}$ \\
\hline
\endfirsthead
\caption{continued.}\\
\hline\hline
Name & $z$ & $\mstar$ & $\xun$ & $\col$ & $M_{stellar}$ \\
\hline
\endhead
\hline
\endfoot
03D1ar & $0.002$ & $23.941 \pm 0.033$ & $-0.945 \pm 0.209$ & $0.266 \pm 0.035$ & $10.1 \pm 0.5$ \\

03D1au &$ 0.503$ & $23.002 \pm 0.088$ & $ 1.273 \pm 0.150$ &$  -0.012 \pm 0.030$ & $ 9.5 \pm 0.1$ \\ 
 03D1aw &$ 0.581$ & $23.574 \pm 0.090$ & $ 0.974 \pm 0.274$ &$  -0.025 \pm 0.037$ & $ 9.2 \pm 0.1$ \\ 
 03D1ax &$ 0.495$ & $22.960 \pm 0.088$ & $ -0.729 \pm 0.102$ &$  -0.100 \pm 0.030$ & $ 11.6 \pm 0.1$ \\ 
 03D1bp &$ 0.346$ & $22.398 \pm 0.087$ & $ -1.155 \pm 0.113$ &$  -0.041 \pm 0.027$ & $ 10.8 \pm 0.1$ \\ 
 03D1co &$ 0.678$ & $24.078 \pm 0.098$ & $ 0.619 \pm 0.404$ &$  -0.039 \pm 0.067$ & $ 8.6 \pm 0.3$ \\ 
 03D1dt &$ 0.611$ & $23.285 \pm 0.093$ & $ -1.162 \pm 1.641$ &$  -0.095 \pm 0.050$ & $ 9.7 \pm 0.1$ \\ 
 03D1ew &$ 0.866$ & $24.354 \pm 0.106$ & $ 0.376 \pm 0.348$ &$  -0.063 \pm 0.068$ & $ 8.5 \pm 0.8$ \\ 
 03D1fc &$ 0.331$ & $21.861 \pm 0.086$ & $ 0.650 \pm 0.119$ &$  -0.018 \pm 0.024$ & $ 10.4 \pm 0.0$ \\ 
 03D1fq &$ 0.799$ & $24.510 \pm 0.102$ & $ -1.057 \pm 0.407$ &$  -0.056 \pm 0.065$ & $ 10.7 \pm 0.1$ \\ 
 03D3aw &$ 0.450$ & $22.667 \pm 0.092$ & $ 0.810 \pm 0.232$ &$  -0.086 \pm 0.038$ & $ 10.7 \pm 0.0$ \\ 
 03D3ay &$ 0.371$ & $22.273 \pm 0.091$ & $ 0.570 \pm 0.198$ &$  -0.054 \pm 0.033$ & $ 10.2 \pm 0.1$ \\ 
 03D3ba &$ 0.292$ & $21.961 \pm 0.093$ & $ 0.761 \pm 0.173$ &$  0.116 \pm 0.035$ & $ 10.2 \pm 0.1$ \\ 
 03D3bl &$ 0.356$ & $22.927 \pm 0.087$ & $ 0.056 \pm 0.193$ &$  0.205 \pm 0.030$ & $ 10.8 \pm 0.1$ \\ 
 03D3cd &$ 0.461$ & $22.575 \pm 0.096$ & $ 1.862 \pm 0.565$ &$  -0.043 \pm 0.038$ & $ 9.3 \pm 0.2$ \\ 
 03D4ag &$ 0.284$ & $21.257 \pm 0.087$ & $ 0.937 \pm 0.105$ &$  -0.085 \pm 0.023$ & $ 10.6 \pm 0.1$ \\ 
 03D4at &$ 0.632$ & $23.739 \pm 0.093$ & $ 0.209 \pm 0.330$ &$  -0.051 \pm 0.067$ & $ 8.8 \pm 0.1$ \\ 
 03D4au &$ 0.466$ & $23.790 \pm 0.090$ & $ 0.377 \pm 0.333$ &$  0.122 \pm 0.043$ & $ 9.5 \pm 0.1$ \\ 
 03D4cj &$ 0.269$ & $21.058 \pm 0.086$ & $ 1.151 \pm 0.085$ &$  -0.080 \pm 0.023$ & $ 6.0 \pm 5.0$ \\ 
 03D4cx &$ 0.947$ & $24.460 \pm 0.115$ & $ -0.096 \pm 0.673$ &$  0.057 \pm 0.065$ & $ 11.0 \pm 0.2$ \\ 
 03D4cy &$ 0.925$ & $24.706 \pm 0.125$ & $ 0.863 \pm 0.640$ &$  -0.058 \pm 0.072$ & $ 9.7 \pm 0.2$ \\ 
 03D4cz &$ 0.693$ & $24.032 \pm 0.100$ & $ -1.764 \pm 0.385$ &$  -0.077 \pm 0.085$ & $ 10.4 \pm 0.2$ \\ 
 03D4dh &$ 0.625$ & $23.387 \pm 0.091$ & $ 1.128 \pm 0.184$ &$  -0.043 \pm 0.050$ & $ 9.4 \pm 0.2$ \\ 
 03D4di &$ 0.897$ & $24.333 \pm 0.110$ & $ 1.454 \pm 0.416$ &$  -0.084 \pm 0.064$ & $ 9.9 \pm 0.1$ \\ 
 03D4dy &$ 0.608$ & $23.245 \pm 0.091$ & $ 1.138 \pm 0.183$ &$  -0.098 \pm 0.036$ & $ 5.3 \pm 52.9$ \\ 
 03D4fd &$ 0.789$ & $24.222 \pm 0.100$ & $ 0.784 \pm 0.472$ &$  -0.043 \pm 0.066$ & $ 10.0 \pm 0.2$ \\ 
 03D4gf &$ 0.578$ & $23.324 \pm 0.090$ & $ 0.509 \pm 0.312$ &$  -0.036 \pm 0.038$ & $ 7.6 \pm 0.3$ \\ 
 03D4gg &$ 0.590$ & $23.431 \pm 0.093$ & $ 1.001 \pm 0.435$ &$  0.012 \pm 0.040$ & $ 10.2 \pm 0.0$ \\ 
 04D1aj &$ 0.720$ & $23.899 \pm 0.096$ & $ 0.391 \pm 0.357$ &$  -0.034 \pm 0.067$ & $ 6.0 \pm 5.0$ \\ 
 04D1dc &$ 0.210$ & $21.057 \pm 0.086$ & $ -1.236 \pm 0.062$ &$  -0.003 \pm 0.025$ & $ 10.6 \pm 0.0$ \\ 
 04D1de &$ 0.767$ & $24.132 \pm 0.097$ & $ 0.906 \pm 0.244$ &$  -0.119 \pm 0.056$ & $ 9.6 \pm 0.2$ \\ 
 04D1ff &$ 0.859$ & $24.246 \pm 0.102$ & $ 0.728 \pm 0.277$ &$  0.047 \pm 0.058$ & $ 8.7 \pm 0.3$ \\ 
 04D1hd &$ 0.368$ & $22.157 \pm 0.086$ & $ 0.789 \pm 0.068$ &$  -0.092 \pm 0.022$ & $ 8.2 \pm 0.1$ \\ 
 04D1hx &$ 0.559$ & $23.700 \pm 0.090$ & $ 0.285 \pm 0.192$ &$  0.112 \pm 0.035$ & $ 9.6 \pm 0.1$ \\ 
 04D1hy &$ 0.849$ & $24.295 \pm 0.101$ & $ 1.151 \pm 0.276$ &$  -0.054 \pm 0.055$ & $ 8.1 \pm 0.8$ \\ 
 04D1iv &$ 0.996$ & $24.618 \pm 0.112$ & $ 1.273 \pm 0.343$ &$  -0.131 \pm 0.051$ & $ 9.0 \pm 0.5$ \\ 
 04D1jd &$ 0.777$ & $24.398 \pm 0.098$ & $ 0.135 \pm 0.287$ &$  0.099 \pm 0.070$ & $ 10.4 \pm 0.1$ \\ 
 04D1jg &$ 0.583$ & $23.273 \pm 0.090$ & $ 0.265 \pm 0.159$ &$  -0.105 \pm 0.034$ & $ 11.3 \pm 0.2$ \\ 
 04D1kj &$ 0.584$ & $23.336 \pm 0.089$ & $ 0.218 \pm 0.126$ &$  -0.063 \pm 0.030$ & $ 9.9 \pm 0.1$ \\ 
 04D1ks &$ 0.797$ & $24.129 \pm 0.098$ & $ 0.699 \pm 0.263$ &$  0.094 \pm 0.062$ & $ 8.0 \pm 0.7$ \\ 
 04D1oh &$ 0.589$ & $23.396 \pm 0.091$ & $ -0.067 \pm 0.207$ &$  -0.069 \pm 0.033$ & $ 9.2 \pm 0.2$ \\ 
 04D1ow &$ 0.913$ & $24.344 \pm 0.105$ & $ -0.044 \pm 0.286$ &$  -0.161 \pm 0.052$ & $ 9.4 \pm 1.4$ \\ 
 04D1pc &$ 0.769$ & $24.546 \pm 0.097$ & $ -0.334 \pm 0.339$ &$  0.085 \pm 0.064$ & $ 10.4 \pm 0.2$ \\ 
 04D1pd &$ 0.948$ & $24.677 \pm 0.112$ & $ 0.183 \pm 0.422$ &$  0.041 \pm 0.061$ & $ 10.2 \pm 0.1$ \\ 
 04D1pg &$ 0.514$ & $23.573 \pm 0.090$ & $ 0.986 \pm 0.196$ &$  0.103 \pm 0.036$ & $ 9.8 \pm 0.2$ \\ 
 04D1pp &$ 0.734$ & $23.985 \pm 0.094$ & $ -1.345 \pm 0.223$ &$  -0.117 \pm 0.053$ & $ 11.4 \pm 0.1$ \\ 
 04D1pu &$ 0.638$ & $24.012 \pm 0.101$ & $ -1.611 \pm 0.377$ &$  0.107 \pm 0.083$ & $ 9.3 \pm 0.2$ \\ 
 04D1qd &$ 0.766$ & $24.244 \pm 0.096$ & $ 0.181 \pm 0.273$ &$  0.012 \pm 0.054$ & $ 10.5 \pm 0.1$ \\ 
 04D1rh &$ 0.435$ & $22.556 \pm 0.087$ & $ 0.828 \pm 0.210$ &$  -0.057 \pm 0.026$ & $ 9.9 \pm 0.1$ \\ 
 04D1rx &$ 0.983$ & $24.760 \pm 0.114$ & $ 0.595 \pm 0.449$ &$  -0.099 \pm 0.059$ & $ 9.8 \pm 0.1$ \\ 
 04D1sa &$ 0.584$ & $23.564 \pm 0.092$ & $ -0.443 \pm 0.289$ &$  -0.066 \pm 0.034$ & $ 10.6 \pm 0.1$ \\ 
  & $\vdots$ & $\vdots$ & $\vdots$ & $\vdots$ & $\vdots$\\\end{longtable}
\noindent
{\bf Notes.} An electronic version of this table, including the covariances, is available at the {\it Centre de Donn\'ees astronomiques de Strasbourg} (CDS). The SALT2 fit parameters can also be downloaded at \url{http://supernovae.in2p3.fr/sdss_snls_jla/ReadMe.html}.\\
\twocolumn

\end{document}